\RequirePackage{fix-cm}

\documentclass[final]{svjour3}                     
\pdfoutput=1
\smartqed  
\usepackage{graphicx}
\usepackage{epstopdf} 
\usepackage[numbers]{natbib} 
\usepackage{graphicx}
\usepackage{placeins} 
\usepackage{amsmath} 
\usepackage{subfig}

\begin{document}

\title{Translational Motion Compensation for Soft Tissue Velocity Images 
}


\author{Christina Koutsoumpa \and Jennifer Keegan \and David Firmin \and Guang-Zhong Yang \and Duncan Gillies
}


\institute{C. Koutsoumpa and G.-Z. Yang \at
              The Hamlyn Centre for Robotic Surgery, Imperial College London, United Kingdom \\
              \email{c.koutsoumpa@imperial.ic.ac.uk}                        
             \and
           	D. Gillies \at Department of Computing, Imperial College London, United Kingdom
			\and
           	J. Keegan and D. Firmin\at
           	Cardiovascular Biomedical Research Unit, Royal Brompton Hospital, London, United Kingdom               
}

\date{Submitted: 17 August 2018 }

\maketitle

\begin{abstract}
\textit{Purpose}
Advancements in MRI Tissue Phase Velocity Mapping (TPM) allow for the acquisition of higher quality velocity cardiac images providing better assessment of regional myocardial deformation for accurate disease diagnosis, pre-operative planning and post-operative patient surveillance. Translation of TPM velocities from the scanner's reference coordinate system to the regional cardiac coordinate system requires decoupling of translational motion and motion due to myocardial deformation. Despite existing techniques for respiratory motion compensation in TPM, there is still a remaining translational velocity component due to the global  motion of the beating heart. To compensate for translational motion in cardiac TPM, we propose an image-processing method, which we have evaluated on synthetic data and applied on in vivo TPM data.
\newline
\textit{Methods}
Translational motion is estimated from a suitable region of velocities automatically defined in the left-ventricular volume. The region is generated by dilating the medial axis of myocardial masks in each slice and the translational velocity is estimated by integration in this region. The method was evaluated on synthetic data and in vivo data corrupted with a translational velocity component (200\% of the maximum measured velocity). Accuracy and robustness were examined and the method was applied on 10 in vivo datasets.
\newline
\textit{Results}
The results from synthetic and in vivo corrupted data show excellent performance with an estimation error less than 0.3\% and high robustness in both cases. The effectiveness of the method is confirmed with visual observation of results from the 10 datasets.
\newline
\textit{Conclusion}
The proposed method is accurate and suitable for translational motion correction of the left ventricular velocity fields. The current method for translational motion compensation could be applied to any annular contracting (tissue) structure. 
\end{abstract}

\section{Introduction}

Estimation of tissue motion, deformation and position often plays an important role not only in the diagnosis of a tissue condition but also in the effectiveness of a computer assisted intervention (CAI) system. Medical imaging provides the necessary visual feedback for spatial navigation and guidance in CAI. Medical images can be acquired pre-operatively with CT or MRI to obtain a detailed visualisation of a pathology, which is then registered on the corresponding intra-operative data to transfer the detailed visual information to clinicians during the intervention. In this process, organ shift and soft tissue deformation that may occur before, at the start or during interventions need to be taken into consideration to ensure accuracy in the navigation and guidance \cite{Mountney2010}. This consideration is particularly important in abdominal and cardiothoracic surgery, where respiration and heart beating involve significant tissue motion that affects not only the participating organs but also the surrounding tissues.  Previously presented strategies for the compensation of soft tissue motion in CAI include marker tracking, gating and real-time image registration \cite{Meinzer2008, Yip2012} techniques, which are either invasive, complex or of limited accuracy. The accurate non-invasive soft tissue motion estimation and compensation in an acceptably short time for real-time applications still remains a challenge. 

Among the established medical imaging techniques for measuring tissue motion and deformation, Tissue Phase Velocity Mapping (TPM) presents significant advantages as it provides direct measurements of tissue velocities in three orthogonal directions with high spatial and temporal resolution \cite{Nayak2015}. TPM yields velocities in reference to the scanner's coordinate system and, therefore, it captures motion that is attributed not only to tissue deformation but also to other sources of motion such as bulk movement of the heart due to beating, respiration and patient movement in the scanner. Moreover TPM data is affected by phase errors attributed to gradient eddy current residuals, concomitant field terms and gradient field distortions that appear as velocity offsets \cite{Nayak2015, Ferreira2013}. Velocity offsets vary smoothly across the image introducing an apparent translational velocity component to the measured TPM velocities.

For accurate assessment of myocardial deformation, correction of TPM for actual and apparent translational velocity is necessary. To this end, we propose a novel image processing translational motion compensation method, in two versions depending on the clinical study. The method is evaluated for effectiveness and robustness on synthetic, semi-synthetic velocity datasets and in vivo TPM datasets from 10 healthy volunteers. Results indicate excellent performance and robustness of the proposed method. Its applicability can be extended to any annular structure of soft tissue.

\section{Methods}

\subsection{Theoretical analysis of velocities}
Let $\vec{U} = \vec{U} (X,Y,Z)$ be the velocity of a tissue voxel measured with TPM at a position $(X,Y)$, in the  $Z-th$ slice and $T-{th}$ frame of the cardiac cycle.  $\vec{U}$ can be analysed into the following components
\begin{equation}\label{eq:1}
\vec{U} =\vec{U}_{motion}+\vec{U}_{offset}+ \vec{U}_{noise}
\end{equation}
where $\vec{U}_{motion}$ is the voxel's actual velocity expressed in the scanner's reference system,$\vec{U}_{offset}$ is the apparent velocity due to phase offset errors and  $\vec{U}_{noise}$ is the noise in velocity measurements. $\vec{U}_{noise}$ can be compensated using the TV Restoration method presented in \cite{Koutsoumpa2015} and it is not considered further in the study. Since offset errors vary slowly across the image, they account for an apparent translational velocity in the tissue velocity measurements which is also compensated by our method. For the moment, it is assumed that $\vec{U}_{offset}$ term is incorporated into the translational motion or alternatively dismissed, thus
\begin{equation}\label{eq:2}
\vec{U}\approx\vec{U}_{motion}+\vec{U}_{offset}\approx\vec{U}_{motion} 
\end{equation}

Tissue velocity expressed in a local coordinate system (rather than the scanner's reference system) is a more intuitive representation of the regional tissue deformation and this is what is defined as \textit{velocity from deformation} in the present study. The current motion compensation method relies on the conversion of velocities from the scanner's reference coordinate system to a local coordinate system adjacent to the heart according to the equation 

\begin{equation}\label{eq:3}
\vec{U}_{motion}=\vec{U}_{transl}+\vec{u} 
\end{equation}

where $\vec{U}_{motion}=\vec{U}_{motion}(X,Y,Z)$ is the voxel's velocity expressed in the scanner's reference system, $\vec{u}$ is the voxel's velocity expressed in the local LV system and $\vec{U}_{transl}$ is the relative velocity of the local system (LV) with respect to the reference system (scanner). In other words, the goal of the motion compensation method is to estimate the translational velocity component $\vec{U}_{transl}$ and subtract it from the initial velocity measurements $\vec{U}_{motion}$.

The definition of $\vec{u}$ is not unequivocal as it depends on the definition of the local LV coordinate system. The local LV system, in turn, is defined based on what the clinician is trying to observe. In general the local LV system is an orthogonal coordinate system aligned with the short axis of the LV following the LV bulk motion. For clinical studies where deformation is examined across the entire volume simultaneously, a local LV system centered around the centre of mass $\vec{M}$ of the LV (V-local system) is more convenient. On the other hand, in clinical studies where deformation is examined slice by slice a local LV system centered around the centre of mass $\vec{m}(Z)$ of that specific slice Z (S-local system) is recommended.


In the V-local system $\vec{U}_{transl}=\vec{U}^{(V)}_{transl}$ expresses the bulk translational motion of the LV as seen from the scanner and it coincides with the velocity of $\vec{M}$ in the scanner's reference system. Likewise, in the S-local system $\vec{U}_{transl}=\vec{U}^{(Sz)}_{transl}$ expresses the bulk translational motion of the $Z-th$ slice as seen from the scanner and it coincides with the velocity of $\vec{m(Z)}$ in the scanner's reference system. The difference between $\vec{U}^{(V)}_{transl}$ and $\vec{U}^{(Sz)}_{transl}$ is the relative velocity of $\vec{m}(z)$ with respect to $\vec{M}$. Due to the assumption made with equation Eq.\ref{eq:2}, the difference $\vec{U}^{(V)}_{transl}-\vec{U}^{(Sz)}_{transl}$ may incorporate differences in the offset errors across different slices as offset errors may vary across the longitudinal LV axis. 

\subsection{Velocity symmetry}
The deformation of the normal LV is governed by a certain degree of symmetry, as it can be observed from previous studies on left ventricular shape and function. Short axis LV segments (according to the American Heart Association - AHA model) show pairwise similarity in velocity time courses \cite{Codreanu2014} and systolic deformation parameters (strain, strain rate, velocity, displacement) \cite{Augustine2013}. Likewise AHA velocity plots \cite{Jung2012} manifest pairwise segmental similarity indicating planar or cylindrical symmetry of myocardial deformation, a hypothesis that is adequately supported by finer segmental velocity plots \cite{Foll2009} and pixel-wise velocity maps \cite{Jung2006c}. Velocities in the LV base and mid-wall match sufficiently the assumption of planar symmetry while apical velocities indicate either cylindrical or planar symmetry, (with poor agreement between the anterior and inferior segments.)

For this study the assumption of myocardial velocity field symmetry is held and for this, two models are considered: \textbf{Model A} is a model of cylindrical velocity vector symmetry around the LV short axis (lines interpolating the centres of mass $\vec{m}(Z)$. \textbf{Model B} is a combined model of i) planar velocity magnitude symmetry with respect to the bisectal plane of the 3D angle formed between the atrioventricular grooves and short axis of the LV and ii) cylindrical velocity directions symmetry similar to model A.


\subsection{Motion correction method}
Assuming either of the two types of velocity symmetry, $\vec{U}_{transl}$ is estimated by integration of $\vec{U}_{motion}$ in an appropriate region $V$ contained in the LV and Eq.\ref{eq:3} becomes

\begin{equation}\label{eq:7}
\int_{V}\vec{U}_{motion}dV=\int_{V}\vec{U}_{transl}dV+\int_{V}\vec{u}dV
\end{equation}

With appropriate selection of the region of integration the $\int_{V}\vec{u}dV$ term is nulled (see proof below) and since $\vec{U}_{transl}$ is assumed constant across the entire region $V$, $\vec{U}_{transl}$ is estimated as  
\begin{equation}\label{eq:8}
\vec{U}_{transl}=\dfrac{1}{V}\int_{V}\vec{U}_{motion}dV=mean_{\{V\}}(\vec{U}) 
\end{equation}
and the velocity from deformation is
\begin{equation}\label{eq:9}
\vec{u}=\vec{U}_{motion}-\vec{U}_{transl}
\end{equation}
 
\paragraph{Region of integration.}
In a LV model of the shape of a perfect hollow truncated ellipsoid, the region of integration (ROI) is defined as a subregion of the LV of similar shape and smaller width aligned with the LV short axis. The cross sections of this region with the slice planes are rings of various radii and constant  width.  It is important that the ROI is equi-width and equidistant from the LV endocardial and epicardial surfaces in order that the error in integration is minimum. To take into consideration variations of the real LV shape from this LV shape model, the ROI is adjusted to another equiwidth and equidistant region sitting in the LV wall. This region is formed by closed straps around the medial axis of the LV cross sections on each slice, which are extended along the longitudinal direction.  The closed straps (also called \textit{rings} in this work), are computed as the dilated skeleton of the myocardial mask on each slice. 

\paragraph{Proof.}

\begin{figure}
\centering
\subfloat[ROI in myocardial cross section]{\includegraphics[height = 1in]{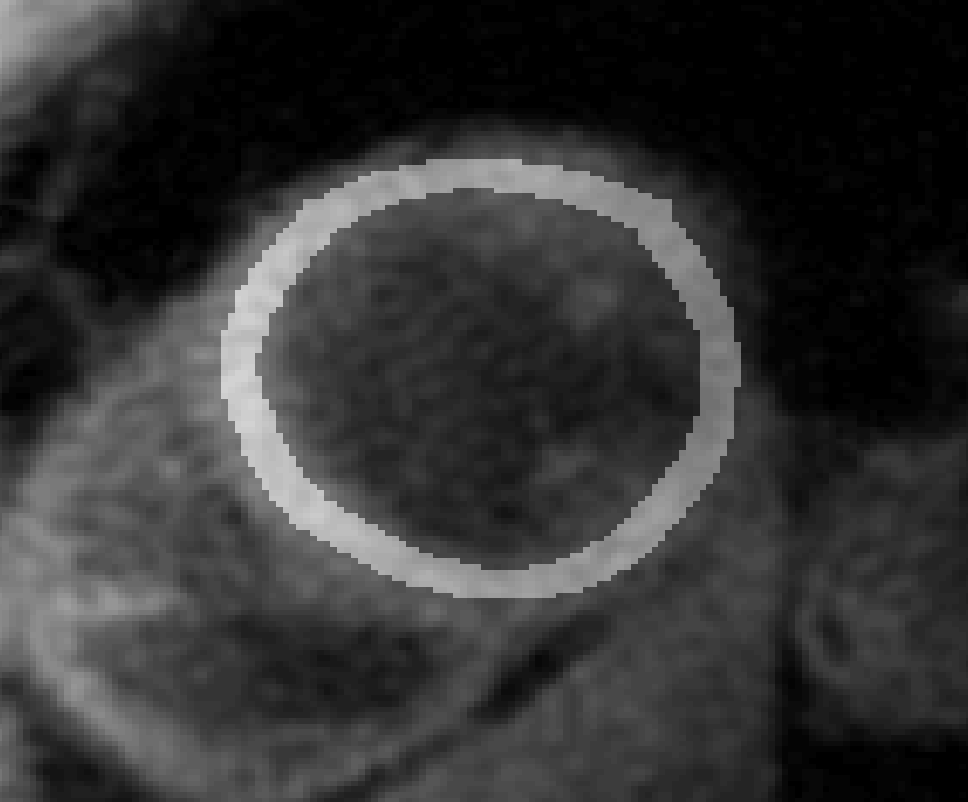}} \quad
\subfloat[mask, ROI and symmetry A]{\includegraphics[height = 1in]{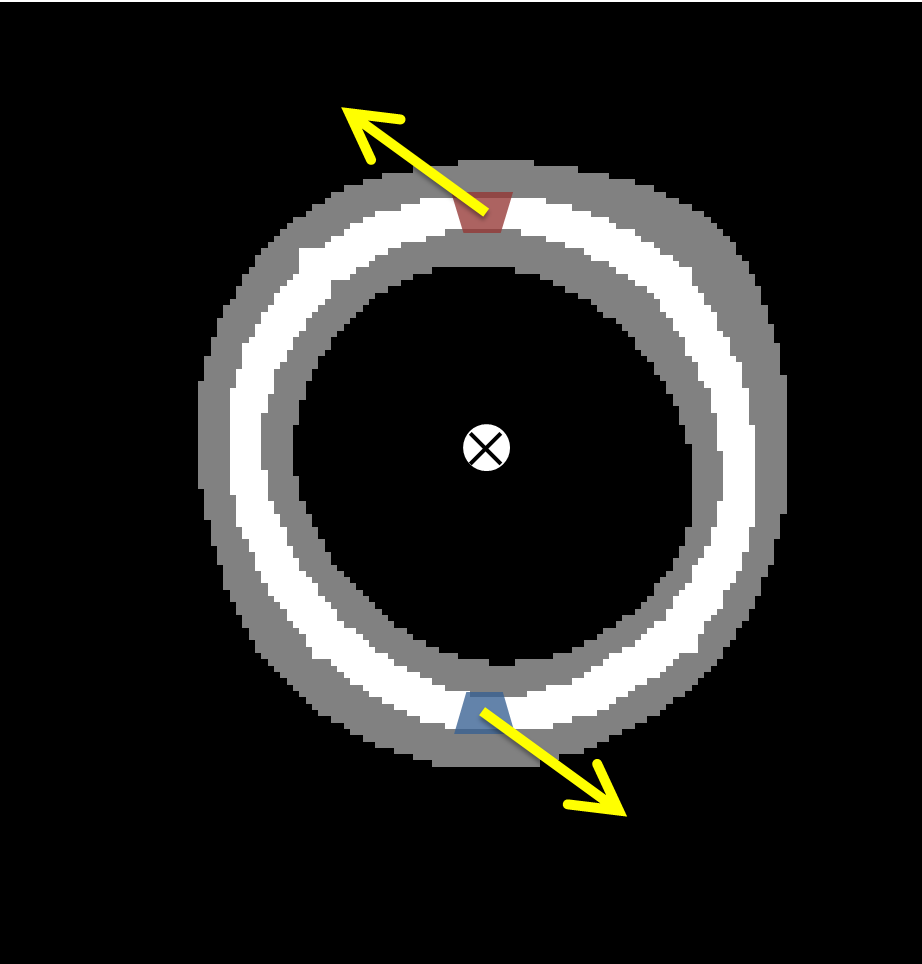}}\quad
\subfloat[mask, ROI and symmetry B]{\includegraphics[trim={0cm 0cm 0cm 0.7cm},clip,height = 1in]{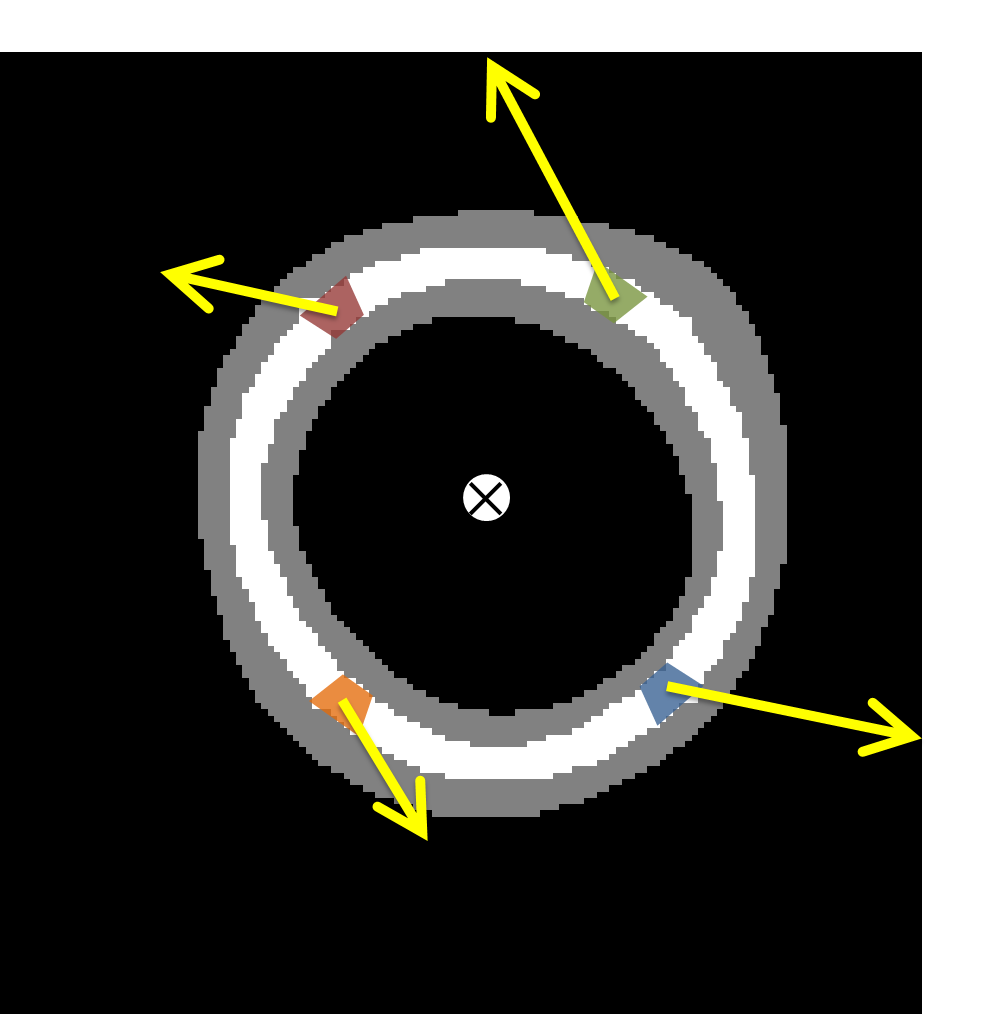}}
\caption{Region of integration (ROI) on a short axis slice (a) superimposed on the corresponding TPM image and (b-c) superimposed on the corresponding mask, showing pairs of symmetric infinitesimal myocardial segments in case of (a) cylindrical and (b) planar symmetry.}
\label{fig:ROI}
\end{figure}

Consider a ring (ROI) from the Z-th slice of a real dataset with area S, as shown in Fig.\ref{fig:ROI}. Let $dS_{1}$ (in blue, Fig.\ref{fig:ROI}a) and $dS_{2}$ (in red, Fig.\ref{fig:ROI}a)be two infinitesimal segments of $S$ that are symmetric about the centre. Then $dS_{1} = wd\theta_{1}$ and $dS_{2} = wd\theta_{2}$, where $w$ is the constant width of the ring by definition and $d\theta_{1}$, $d\theta_{2}$ are the circumferential infinitesimal lengths of $dS1$ and $dS2$ respectively. Because there is neglegable varitation of the mask along the circumference of the ring, it can be considered that $wd\theta_{1} \approx wd\theta_{2}$, hence $dS_{1} \approx dS_{2}$. If $\vec{u}_{1}$ and $\vec{u}_{2}$ are the velocities at $dS_1$ and $dS_2$ respectively and model A symmetry applies, $\vec{u}_{1} = -\vec{u}_{2}$ and  thus  $\vec{u}_1dS_{1} \approx -\vec{u}_2dS_{2}$ or $\vec{u}_1dS_{1}+\vec{u}_2dS_{2} \approx 0$. Integrating the latest relationship, it is proved that $\int_{S}\vec{u}dS\approx0$, for each slice. If ROI is defined in the 3D space $\int_{S}\vec{u}dS=\sum_z(\int_{S_z}\vec{u}dS_z)\approx0$.

In case of model B symmetry, it is also proved that $\int_{S}\vec{u}dS\approx0$. Let $dS_{1}$ (in blue, Fig.\ref{fig:ROI}c), $dS_{2}$ (in green, Fig.\ref{fig:ROI}c, $dS_{3}$ (in red, Fig.\ref{fig:ROI}c) and $dS_{4}$ (in orange, Fig.\ref{fig:ROI}c) be four infinitesimal segments of $S$ with $\{dS_1,dS_2\}$ and $\{dS_3,dS_4\}$ located symmetrically in pairs about the bisector line as defined in model B, $\{dS_1,dS_3\}$ and $\{dS_2,dS_4\}$ located symmetrically in pairs about the centre. As above $dS_{1} \approx dS_{2}\approx dS_{3} \approx dS_{4}$. If $\vec{u}_{1},\vec{u}_{2},\vec{u}_{1},\vec{u}_{2}$ are the corresponding  velocities at $dS_1,dS_2,dS_3,dS_4$ and since model B symmetry applies, the following relationships for the velocity magnitudes $u_{1} = u_{2},u_{3} = u_{4}$ and velocity directions $\hat{u}_{1} = - \hat{u}_{3},\hat{u}_{2} = -\hat{u}_{4}$ occur. For this, the sum $\vec{u}_{1}+\vec{u}_{2}+\vec{u}_{1}+\vec{u}_{2}=0$, hence as above $\vec{u}_1dS_{1}+\vec{u}_2dS_{2}+\vec{u}_3dS_{3}+\vec{u}_4dS_{4} \approx 0$. By integration over the entire $S$ it is proved that $\int_{S}\vec{u}dS\approx0$ and thus $\int_{S}\vec{u}dS=\sum_z(\int_{S_z}\vec{u}dS_z)\approx0$.

\paragraph{Motion correction.}
For this study, two versions of the motion correction method were implemented and evaluated. For the first variation \textbf{method 1}, the velocity from deformation $\vec{u}$ is expressed in the S-local coordinate system on each slice, the ROI is a 2D region within the myocardial mask on each slice, from which the mean velocity is calculated per frame and slice and then subtracted from the initial velocity image. For the second variation \textbf{method 2}, the velocity from deformation $\vec{u}$ is expressed in the V-local coordinate system, the ROI is a 3D region within the myocardium, from which the mean velocity is calculated per frame across all slices and then subtracted from the initial velocity images. 
\section{Experiments}
\subsection{Numerical validation}

To evaluate the effectiveness and robustness of the proposed motion correction algorithms, synthetic and semi-synthetic velocity datasets were constructed and corrupted with a large known translational velocity.

\paragraph{Simulated synthetic data.}
The simulated synthetic velocity datasets were generated with the following specifications: i) the myocardial mask is a hollow disk centred at Cm, ii) the 2D velocity field has the symmetry of model A or model B and is centered at Cv, iii) Cm and Cv coincide, iv) velocity vectors increase progressively from the outer boundary (epicardium) to the inner boundary (endocardium) to model the higher deformation of the endocardium as compared to the epicardium and v) velocity directions evolve periodically over time passing through the phases of anti-clockwise rotations, rotation, clock-wise rotation and expansion. 

Let: 
\begin{equation}\label{eq:synthetic}
\vec{u}=u \hat{u}T
\end{equation}
be the velocity of a pixel in the simulated dataset at time t, $u$ the magnitude of velocity at each pixel, $\hat{u}$ the unit vector in the direction of velocity at each pixel and $T(t)$ a sinuisoidal term attributing periodicity over time $t$. For \textbf{model A} the terms in Eq.\ref{eq:synthetic} become $u=u_r(r)$ and $\hat{u}=\hat{u}(\theta)$, where $u_r(r)$ is reversely proportional to the distance $r$ of each pixel from Cv and $\hat{u}(\theta)$ is a function of the angle $\theta$ formed between the line connecting Cv with each pixel and the horizontal axis. In other words the vectors $ \vec{u}$ are symmetric about the centre of the vector field Cv. For \textbf{model B} the terms in Eq.\ref{eq:synthetic} become  $u=u_r (r) u_\theta (\theta) $ and $\hat{u}=\hat{u}(\theta)$, where $u_\theta (\theta)$ is a magnitude term than varies smoothly along the circumferential direction of the dataset nulling at the positions of the suppositional atrioventricular grooves and $u_r (r)$ and $\hat{u}(\theta)$ are defined as above. In this case the velocity magnitude $u$ is symmetric about the bisector of the angle formed between the centre and the suppositional atrioventricular grooves whereas vector directions $\hat{u}$ maintain the symmetry about the centre Cv.

Translational velocity of large magnitude (200\% of maximum u) was added to the synthetic velocity field to form the \textit{uncorrected} dataset. To examine robustness, \textit{distorted} synthetic datasets were generated from the above synthetic datasets with distortion of the mask symmetry and misalignment between the mask and the velocity field. The motion correction method 1 was then applied.

\paragraph{Realistic semi-synthetic data.}
To take into consideration other deviations from symmetry and unmodelled characteristics of realistic data, the motion correction method was validated (or also evaluated) on a realistic semi-synthetic dataset. The semi-synthetic dataset was created from an in vivo 3D-t TPM velocity dataset that was used as a template on which a known translational velocity was added. Motion correction methods 1 \& 2 were the applied and the corrected velocity fields were compared to the template. 

\subsection{In vivo study}

For in vivo validation, TPM velocity data from 10 healthy volunteers were analysed. Each subject was scanned with the breath-hold spiral PC-MRI sequence described in \cite{Simpson2013a} on a Siemens 3 Tesla scanner (MAGNETOM Skyra, Siemens AG Healthcare Sector, Germany). Nine contiguous short-axis slices were acquired with an in-plane spatial resolution of 1.7 x 1.7mm (reconstructed to 0.85 x 0.85 mm), slice thickness of 8mm, three-directional velocity encoding with in-plane and through-plane velocity sensitivity (VENC) of 20 cm/sec and 30cm/sec respectively. Fifty frames per slice were reconstructed covering a full cardiac cycle and equally spaced through it.  Blood flow and respiration artefact were supressed with a black blood preparation pulse and breath-hold acquisitions correspondingly.

TPM data were processed prior to motion correction. Pre-processing includes alignment of images to a common Cartesian coordinate system, conversion of phase measurements to velocities according to the VENC parameters and restoration of velocities for noise with the TV Restoration-based method described in \cite{Koutsoumpa2015}. For visualization purposes the myocardial region on each image was semi-automatically delineated. The proposed translational motion correction method was then applied. Analysis and visualization of all images were performed with in-house software built in MATLAB (The Mathworks Inc, Natick, MA) for this study. 

\section{Results}
\begin{figure}[h!b]
\captionsetup[subfigure]{labelformat=empty}
\centering
\subfloat[Ground Truth]{\includegraphics[trim={1.1cm 1.1cm 1.1cm 1.1cm},clip,width = 0.77in]{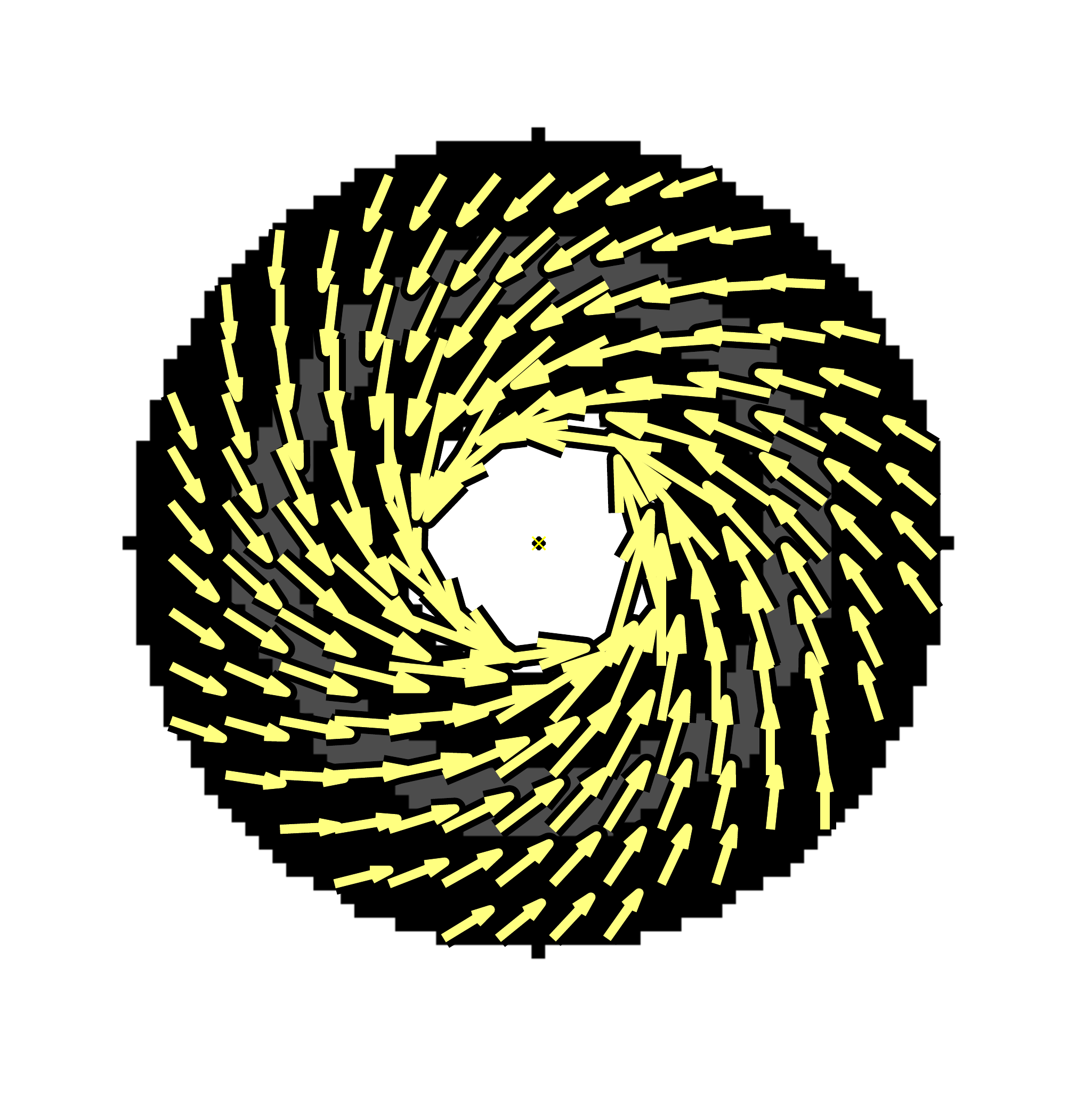}} 
\subfloat[Uncorrected]{\includegraphics[trim={0cm 2.2cm 2.2cm 0cm},clip,width = 0.84in]{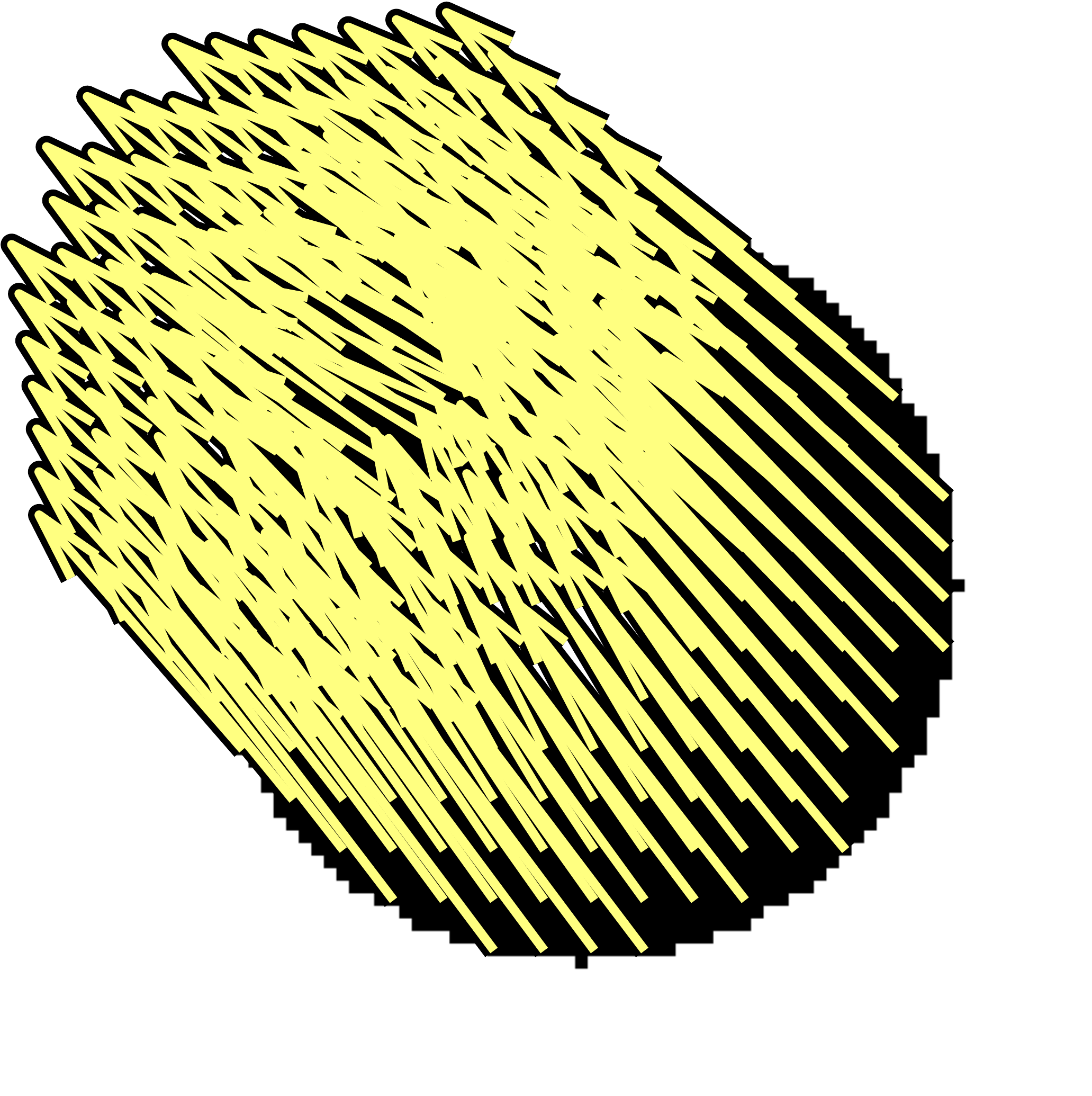}} 
\subfloat[Corrected]{\includegraphics[trim={1.1cm 1.1cm 1.1cm 1.1cm},clip,width = 0.77in]{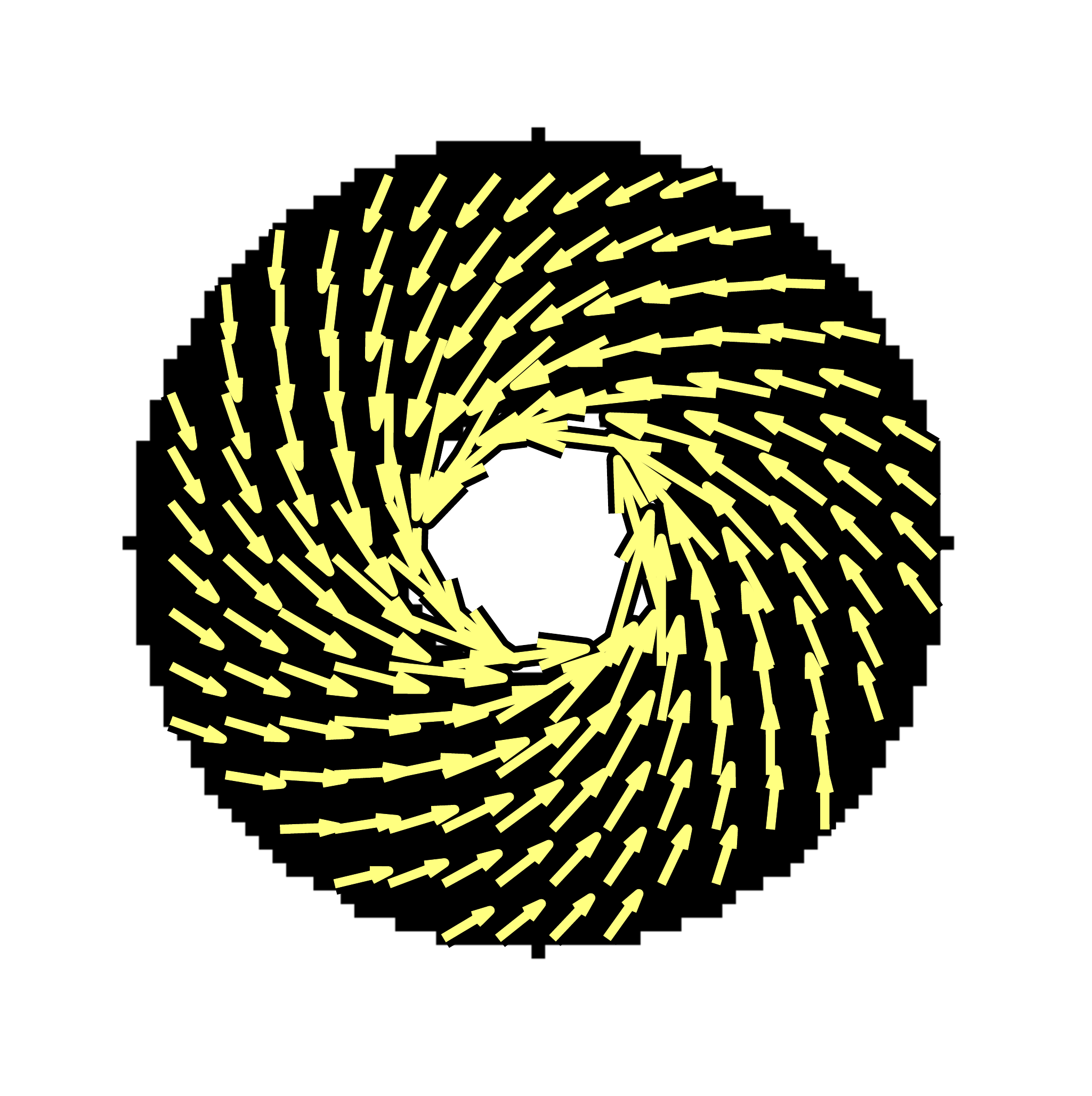}}
\hspace{-4px} \rule{1px}{0.77in} \hspace{-4px}
\subfloat[Ground Truth]{\includegraphics[trim={1.1cm 1.1cm 1.1cm 1.1cm},clip,width = 0.77in]{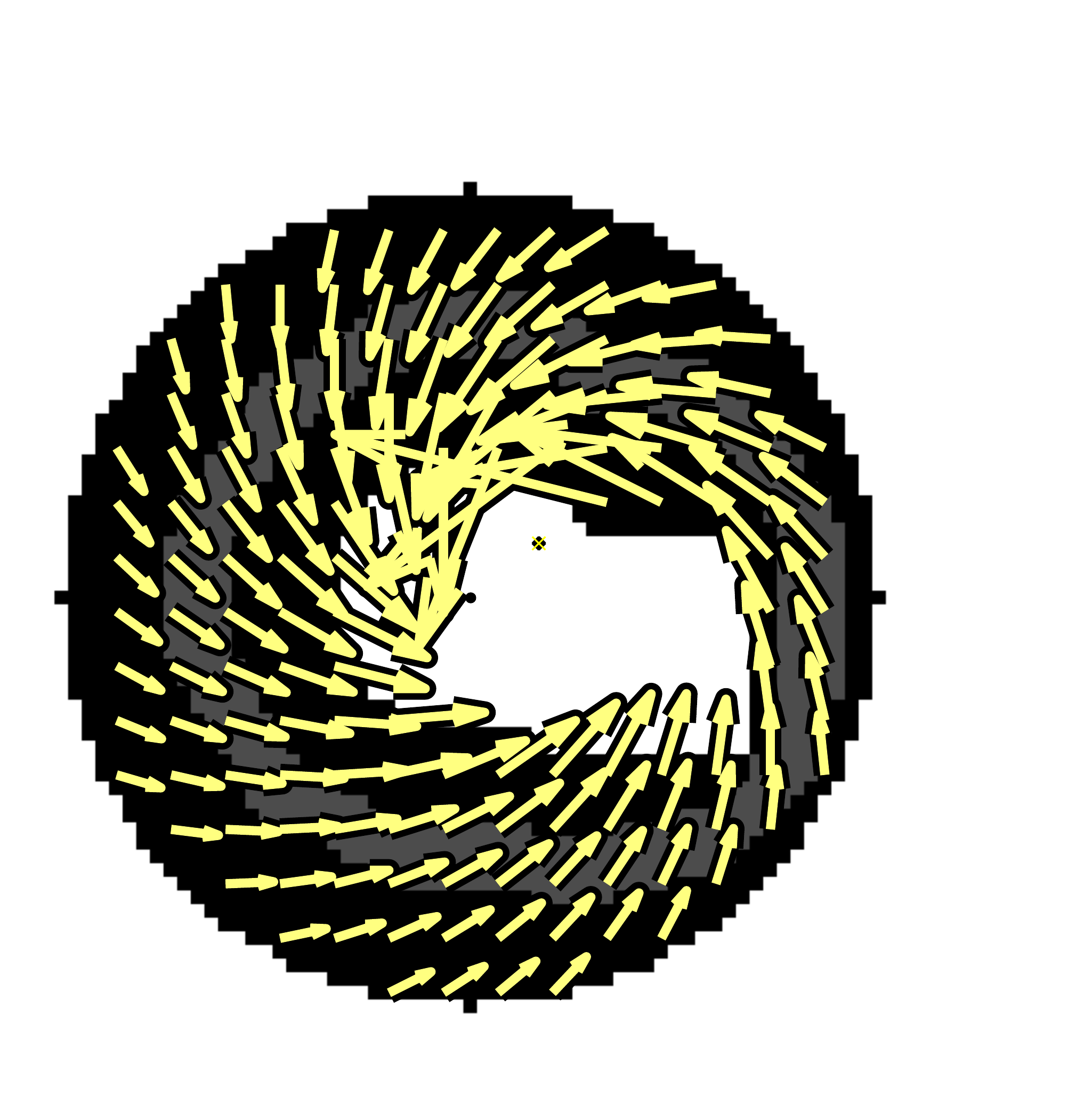}}
\subfloat[Uncorrected]{\includegraphics[trim={0cm 2.2cm 2.2cm 0cm},clip,width = 0.86in]{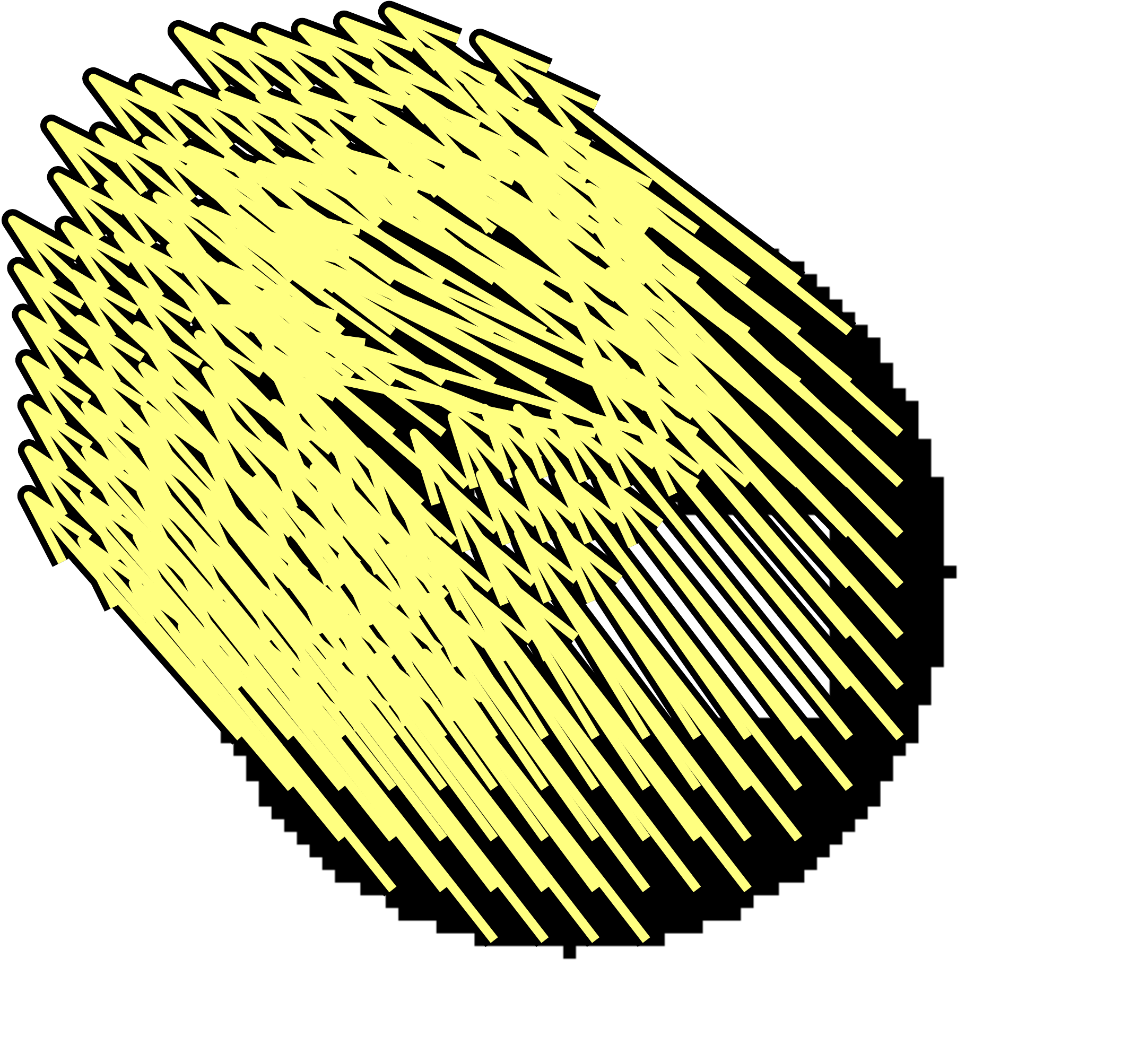}}
\subfloat[ Corrected]{\includegraphics[trim={1.1cm 1.1cm 1.1cm 1.1cm},clip,width = 0.77in]{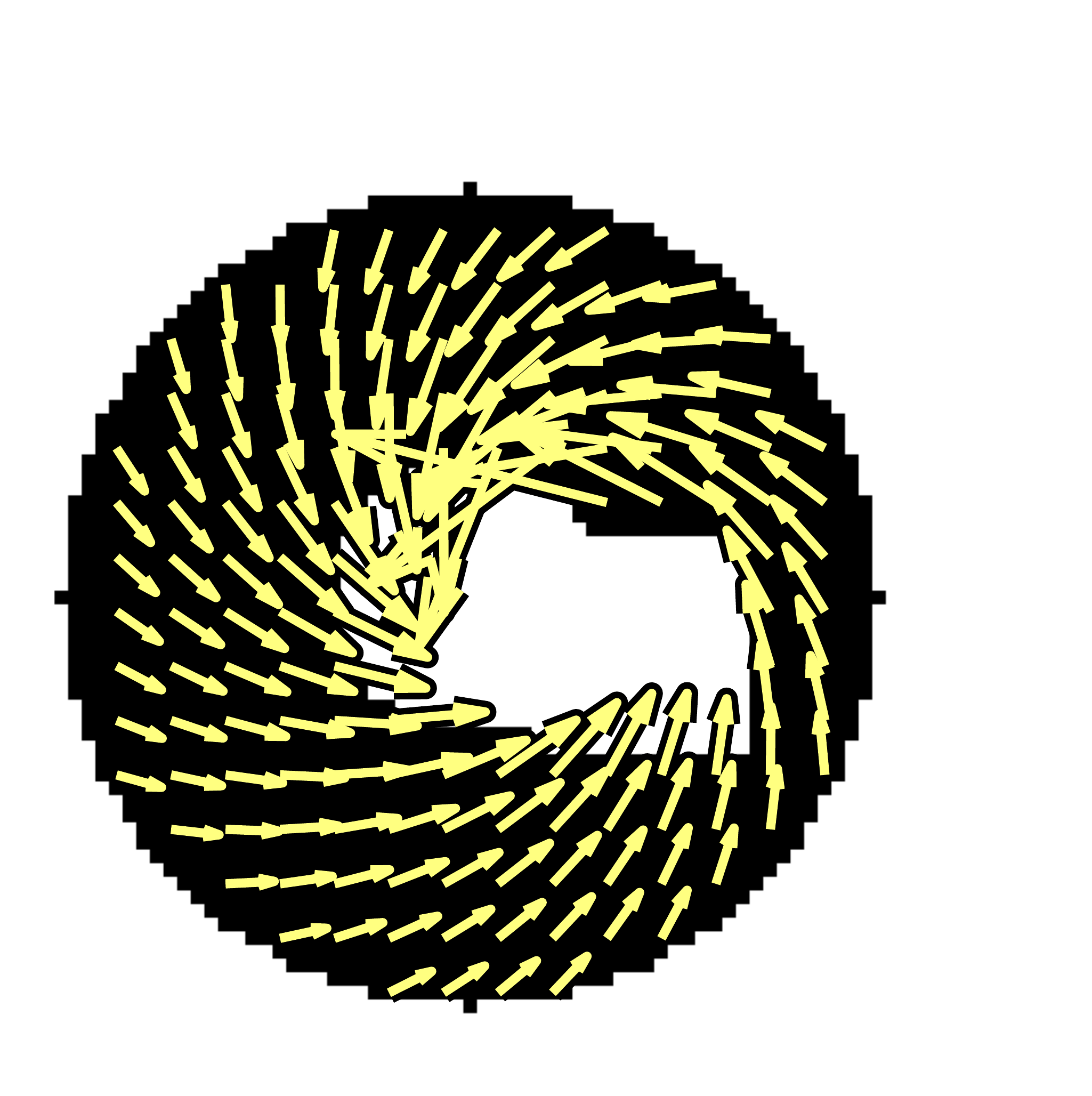}}
\vspace{-2px}
\flushleft (a) Model A, no distortion \hspace{1in}  (b) Model A, distortion

\subfloat[Ground Truth]{\includegraphics[trim={1.1cm 1.1cm 1.1cm 1.1cm},clip,width = 0.77in]{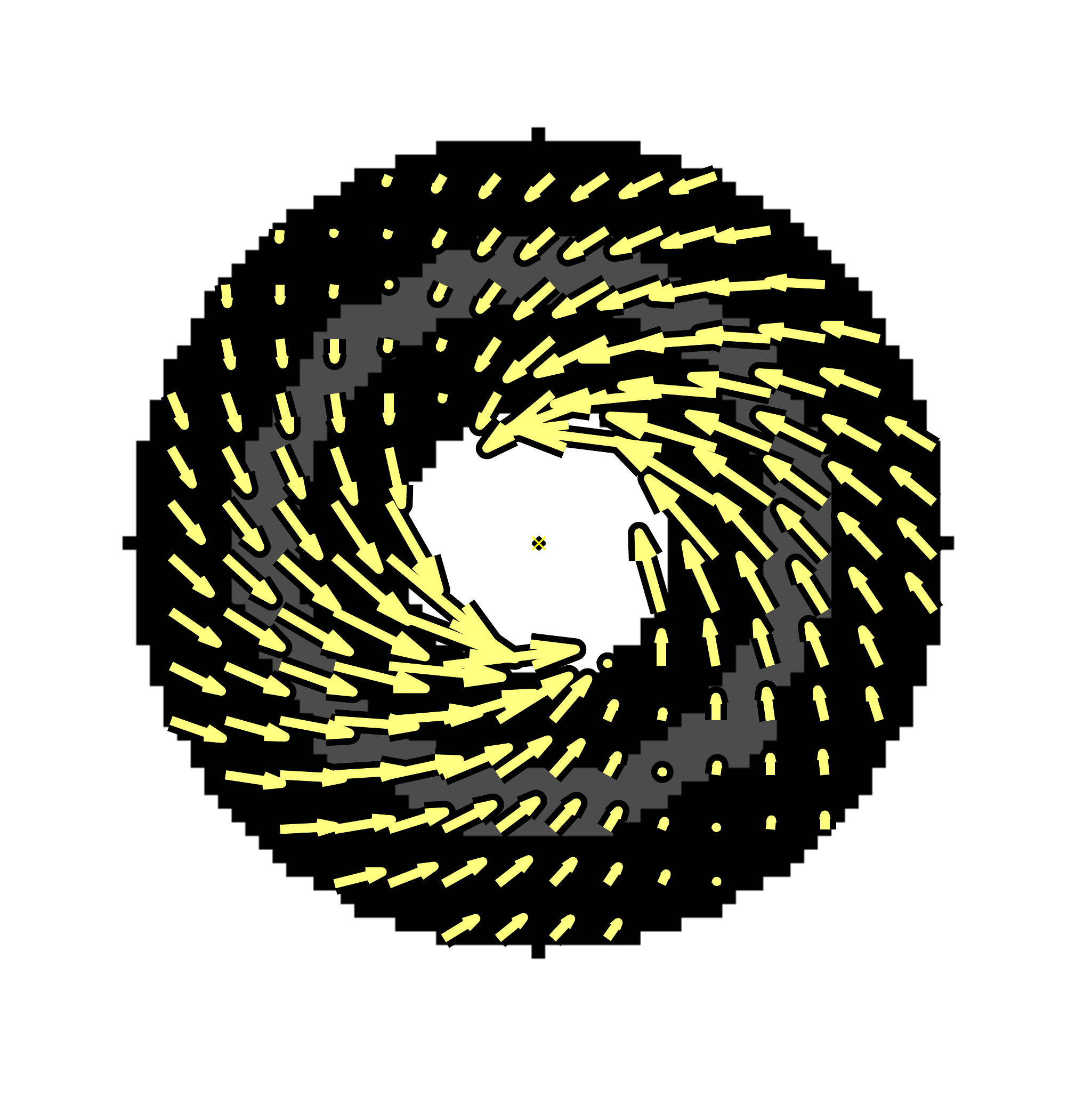}}
\subfloat[Uncorrected]{\includegraphics[trim={0cm 2.2cm 2.2cm 0cm},clip,width = 0.84in]{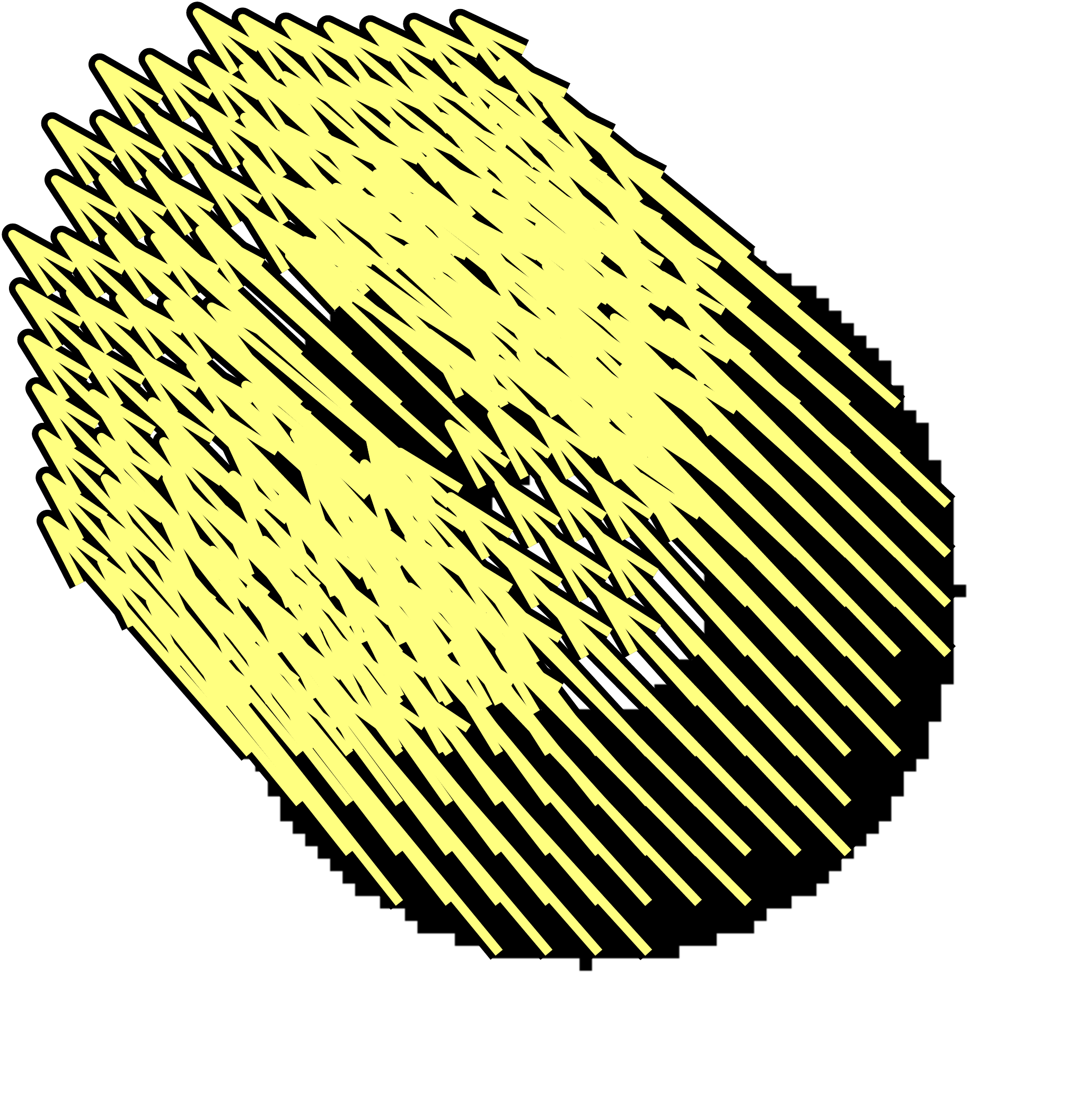}}
\subfloat[Corrected]{\includegraphics[trim={1.1cm 1.1cm 1.1cm 1.1cm},clip,width = 0.77in]{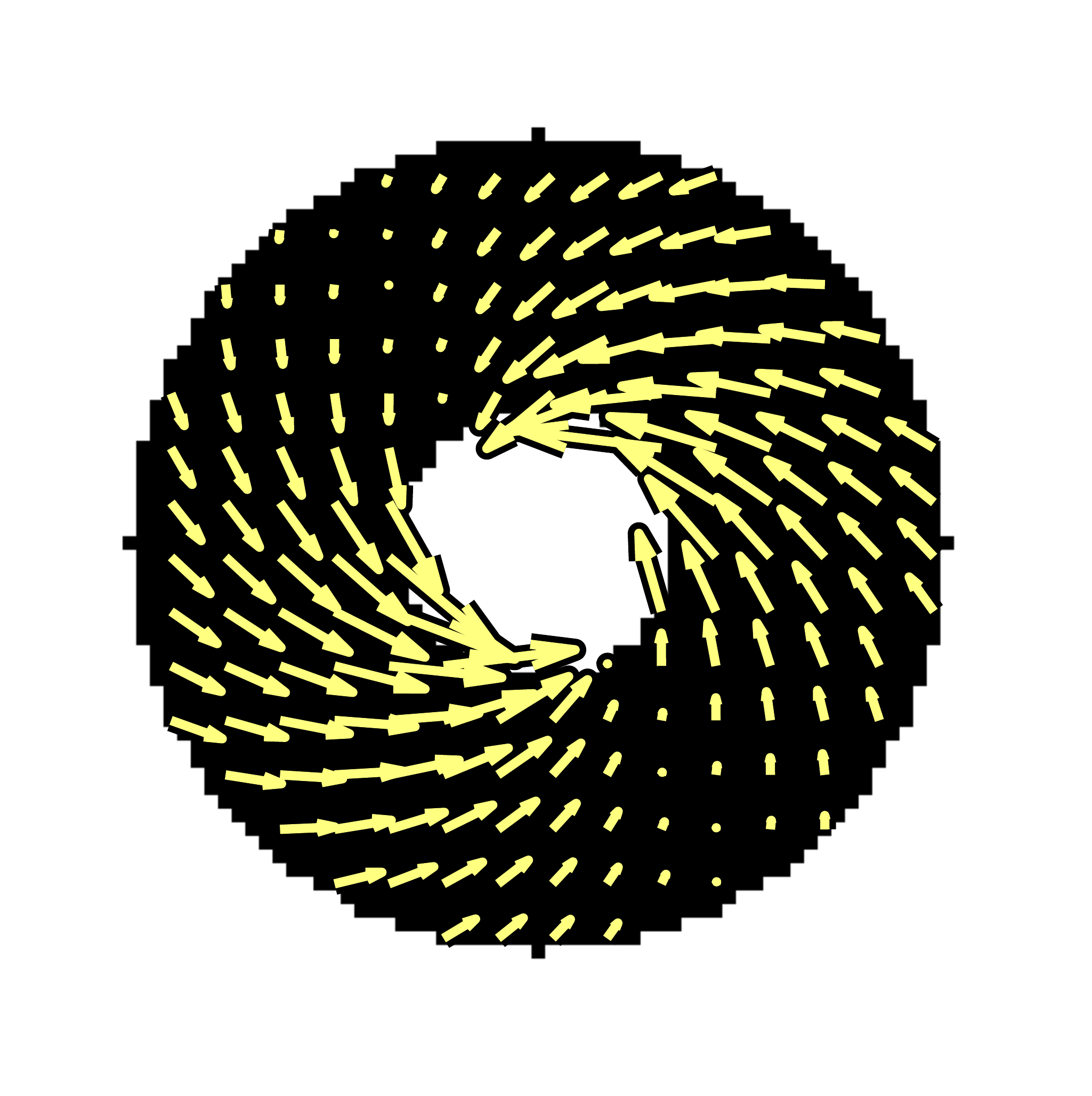}}
\hspace{-4px} \rule{1px}{0.77in} \hspace{-4px}
\subfloat[Ground Truth]{\includegraphics[trim={1.1cm 1.1cm 1.1cm 1.1cm},clip,width = 0.77in]{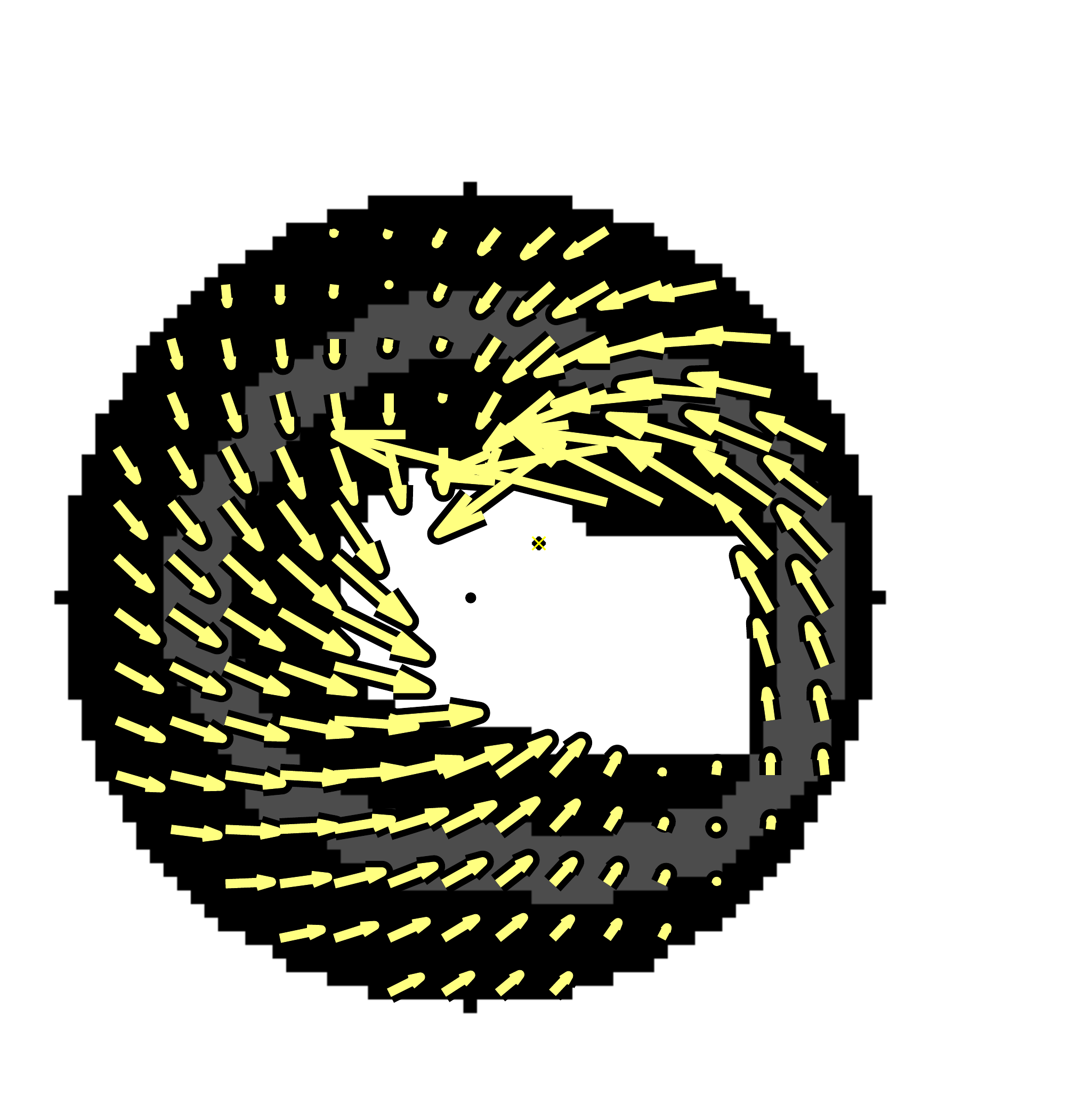}}
\subfloat[Uncorrected]{\includegraphics[trim={0cm 2.2cm 2.2cm 0cm},clip,width = 0.86in]{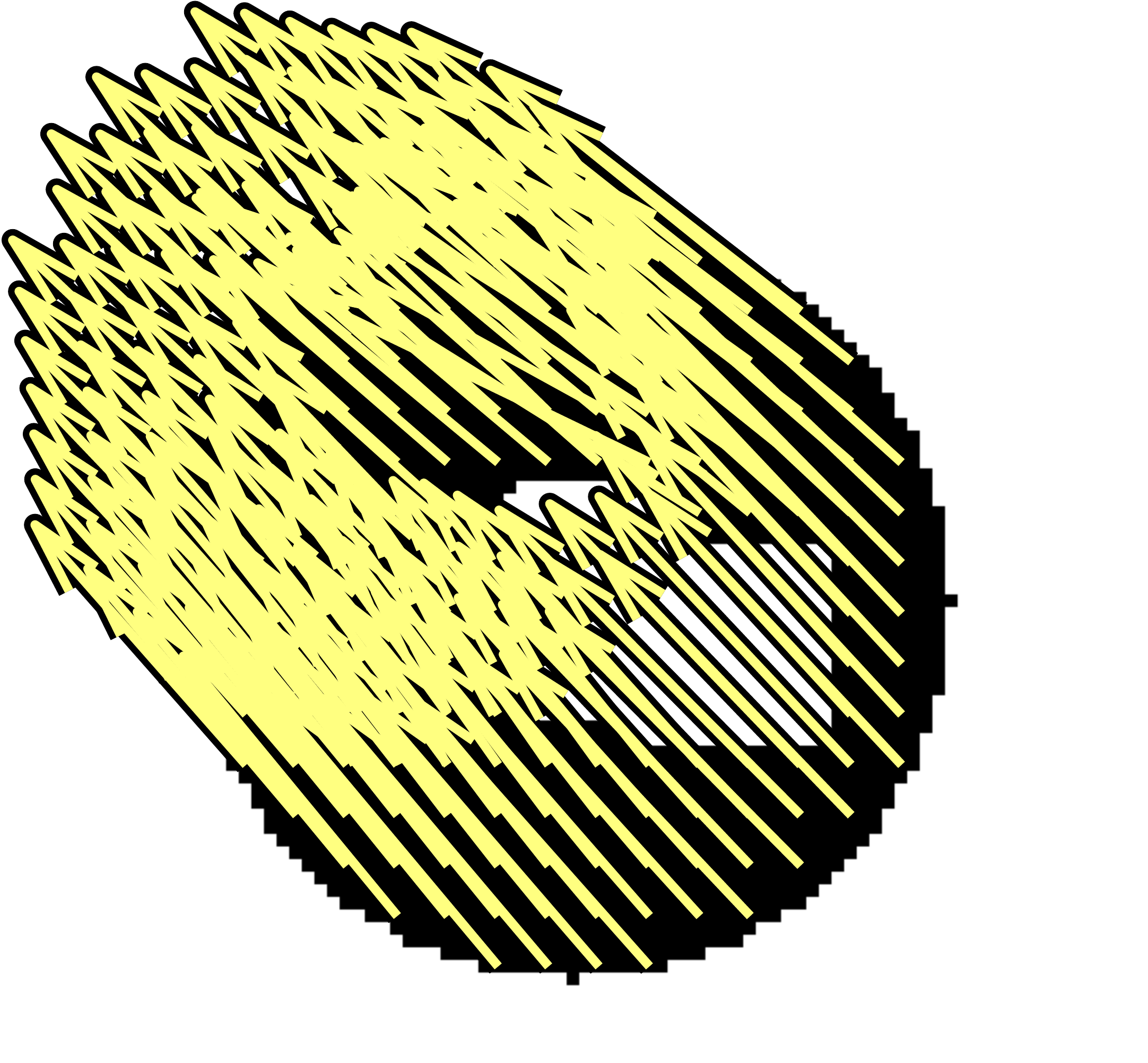}}
\subfloat[ Corrected]{\includegraphics[trim={1.1cm 1.1cm 1.1cm 1.1cm},clip,width = 0.77in]{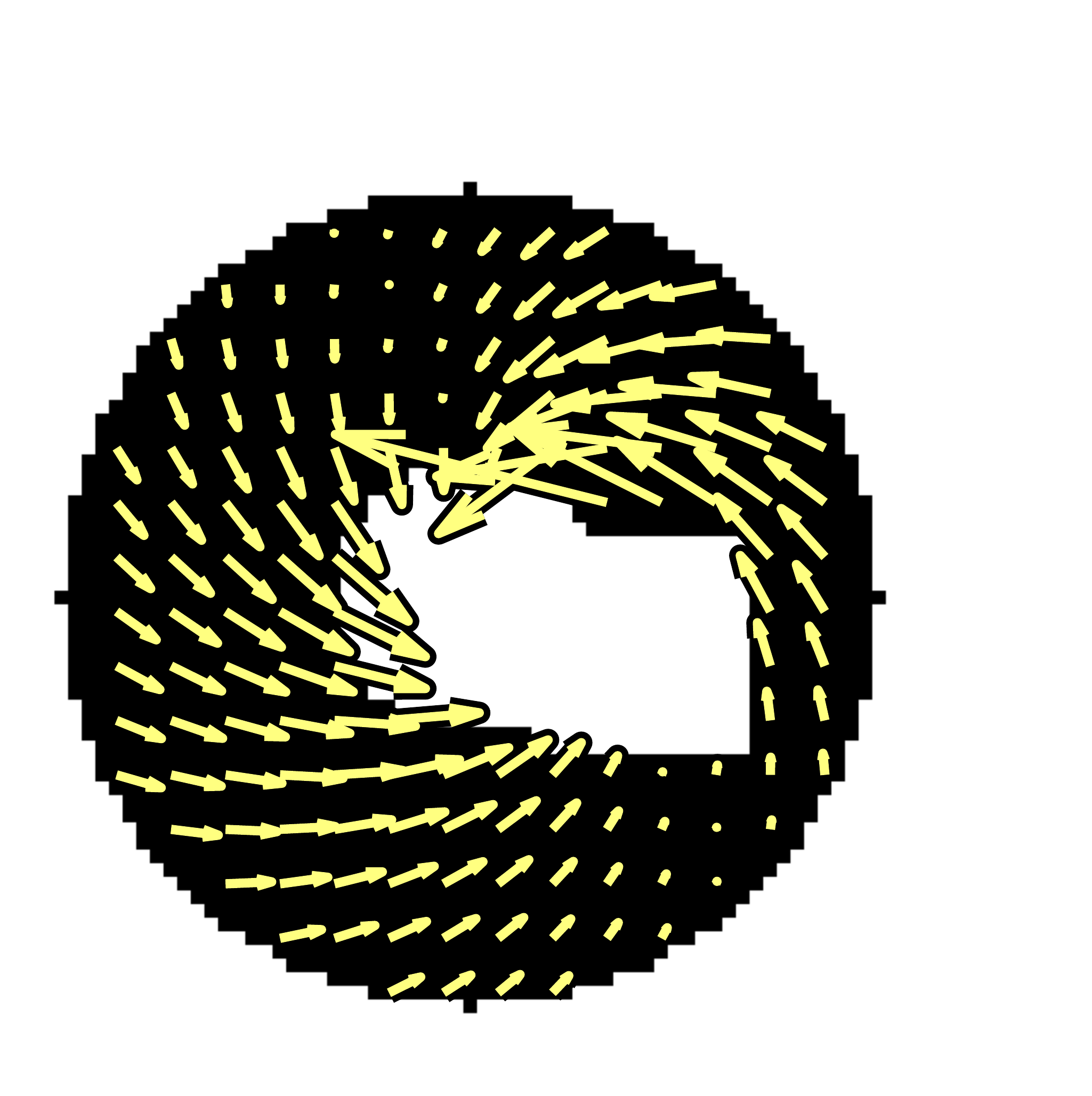}} 
\vspace{-2px}
\flushleft (c) Model B, no distortion \hspace{1in}  (d) Model B, distortion
 
\caption{Evaluation of motion correction method 1 on a 2D synthetic velocity field of (a) model A symmetry (b)model A symmetry with mask distortion, velocity field and mask misalignment, (c)model B symmetry, (d)model B symmetry with mask distortion, velocity field and mask misalignment. Estimation error of the translational velocity in x and y directions is no more than 0.3 \% of actual added translational velocity in all cases.}
\label{fig:syntheticData}
\end{figure}

Motion correction method 1 was first applied on the \textit{uncorrected} simulated synthetic datasets of model symmetry A \& B and their distorted versions, as defined above. Results of motion correction from an intermediate frame of combined contraction and anti-clockwise rotation for the four types of simulated datasets are presented in Fig.\ref{fig:syntheticData}. The corrected images look identical to the ground truth images. As shown in Table \ref{tab:syntheticData}, in all cases the percentage error in estimating the translational velocity does not exceed 0.3\%, despite the large added translational motion (200\% of the maximum underlying velocity) and distortions, indicating excellent effectiveness and robustness of motion correction method 1. 

\begin{figure} 
\centering
\captionsetup[subfigure]{labelformat=empty}
\subfloat[Ground truth]{\includegraphics[trim={3cm 3.5cm 1cm 1cm},clip,width = 1in]{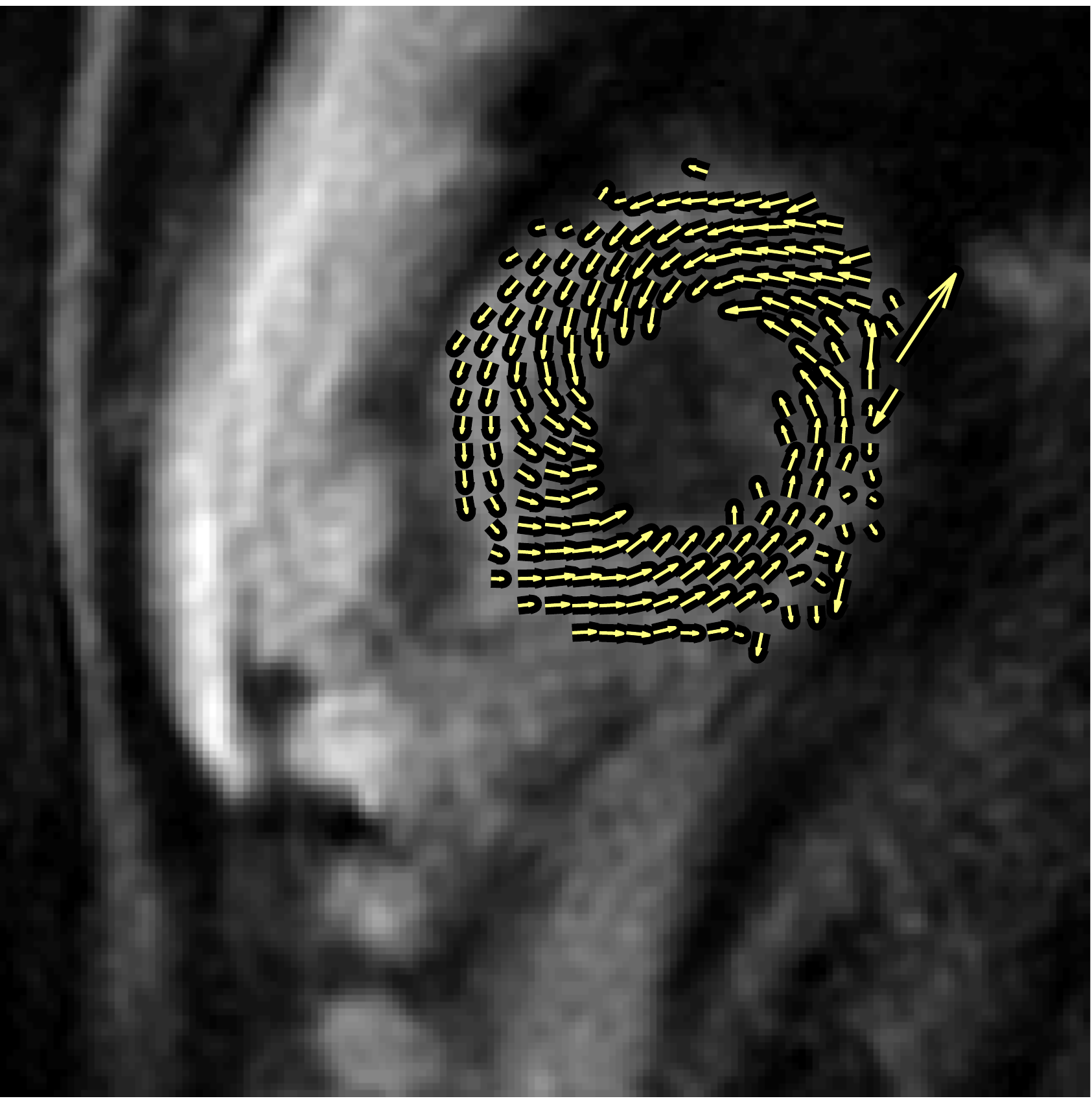}} \hspace{15px}
\subfloat[Uncorrected]{\includegraphics[trim={3cm 3.5cm 1cm 1cm},clip,width = 1in]{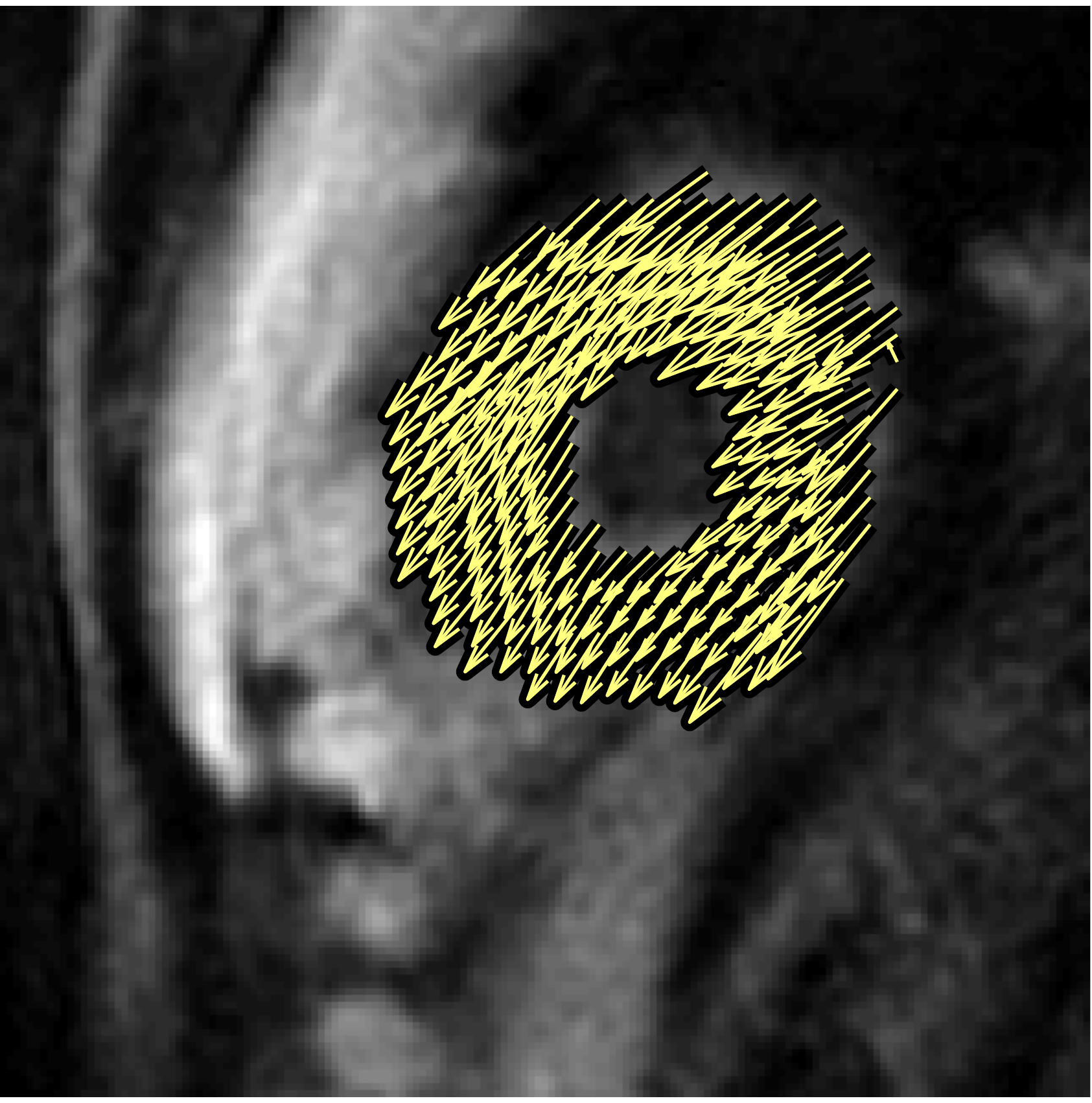}} \hspace{15px}
\subfloat[Corrected]{\includegraphics[trim={3cm 3.5cm 1cm 1cm},clip,width = 1in]{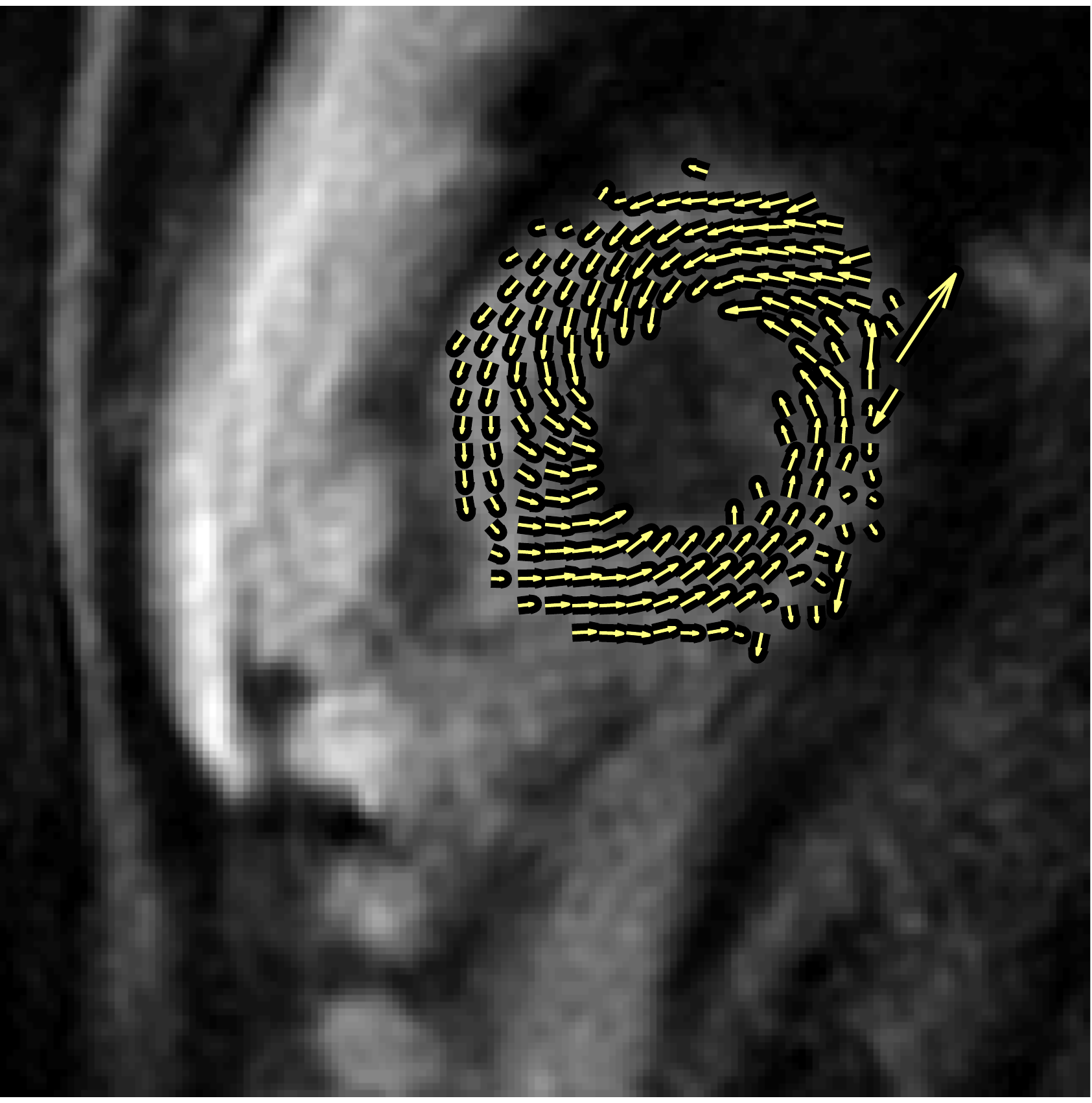}} 
\vspace{-3px}
\flushleft (a) Apical slice at peak anticlockwise rotation 

\vspace{-2px}
\centering
\subfloat[Ground truth]{\includegraphics[trim={3cm 3.5cm 1cm 1cm},clip,width = 1in]{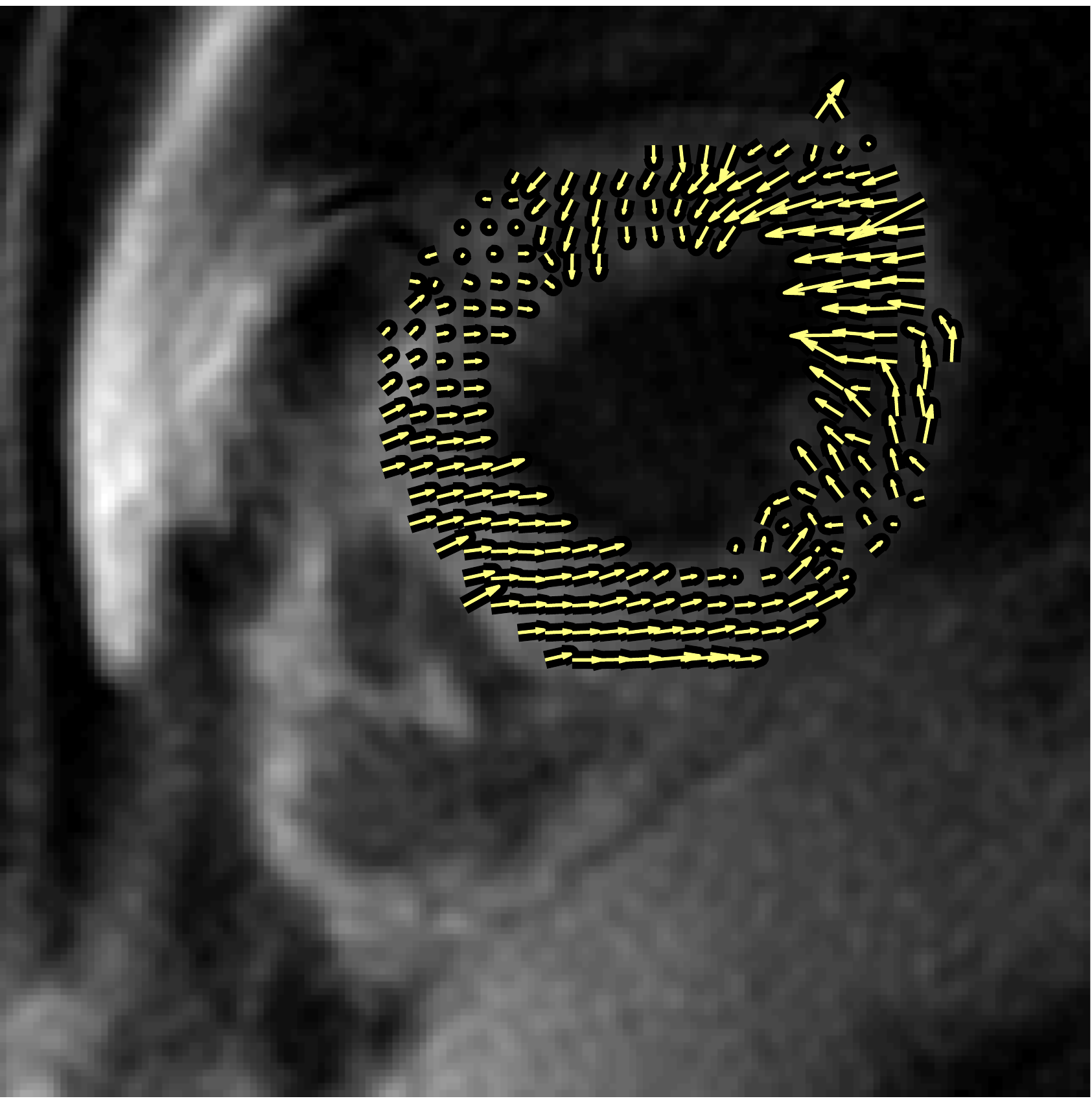}} \hspace{15px}
\subfloat[Uncorrected]{\includegraphics[trim={3cm 3.5cm 1cm 1cm},clip,width = 1in]{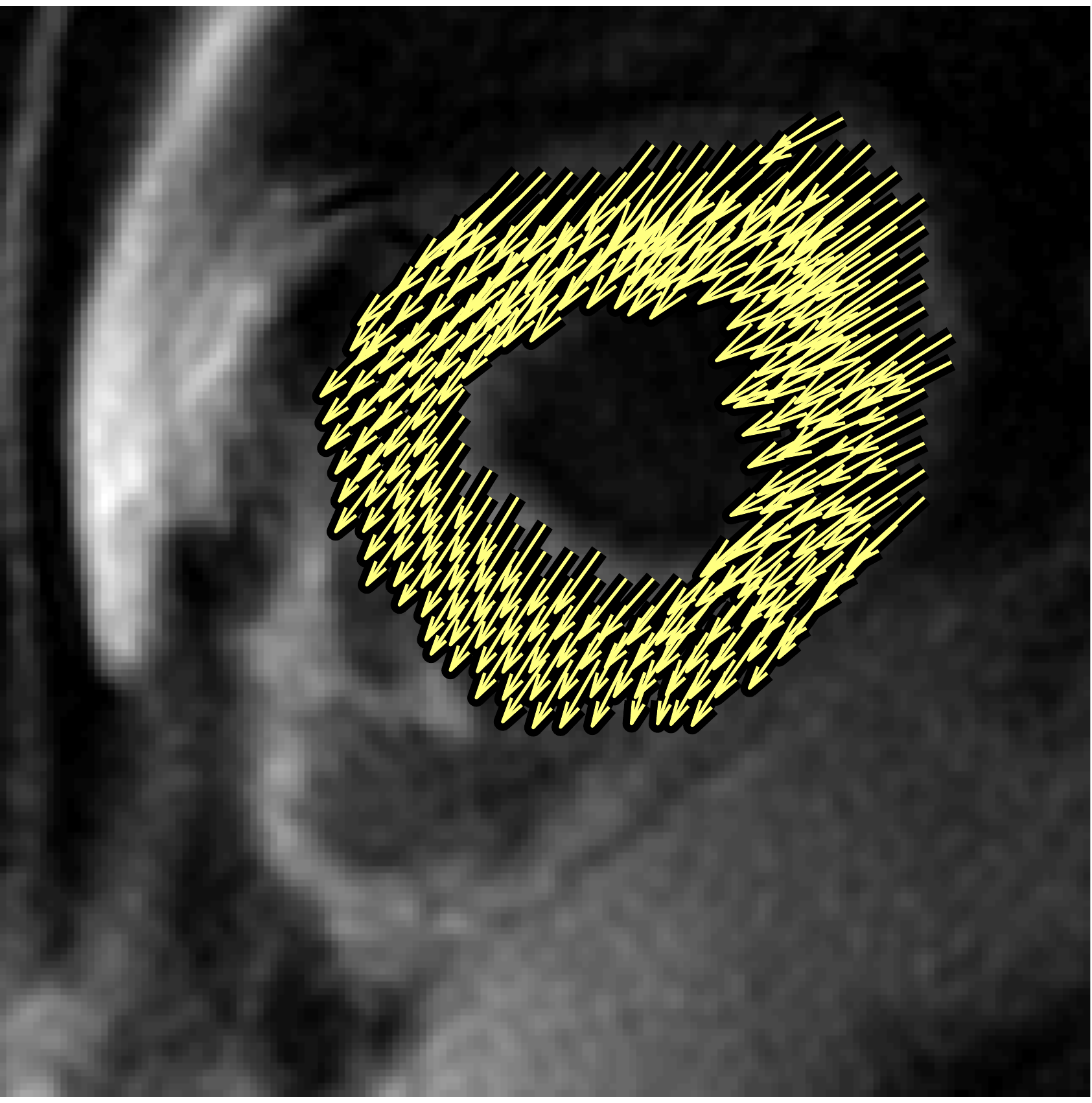}} \hspace{15px}
\subfloat[Corrected]{\includegraphics[trim={3cm 3.5cm 1cm 1cm},clip,width = 1in]{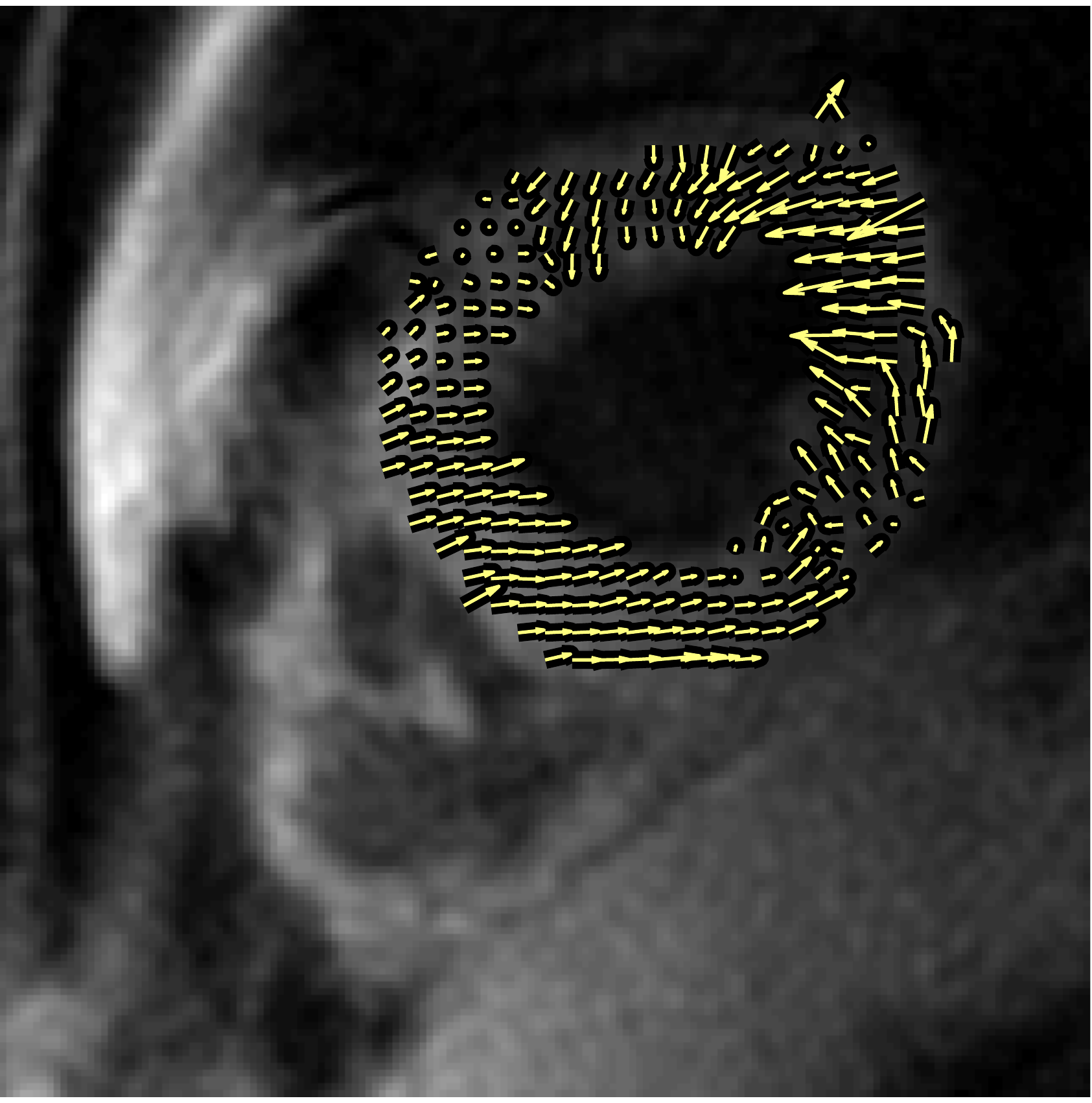}} 

\vspace{-3px}
\flushleft (b) Mid-wall slice at peak anticlockwise rotation 

\vspace{-2px}
\centering
\subfloat[Ground truth]{\includegraphics[trim={3cm 3.5cm 1cm 1cm},clip,width = 1in]{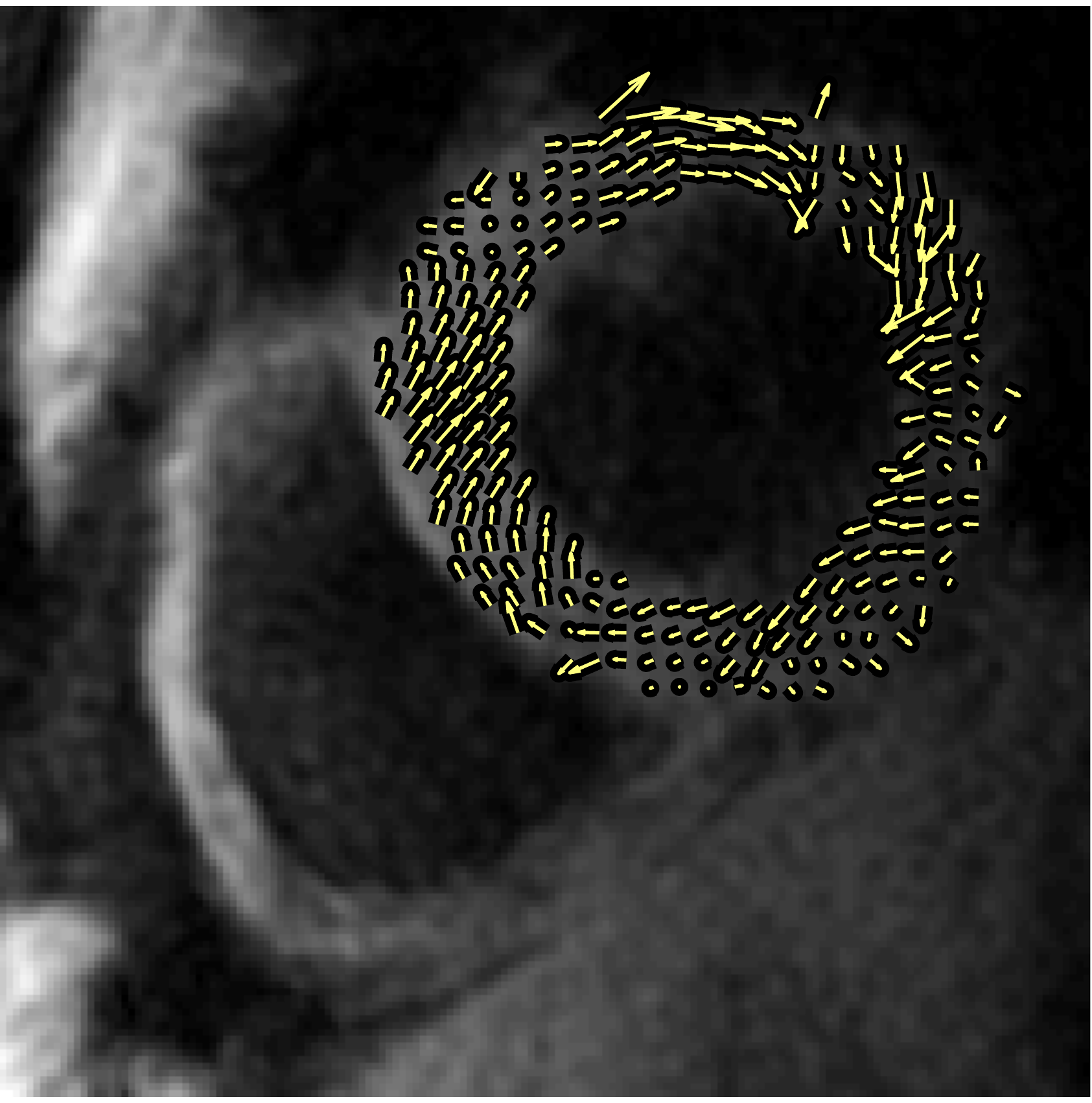}} \hspace{15px}
\subfloat[Uncorrected]{\includegraphics[trim={3cm 3.5cm 1cm 1cm},clip,width = 1in]{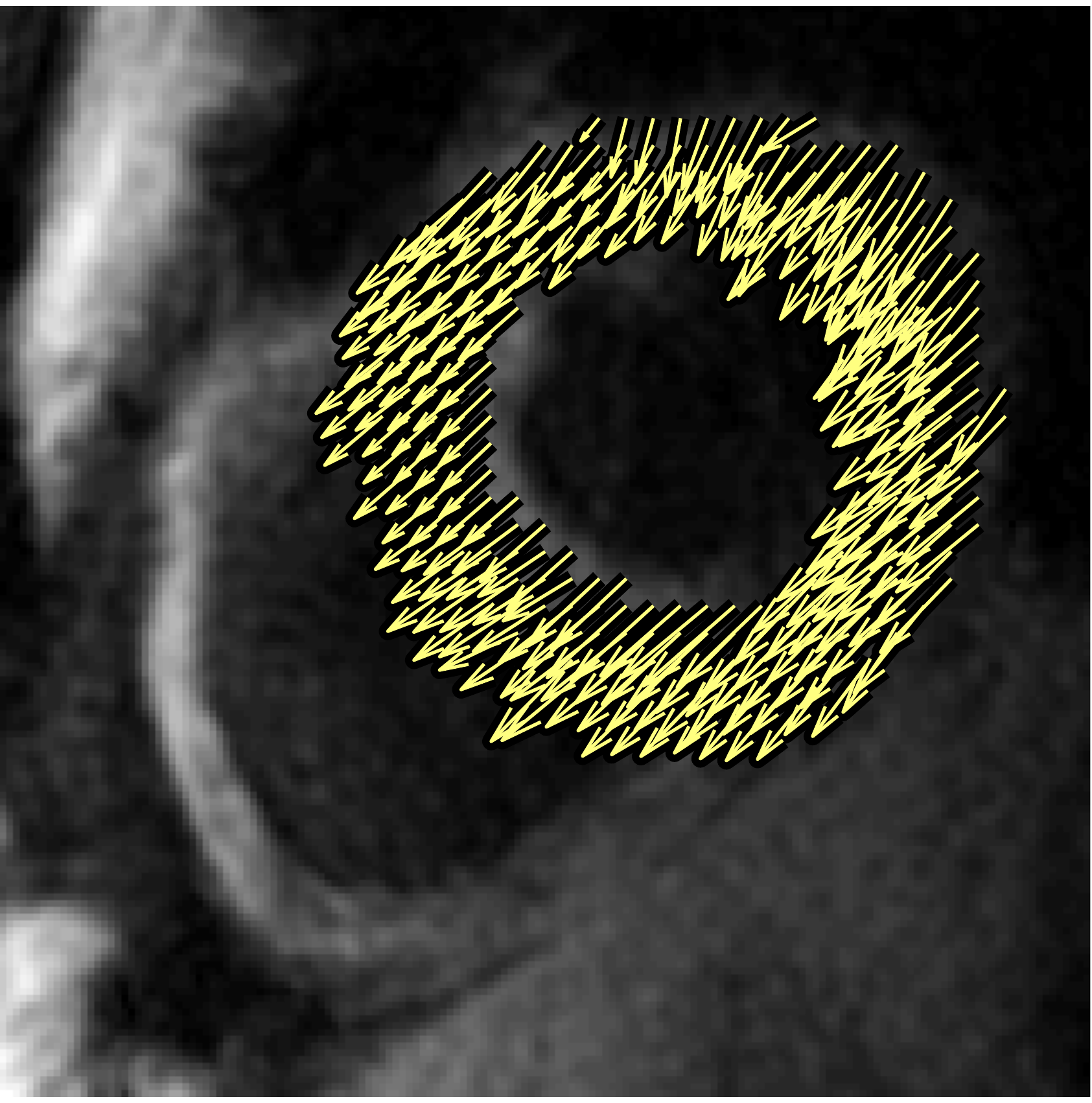}} \hspace{15px}
\subfloat[Corrected]{\includegraphics[trim={3cm 3.5cm 1cm 1cm},clip,width = 1in]{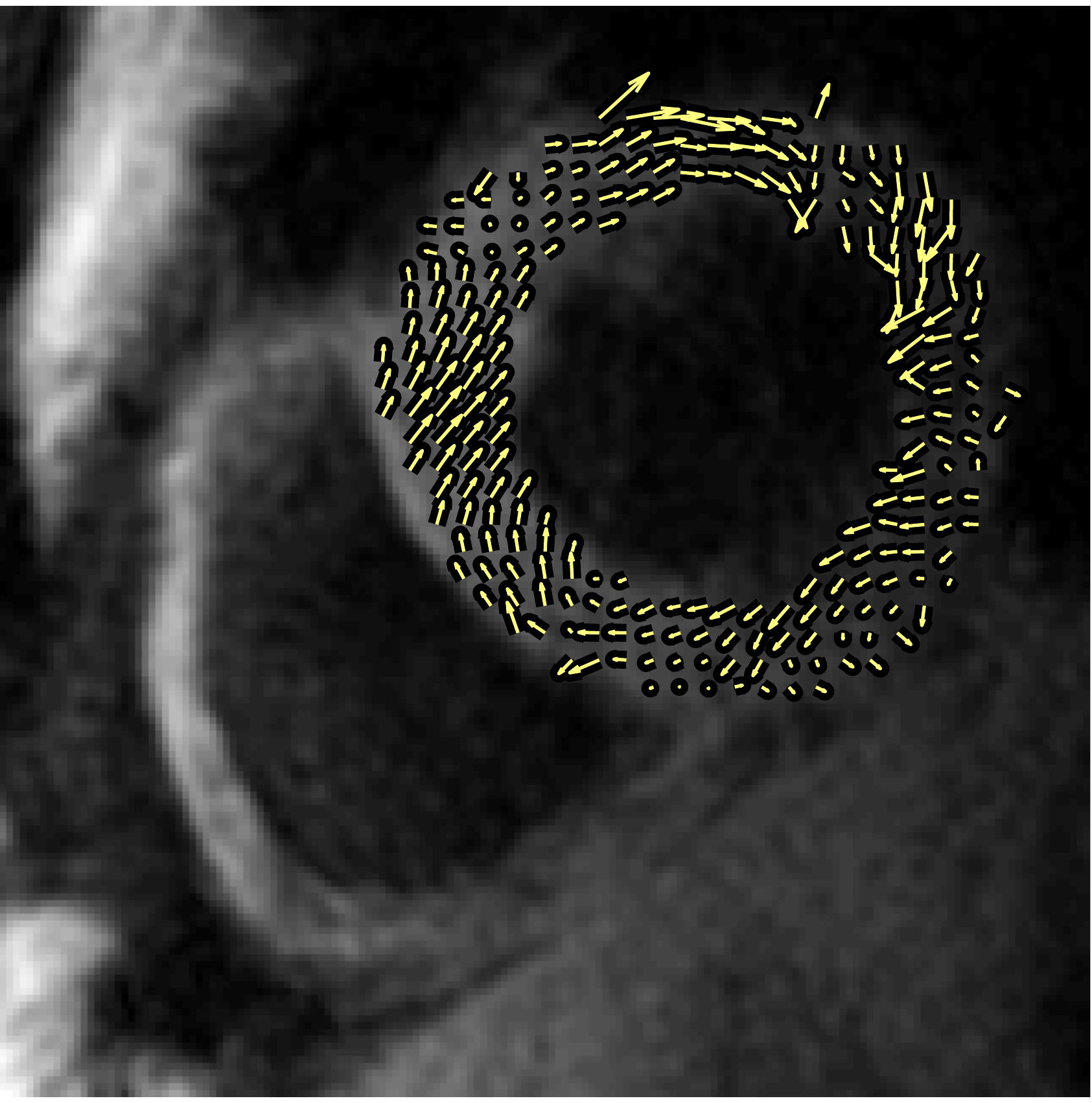}}
\vspace{-3px}
\flushleft (c) Basal slice at peak anticlockwise rotation 

\vspace{-2px}
\caption{In-planar velocity field from an (a) apical (b) mid-ventricular and (c) basal short axis slice of a semi-synthetic TPM velocity stack at peak anti-clockwise rotation without translational motion, with translational motion and corrected for translational motion with method 1.}

\label{fig:semisynthetic}
\end{figure}

Motion correction methods 1 \& 2 were then applied on the realistic semi-synthetic dataset corrupted, as described above. Results of motion correction method 1 from an apical, mid-wall and basal slice of the LV at peak anti-clockwise rotation are presented in Fig.\ref{fig:semisynthetic}. The translational component is successfully removed and no remaining errors are observed. The corrected images look identical to the ground truth images confirming the effectiveness of the method.

\begin{table}
\caption{Error in the estimation of translational velocity in the synthetic velocity field}
\label{tab:syntheticData}       
\begin{tabular}{lllll}
\hline\noalign{\smallskip}
&Model A &Model B&Model A/distorted&Model B/distorted\\
\noalign{\smallskip}\hline\noalign{\smallskip}
x axis \hspace{1cm} &  +0.26 \% &  +0.14 \% & -   0.17 \% & -   0.02 \% \\
y axis \hspace{1cm} & -   0.06 \% & -   0.05 \% &	 +0.23 \% &+0.10 \% \\
\noalign{\smallskip}\hline
\end{tabular}
\end{table}

\begin{figure} 
\captionsetup[subfigure]{labelformat=empty}
\centering
\subfloat[Uncorrected]{\includegraphics[trim={3cm 3.5cm 1cm 1cm},clip,width = 1in]{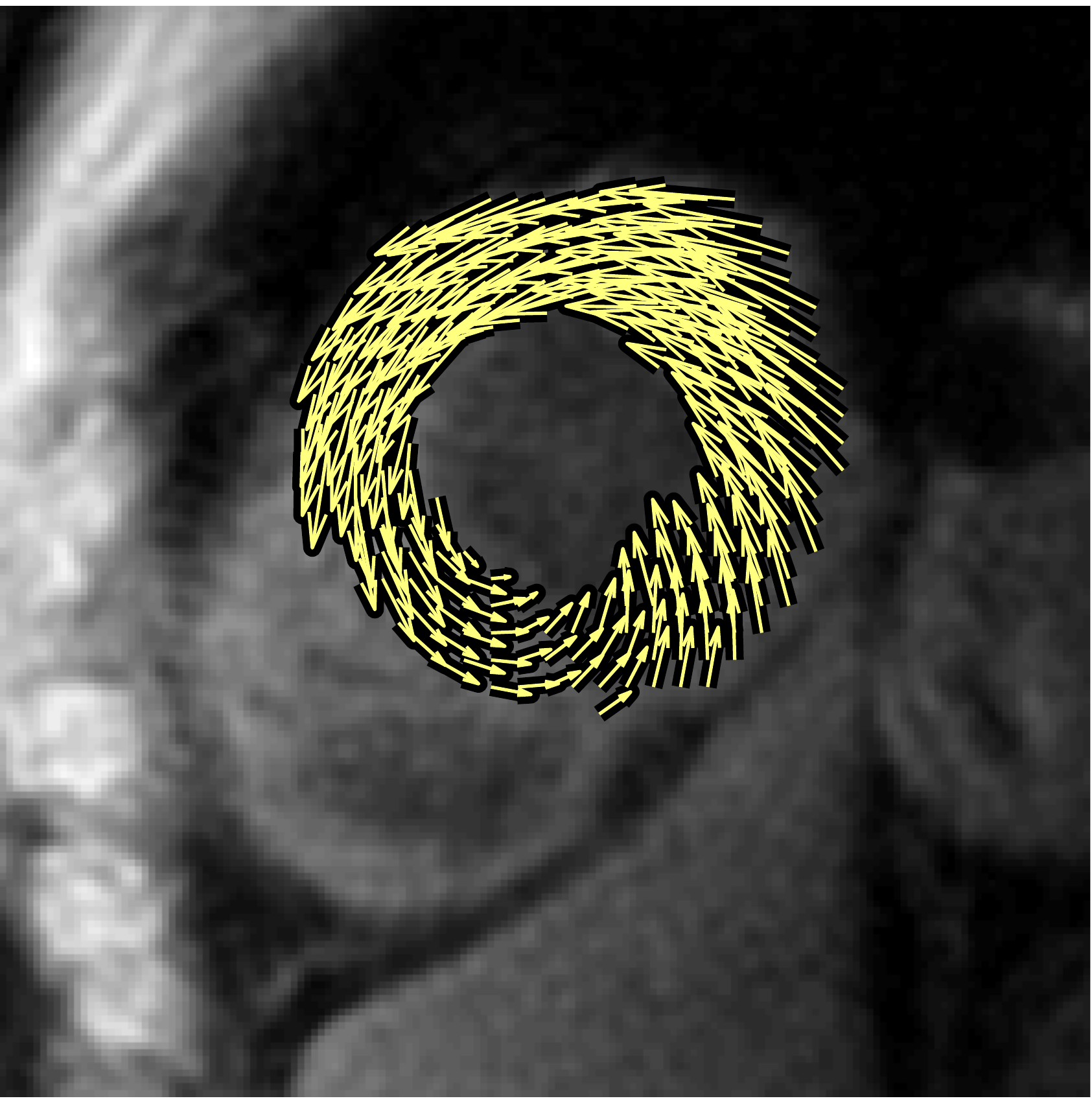}} \quad
\subfloat[Corrected]{\includegraphics[trim={3cm 3.5cm 1cm 1cm},clip,width = 1in]{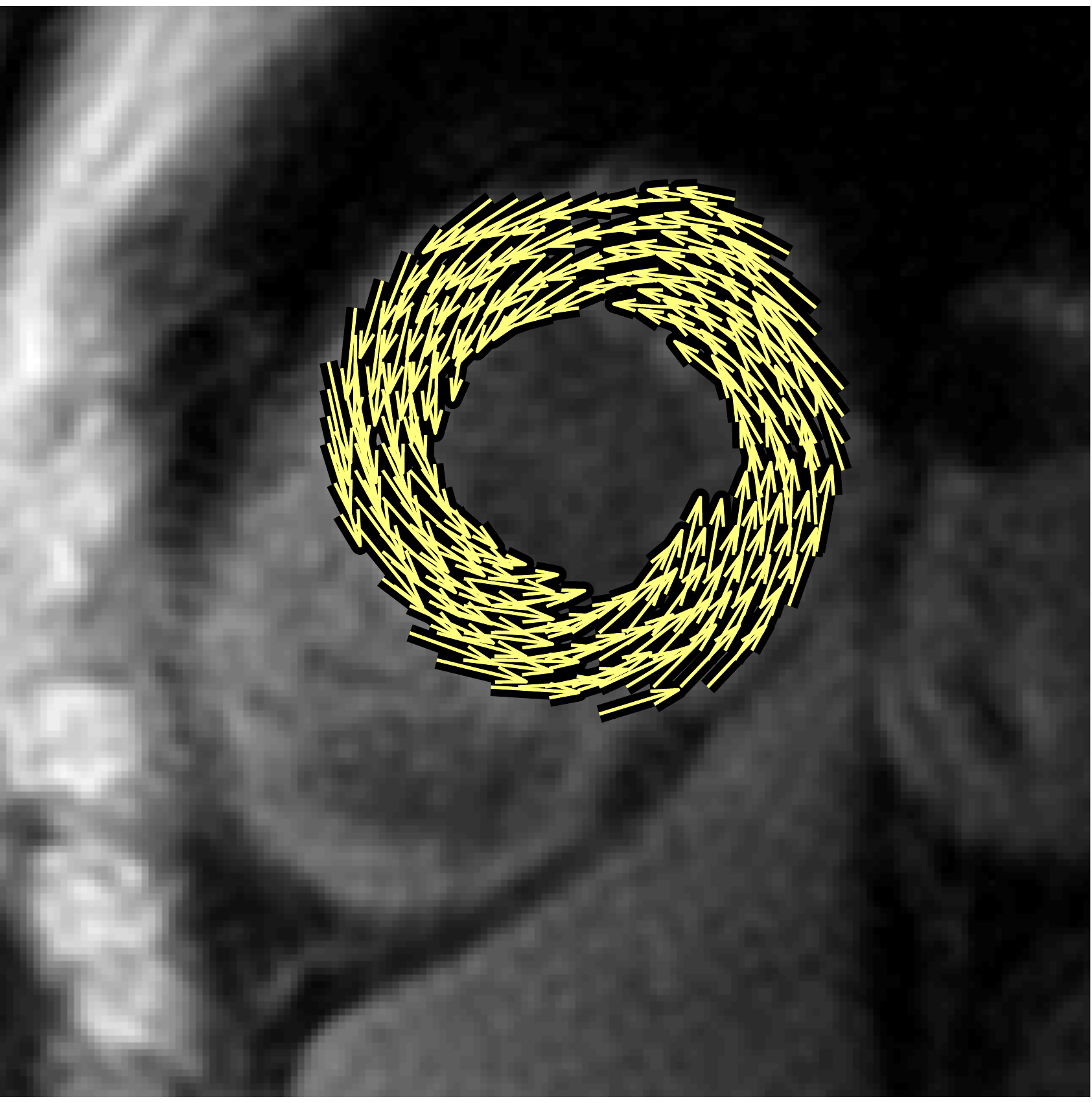}} \qquad
\subfloat[Uncorrected]{\includegraphics[trim={3cm 3.5cm 1cm 1cm},clip,width = 1in]{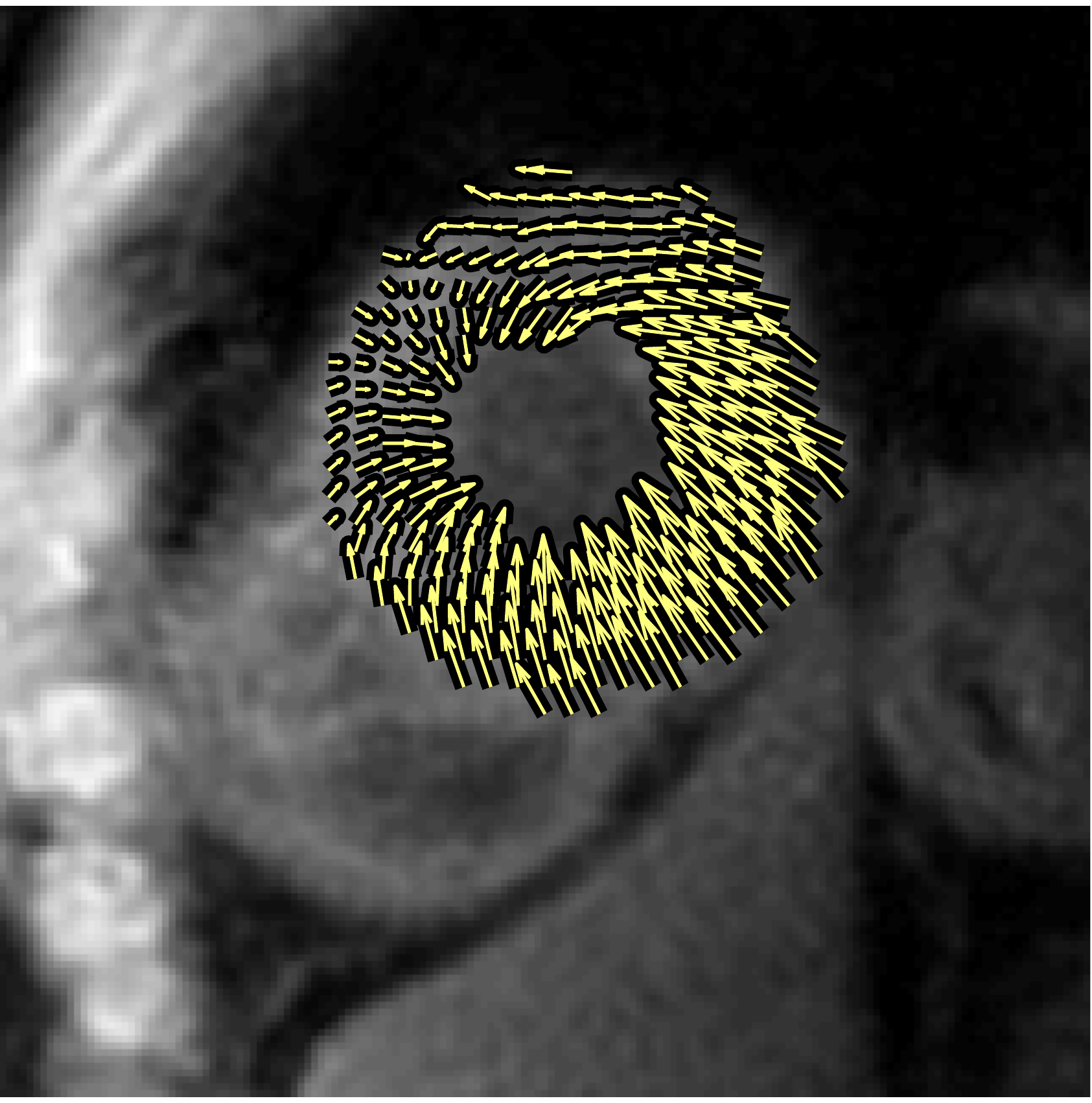}} \quad
\subfloat[Corrected]{\includegraphics[trim={3cm 3.5cm 1cm 1cm},clip,width = 1in]{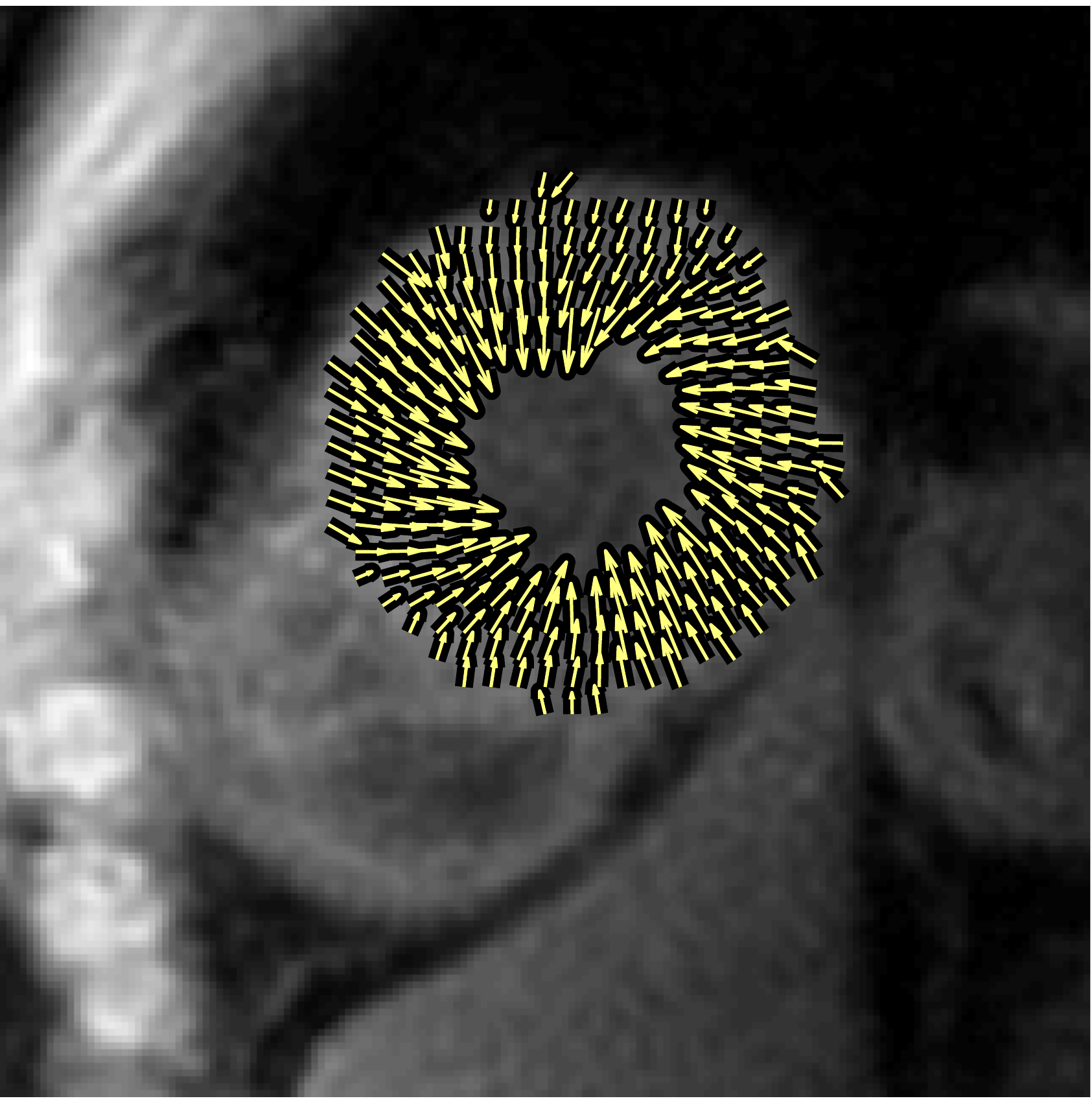}} 
\vspace{-2px}
\flushleft (a) Apical slice at peak anticlockwise rotation (left) and peak contraction (right)

\centering
\subfloat[Uncorrected]{\includegraphics[trim={3cm 3.5cm 1cm 1cm},clip,width = 1in]{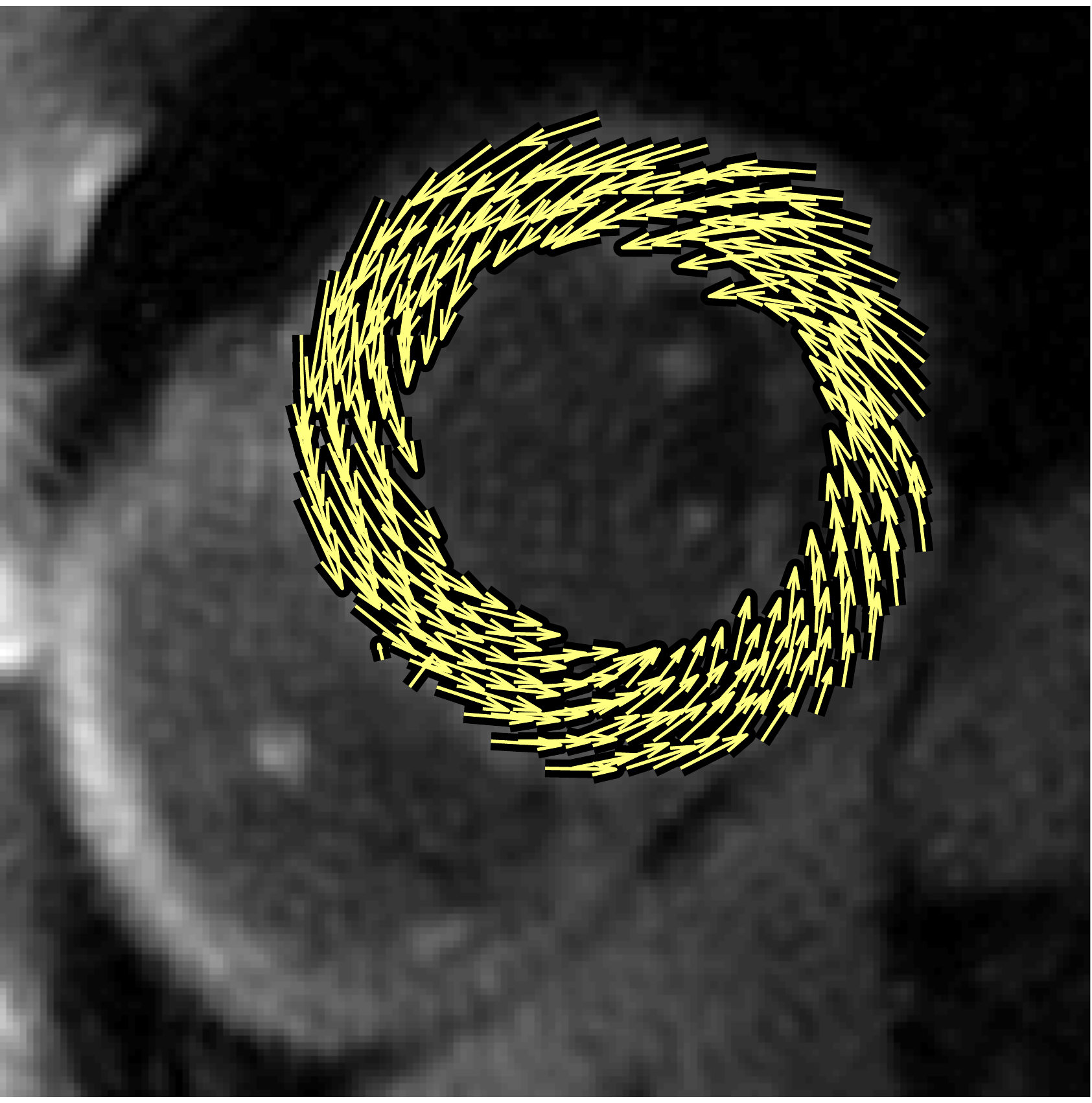}} \quad
\subfloat[Corrected]{\includegraphics[trim={3cm 3.5cm 1cm 1cm},clip,width = 1in]{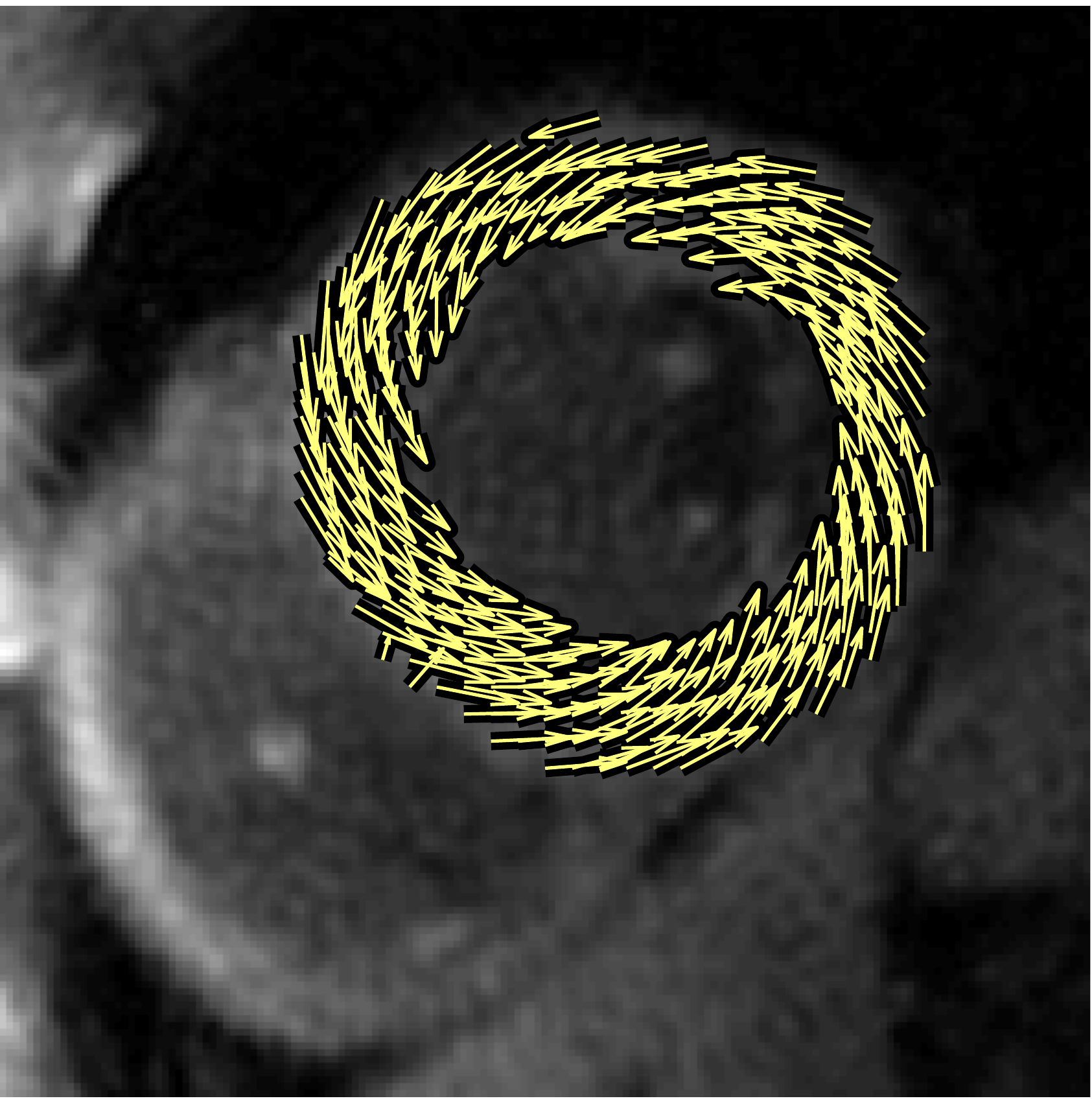}} \qquad
\subfloat[Uncorrected]{\includegraphics[trim={3cm 3.5cm 1cm 1cm},clip,width = 1in]{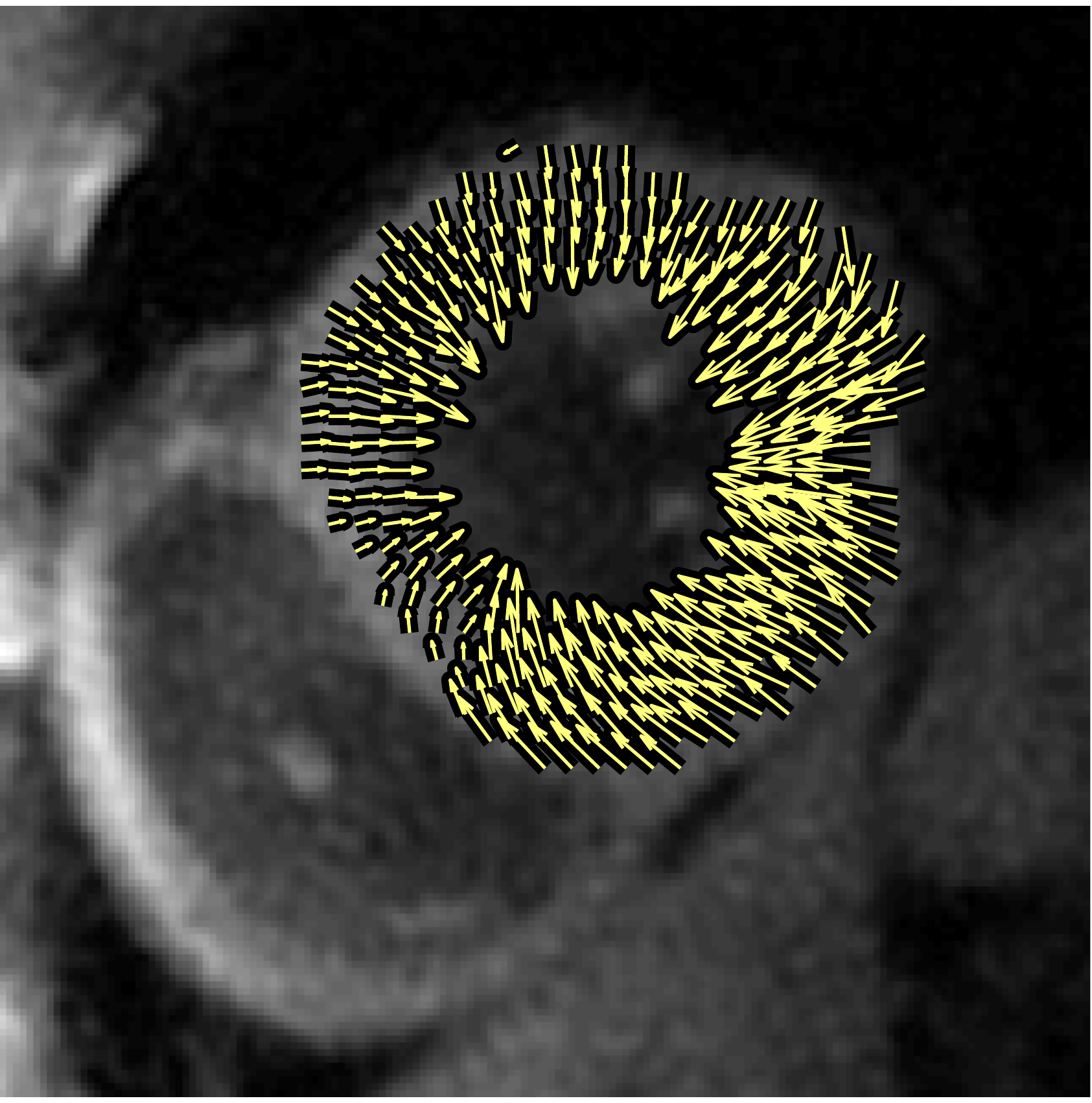}} \quad
\subfloat[Corrected]{\includegraphics[trim={3cm 3.5cm 1cm 1cm},clip,width = 1in]{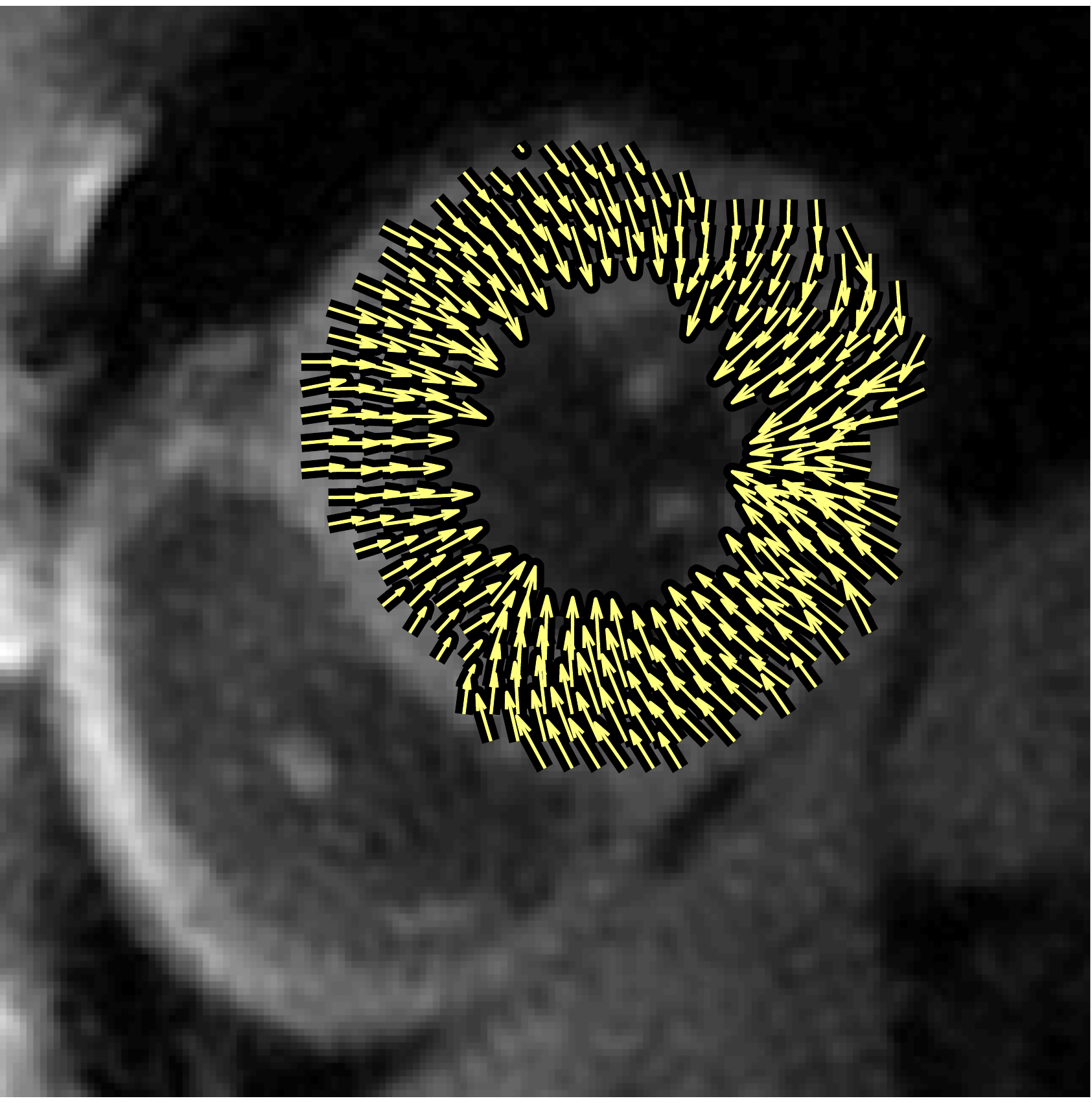}}
\vspace{-2px}
\flushleft (b) Mid-wall slice at peak anticlockwise rtoation (left) and peak contraction (right)

\centering
\subfloat[Uncorrected]{\includegraphics[trim={3cm 3.5cm 1cm 1cm},clip,width = 1in]{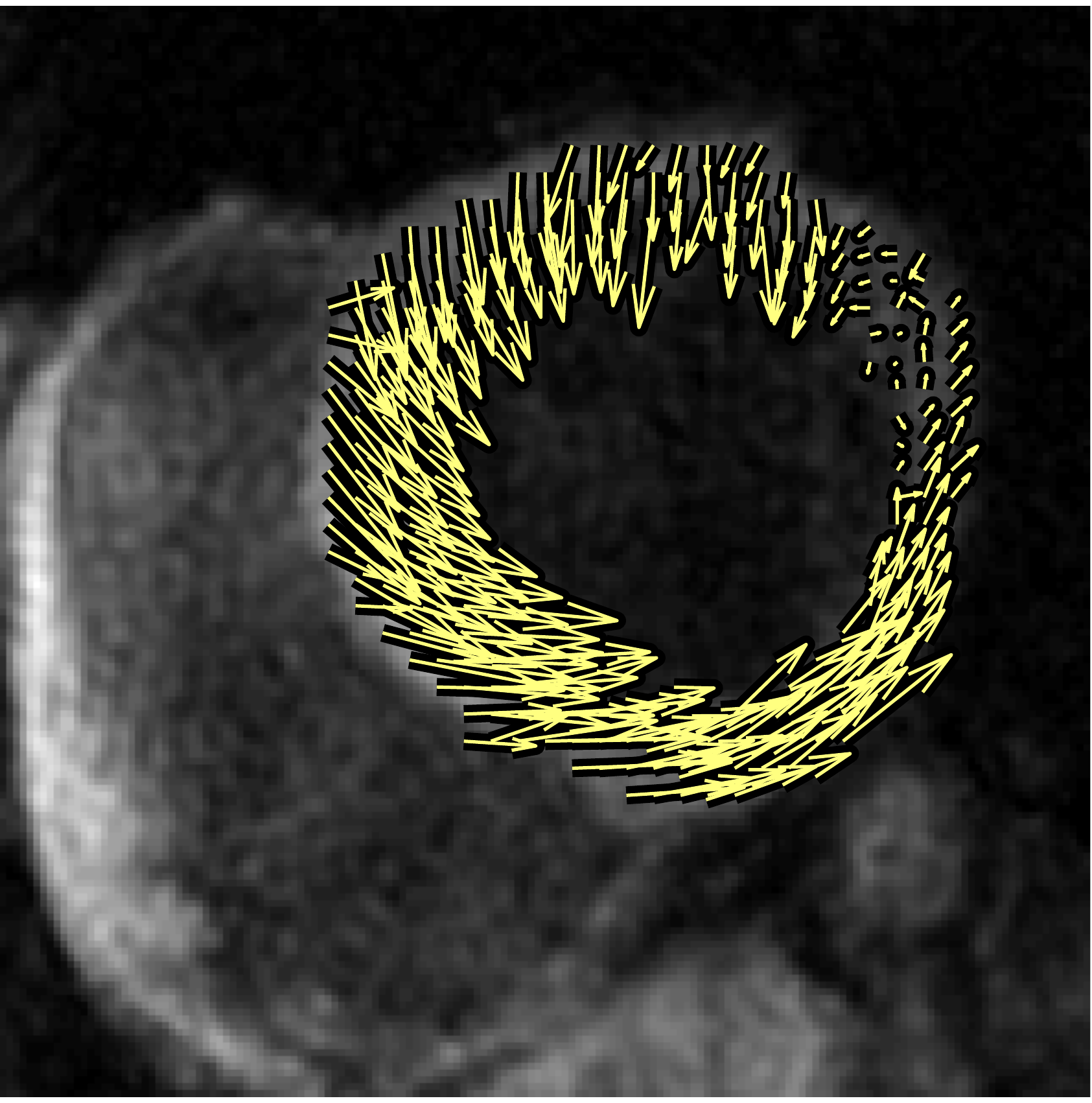}} \quad
\subfloat[Corrected]{\includegraphics[trim={3cm 3.5cm 1cm 1cm},clip,width = 1in]{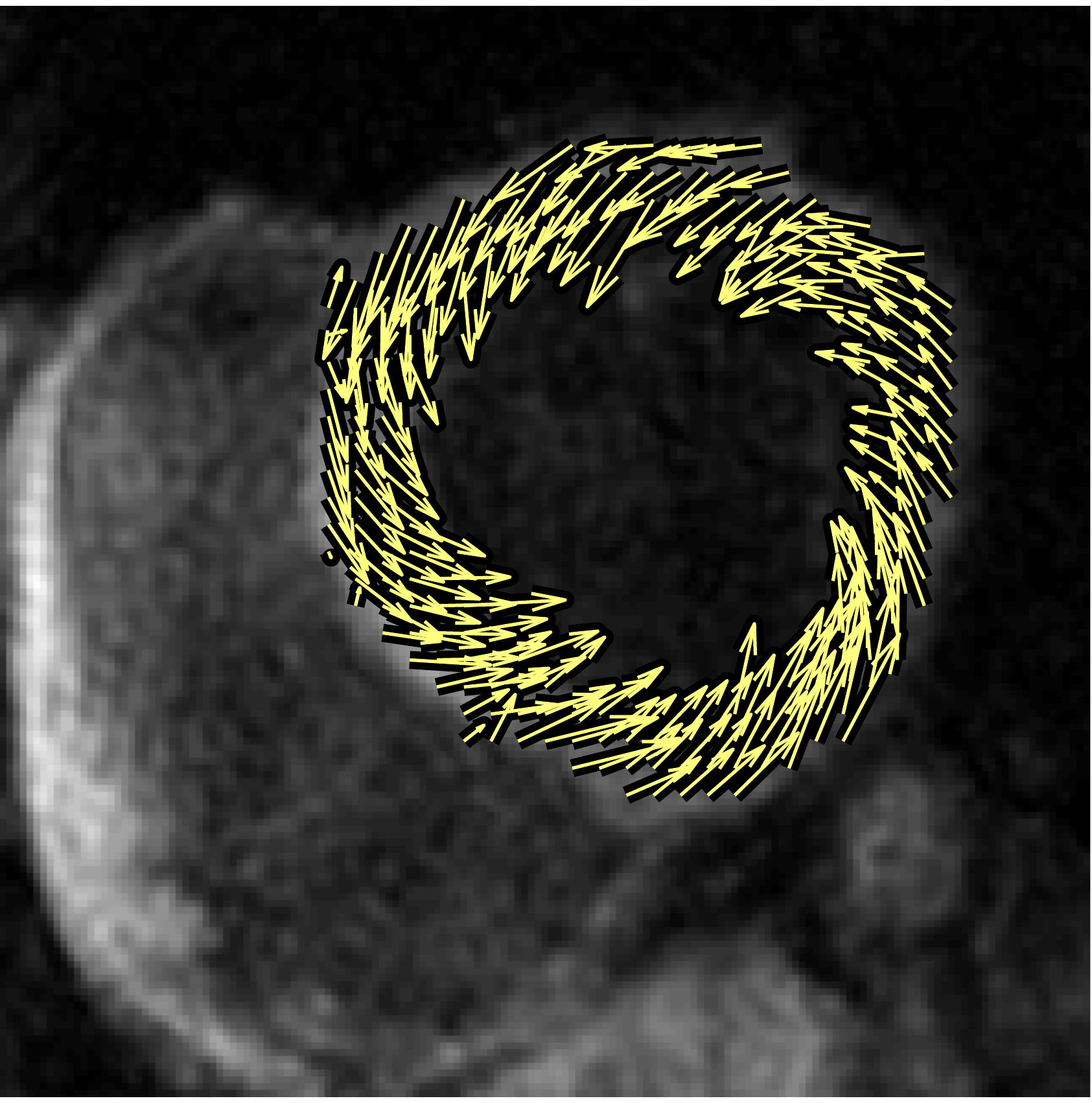}} \qquad
\subfloat[Uncorrected]{\includegraphics[trim={3cm 3.5cm 1cm 1cm},clip,width = 1in]{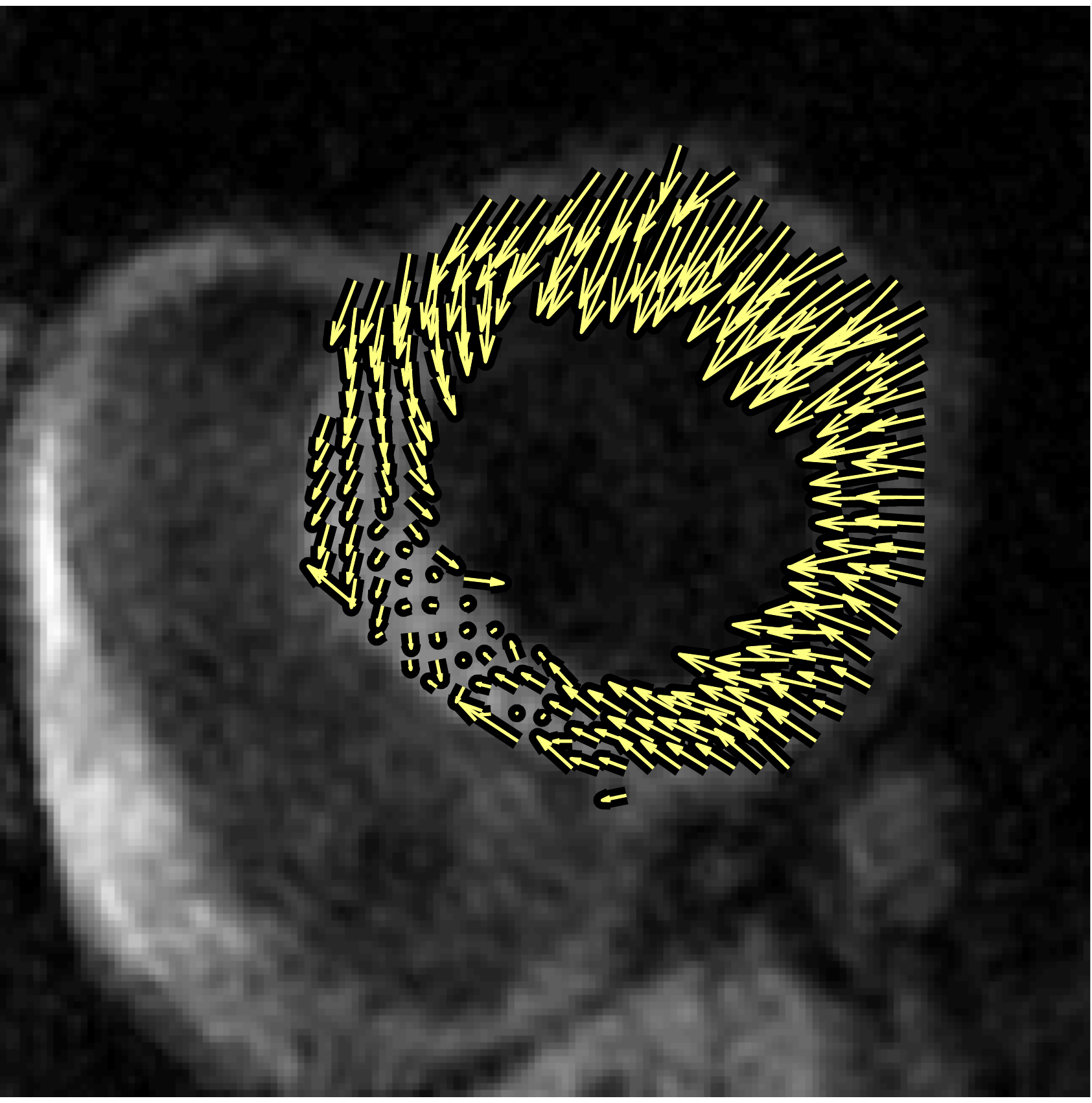}} \quad
\subfloat[Corrected]{\includegraphics[trim={3cm 3.5cm 1cm 1cm},clip,width = 1in]{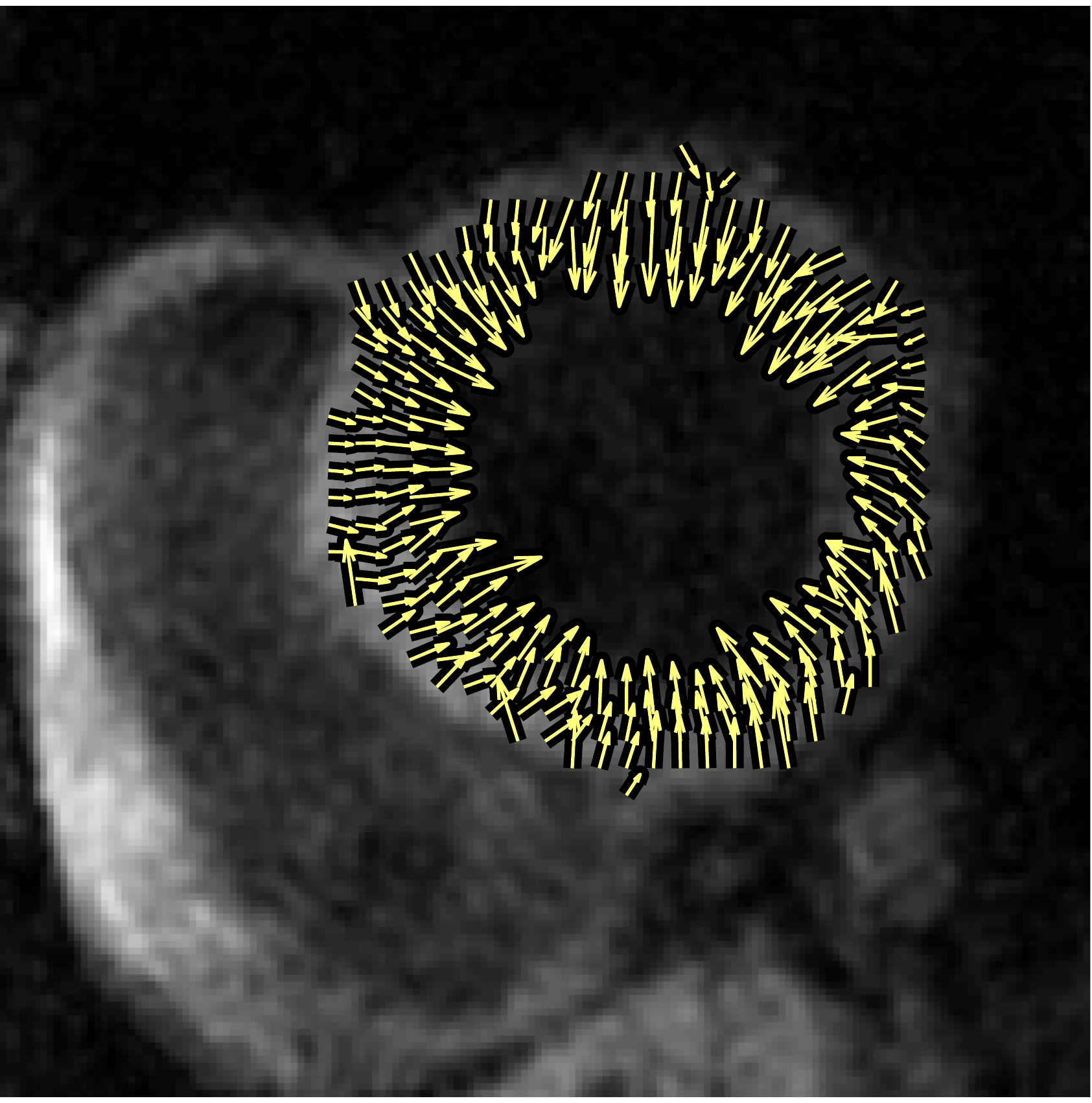}}
\vspace{-2px}
\flushleft (c) Basal slice at peak anticlockwise rotation (left) and peak contraction (right)

\caption{In-planar velocity field from an (a) apical (b) mid-ventricular and (c) basal short axis slice of a TPM velocity stack from a healthy volunteer at peak anti-clockwise rotation and peak contraction before and after translational motion compensation with method 1.}

\label{fig:2D_vectors}
\end{figure}

The proposed motion correction was subsequently applied to 8 slices and 50 frames of each of the 10 in vivo TPM datasets and a sample of this group of results is illustrated in Fig.\ref{fig:2D_vectors}. The figure shows the in-planar myocardial velocity field from an apical, mid-wall and basal slice of the LV at peak anti-clockwise rotation and peak contraction, before and after motion correction with method 1. Regions of enhanced velocities lying opposite to regions of suppressed velocities appear in the uncorrected velocity fields and, thus, indicate the presence of translational velocity on top of the velocity from deformation. Such distortions of velocity are not visible in the corrected images, indicating the effectiveness of the method. 

\begin{figure} 
\subfloat[Uncorrected]{\includegraphics[trim={3cm 3cm 3cm 2cm},clip,width = 1.5in]{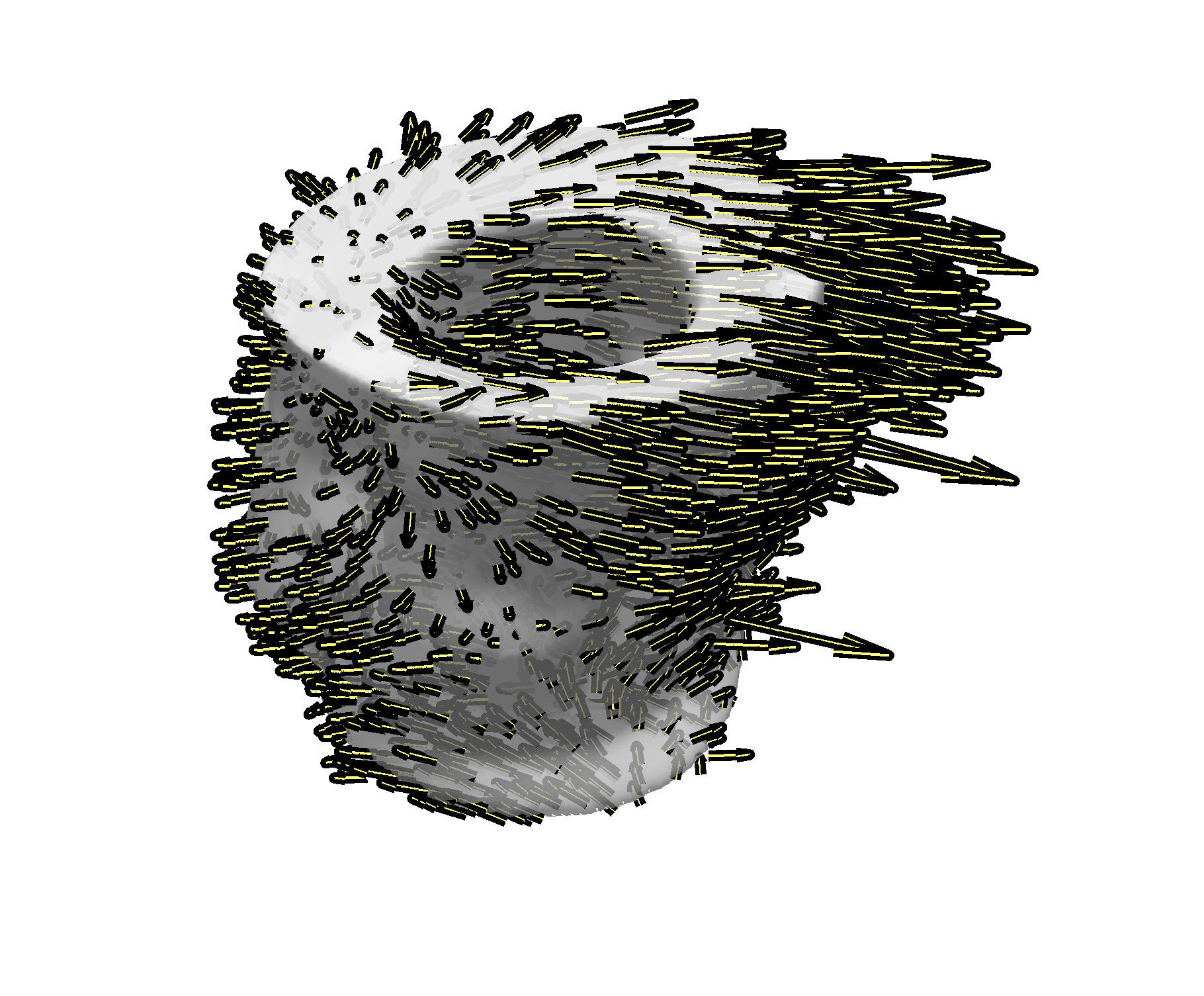}}
\subfloat[Corrected,method 2]{\includegraphics[trim={3cm 3cm 3cm 2cm},clip,width = 1.5in]{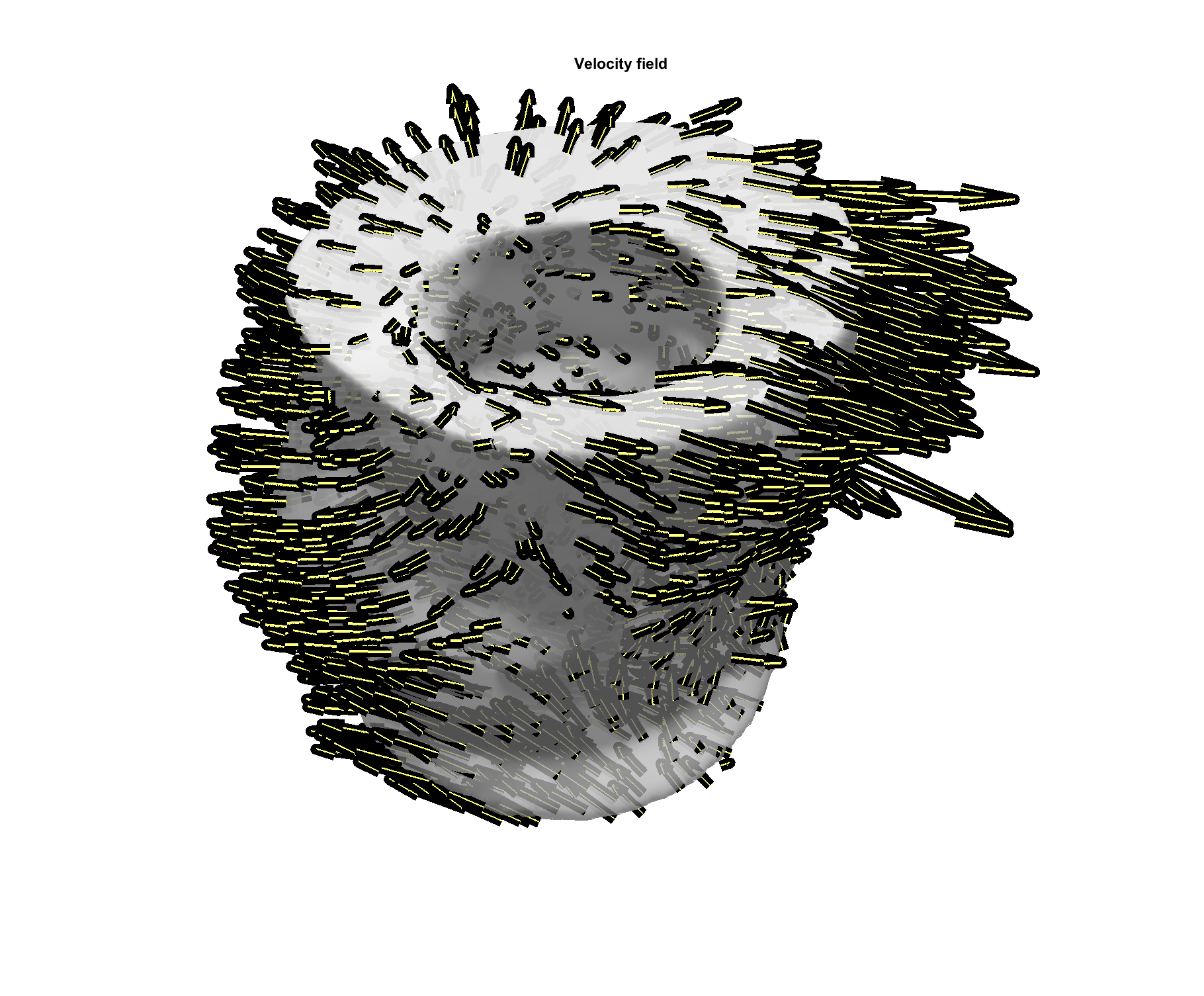}}
\subfloat[Corrected,method 1]{\includegraphics[trim={3cm 3cm 3cm 2cm},clip,width = 1.5in]{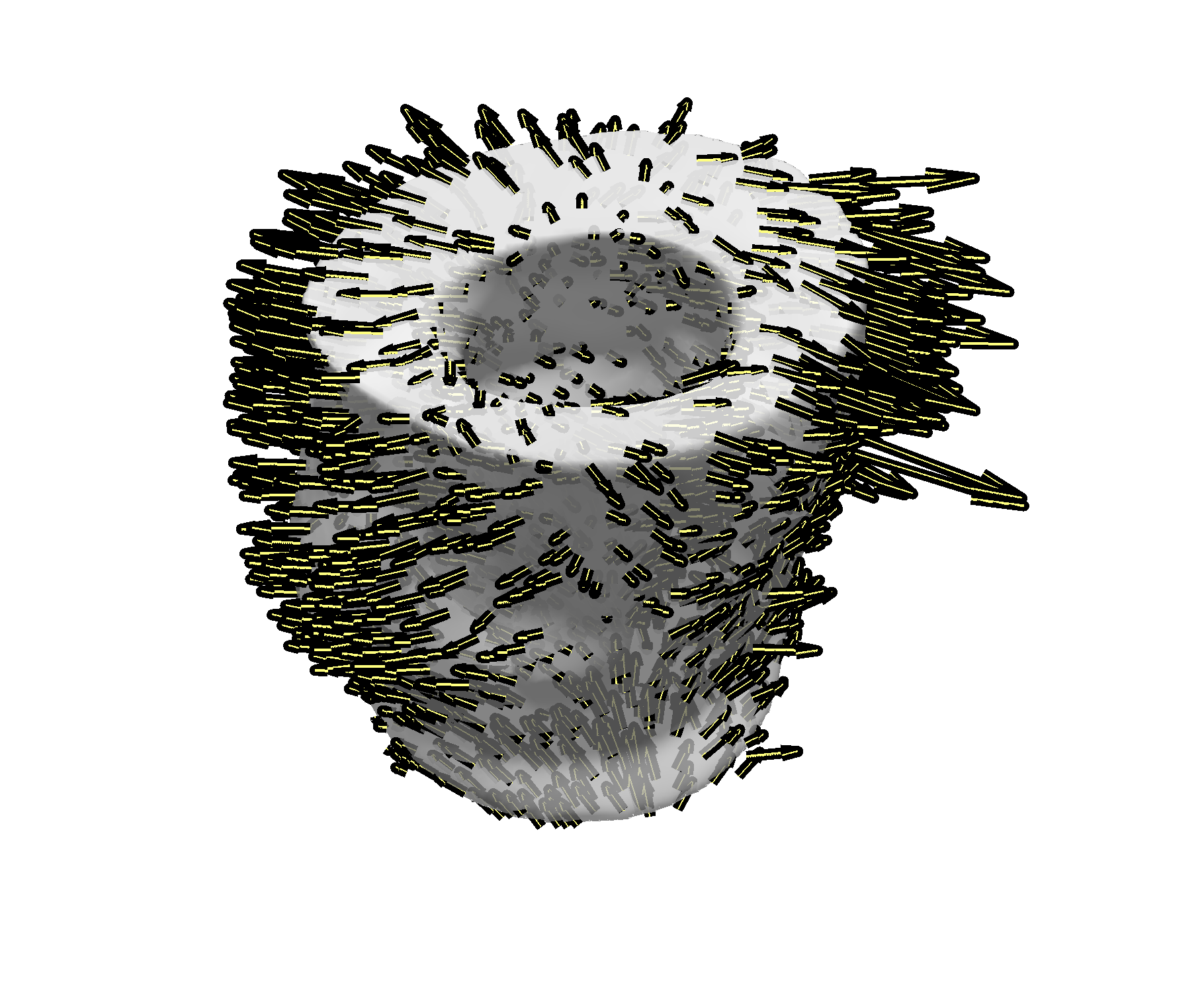}}

\caption{3D velocity field from a full stack of TPM data from a healthy subject at peak expansion (a) before motion correction , (b) after motion correction with method 2 and (c) method 3.}

\label{fig:3D_vectors}
\end{figure}

Fig. \ref{fig:3D_vectors} demonstrates a 3D visualisation of the 3D velocity field from a full stack of TPM at peak expansion (a) before translational motion compensation, (c) after before translational motion compensation with method 1 and (b) method 2. The 3D visualisation of the uncorrected stack shows a bias in the velocity field caused by translational velocity, whereas the corrected stacks look more balanced. This visualisation of results pictures the difference in the two methods: Method 1 yields a stack which looks overall more symmetric as correction has been applied slice by slice taking into account the centre of mass per slice. On the other hand method 2 yields a stack where the base seems to move in opposite direction to the mid-wall and apex, revealing a relative motion between the lower and upper part of the LV. Visualisation of the second stack is looks more realistic and method 2 is recommended when examining deformation from the entire volume. When looking at deformation slice by slice method 1 yields more intuitive results and it is recommended in that case.

\begin{figure} 
\subfloat{\includegraphics[width = 1.65in, height = 0.9in]{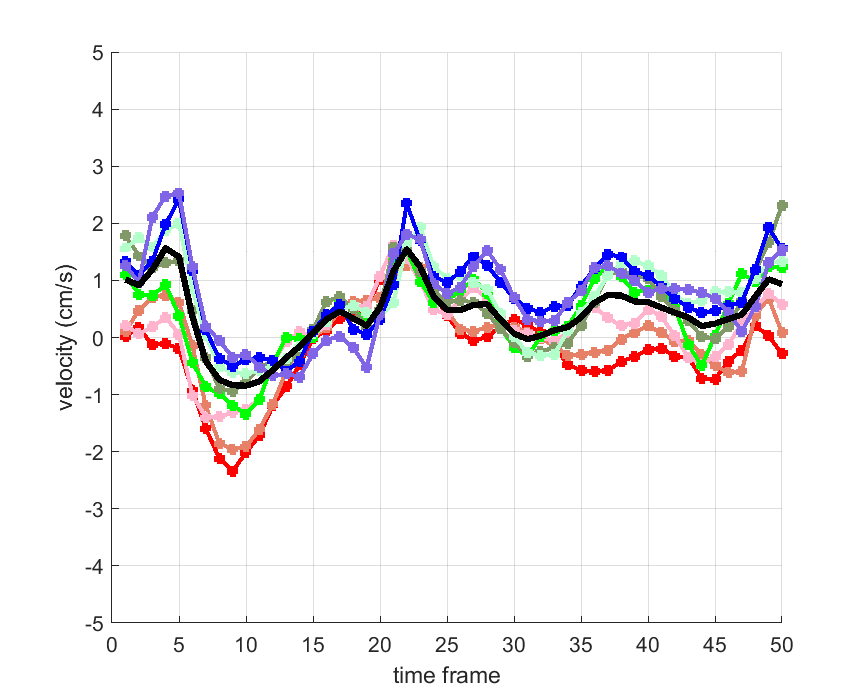}}  
\subfloat{\includegraphics[width = 1.65in, height = 0.9in]{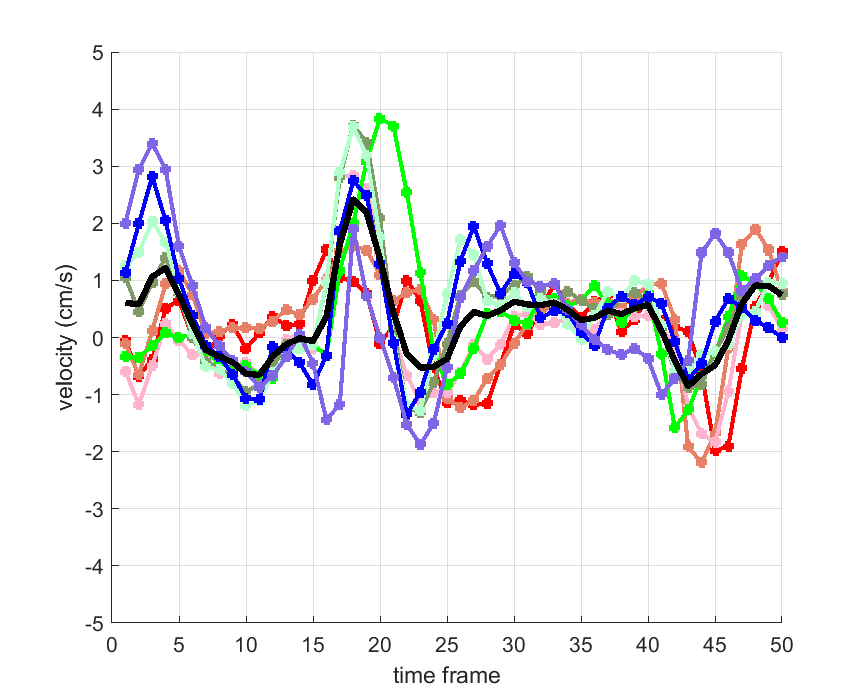}}  
\subfloat{\includegraphics[width = 1.65in, height = 0.9in]{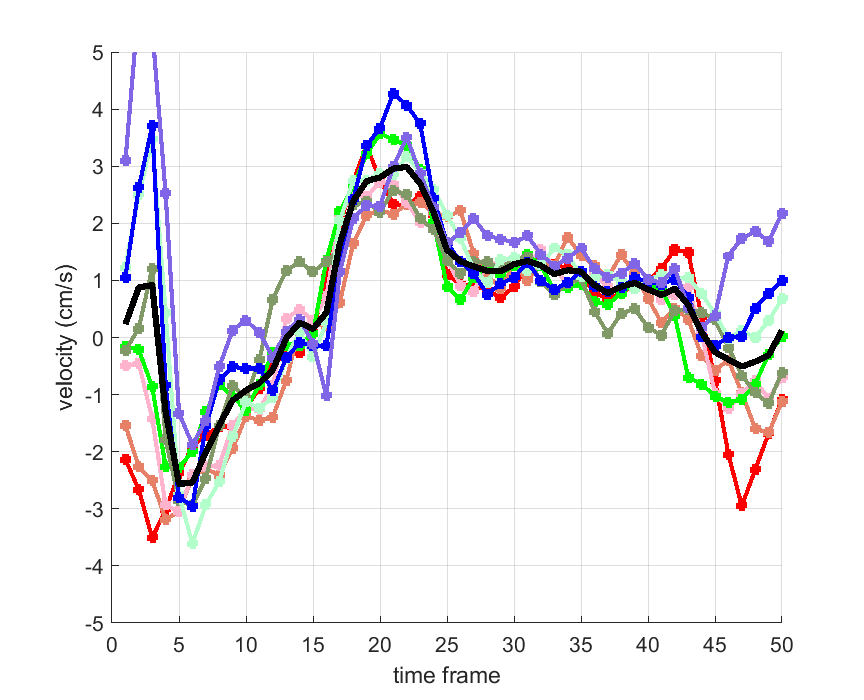}} 
\vspace{-10px}
\subfloat{\includegraphics[width = 1.65in, height = 0.9in]{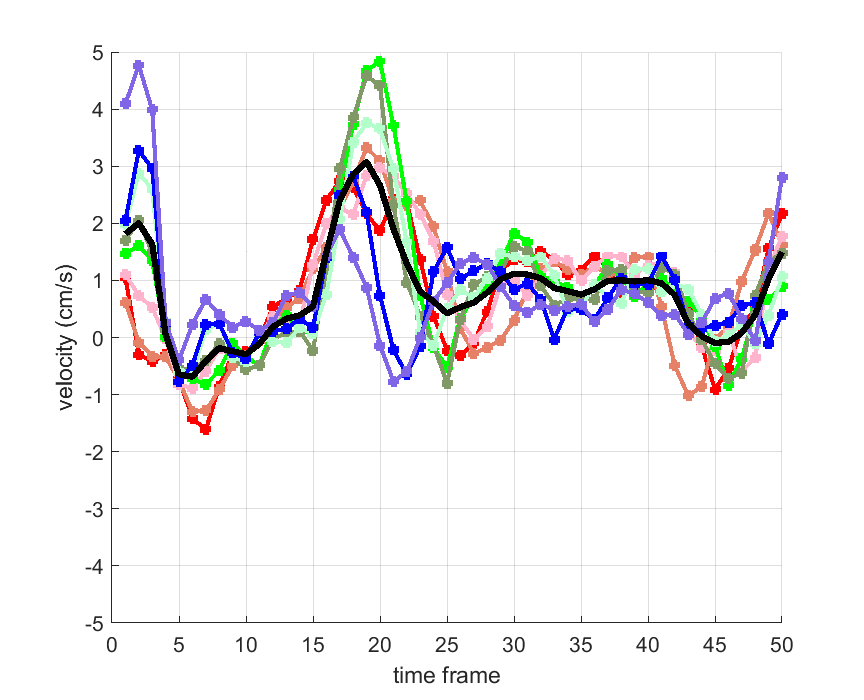}}  
\subfloat{\includegraphics[width = 1.65in, height = 0.9in]{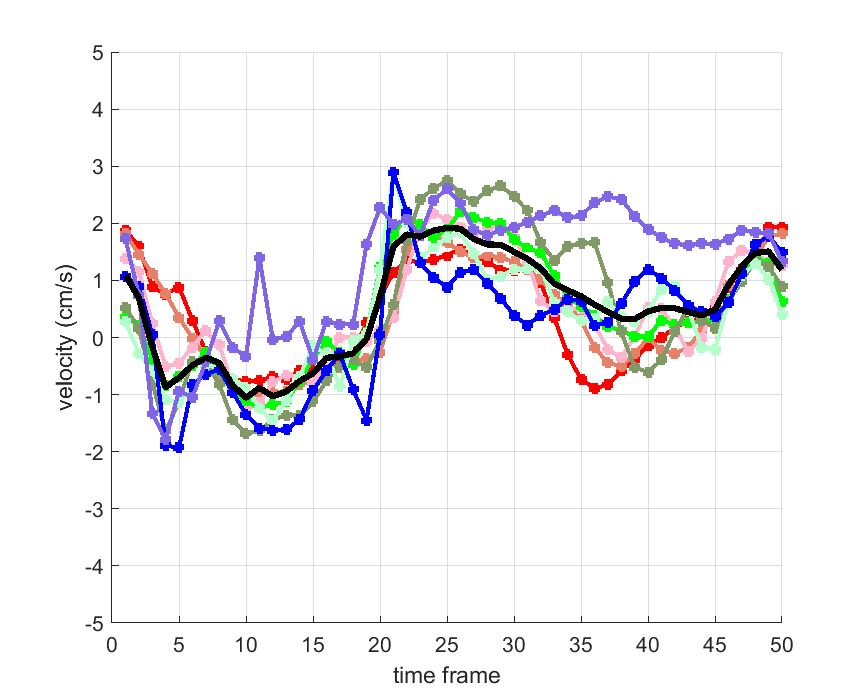}} 
\subfloat{\includegraphics[width = 1.65in, height = 0.9in]{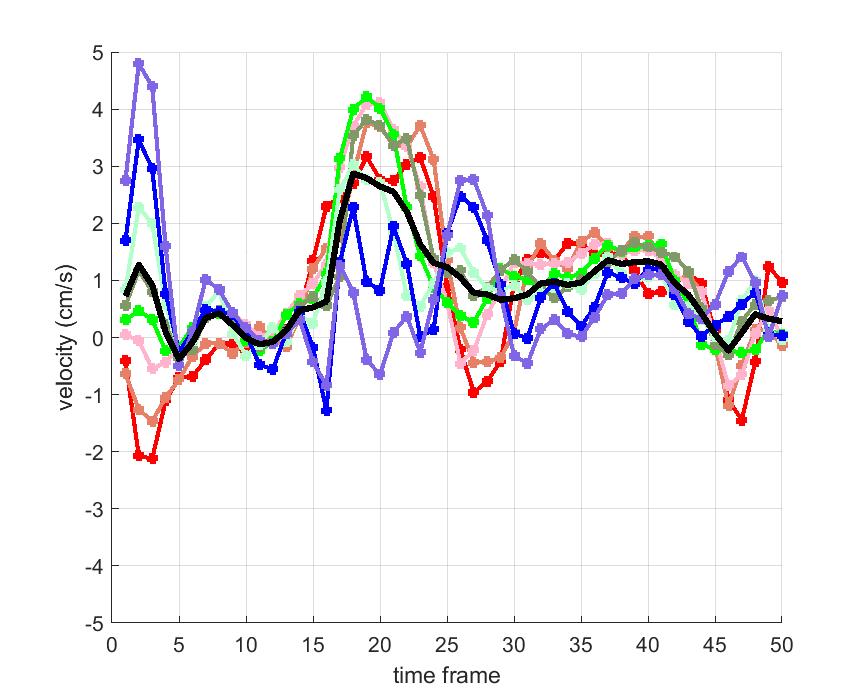}} 
\vspace{-10px}
\subfloat{\includegraphics[width = 1.65in, height = 0.9in]{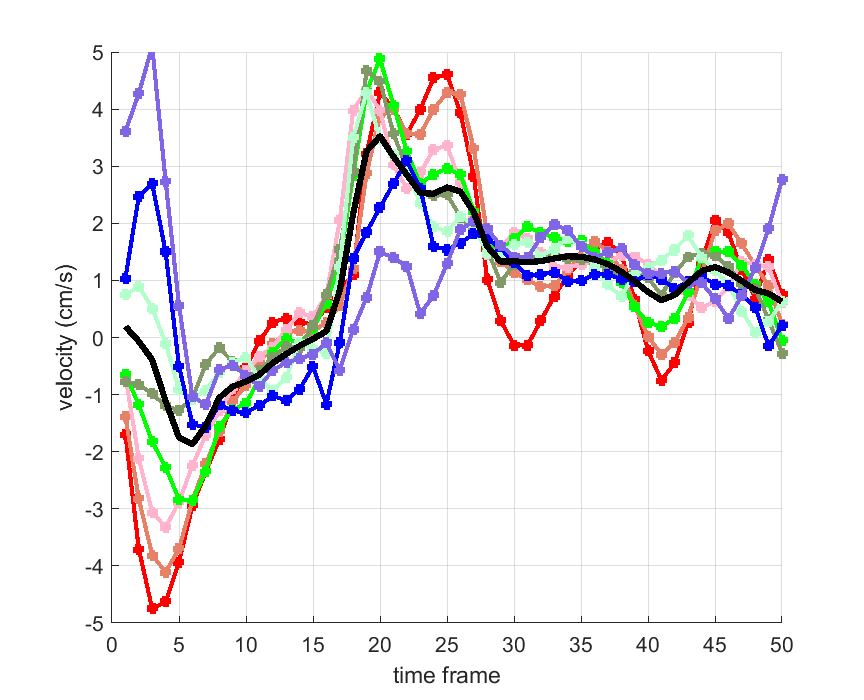}} 
\subfloat{\includegraphics[width = 1.65in, height = 0.9in]{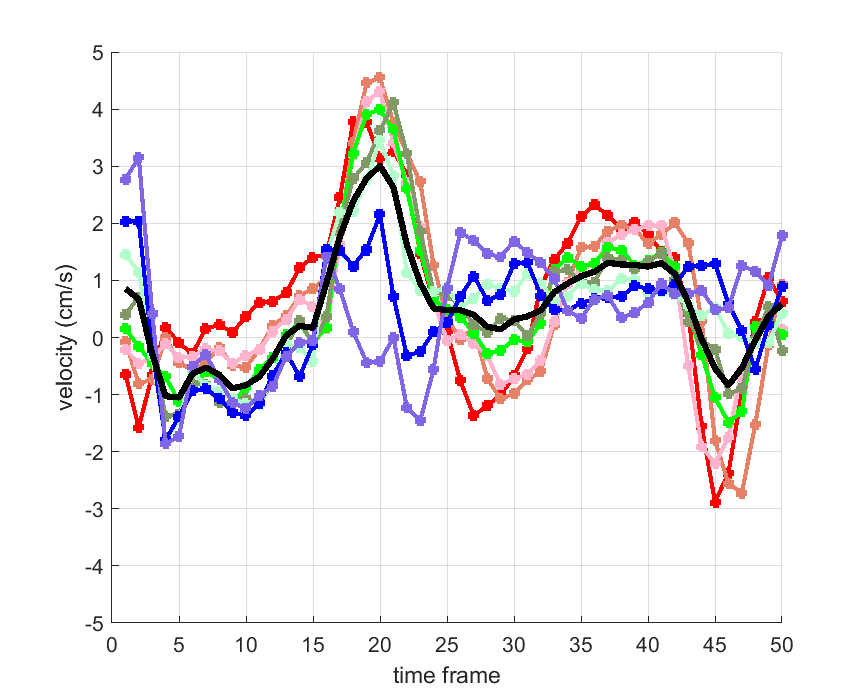}}  
\subfloat{\includegraphics[width = 1.65in, height = 0.9in]{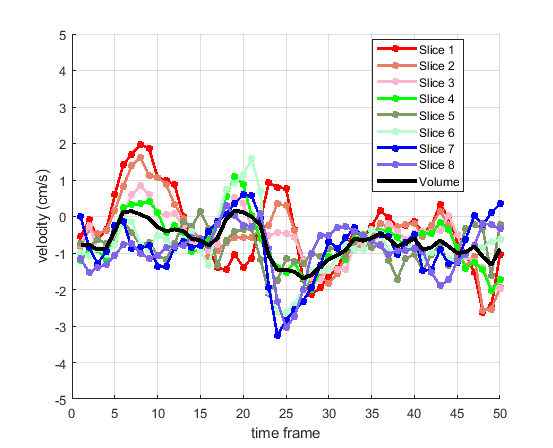}} 

\caption{Time courses of removed translational velocity component per slice with method 1 (coloured lines) and per volume with method 2 (black line) in x direction.}
\label{fig:time_coursesx}
\end{figure}

\begin{figure} 

\subfloat{\includegraphics[width = 1.65in, height = 0.9in]{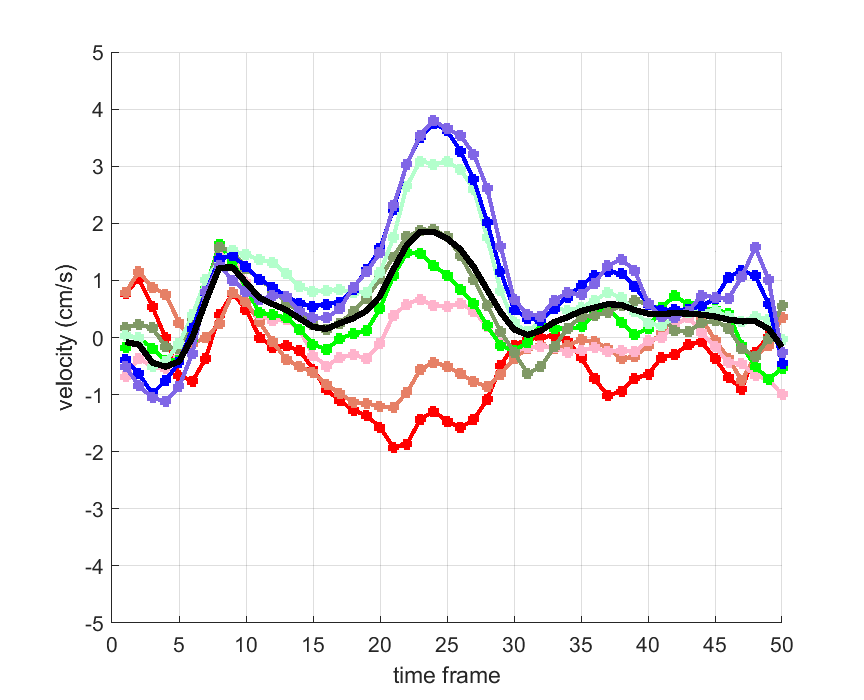}} 
\subfloat{\includegraphics[width = 1.65in, height = 0.9in]{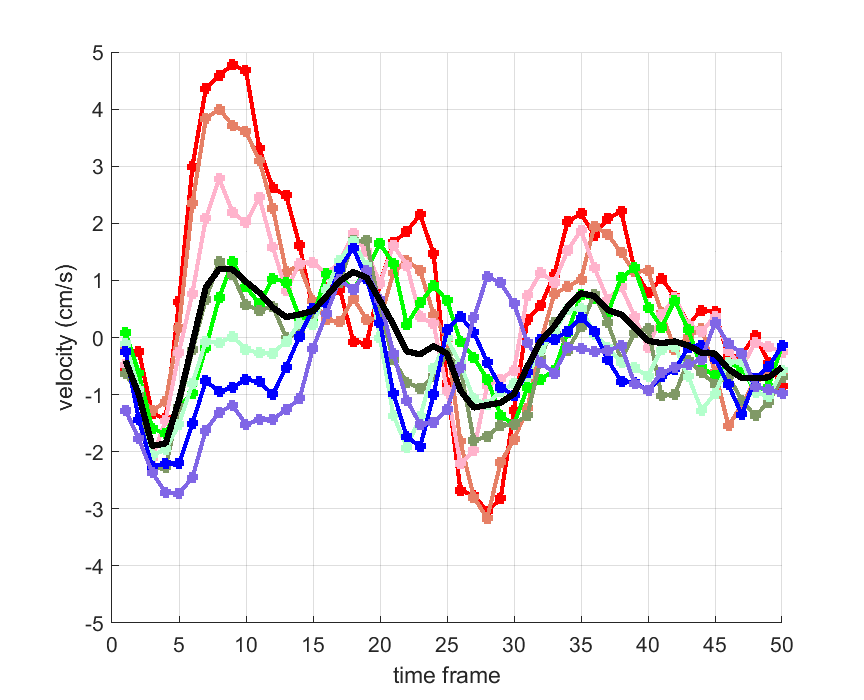}} 
\subfloat{\includegraphics[width = 1.65in, height = 0.9in]{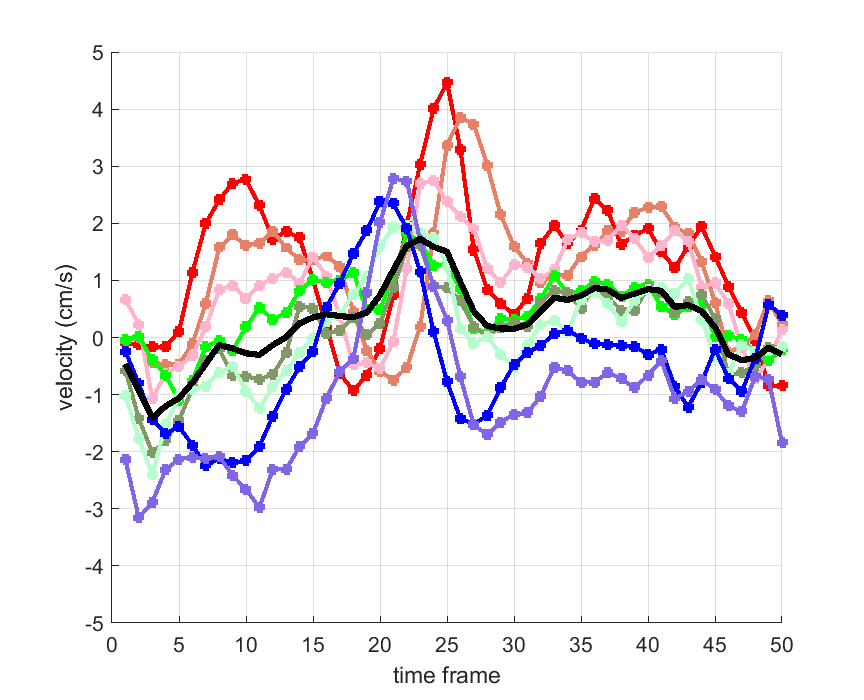}} 
\vspace{-10px}
\subfloat{\includegraphics[width = 1.65in, height = 0.9in]{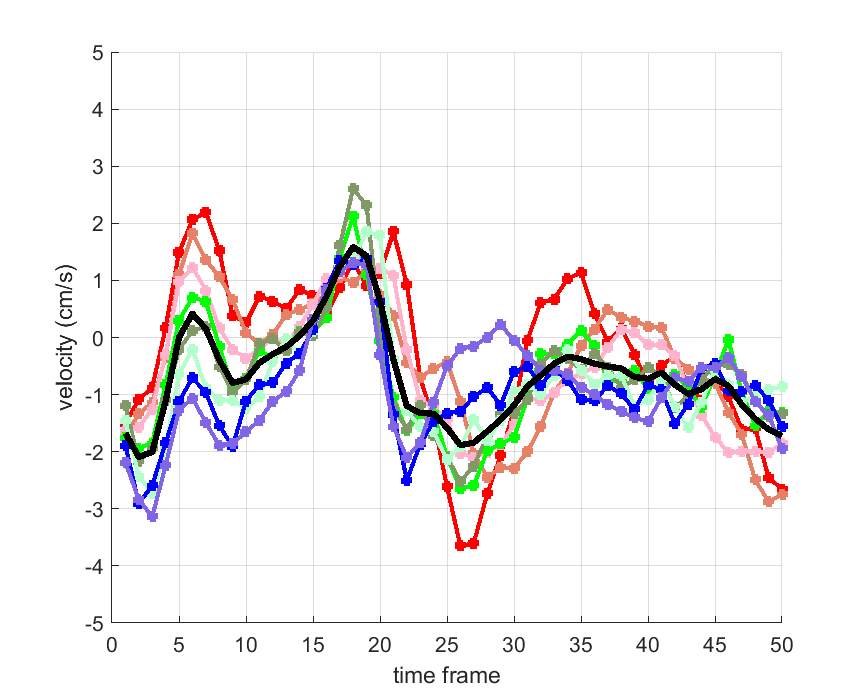}} 
\subfloat{\includegraphics[width = 1.65in, height = 0.9in]{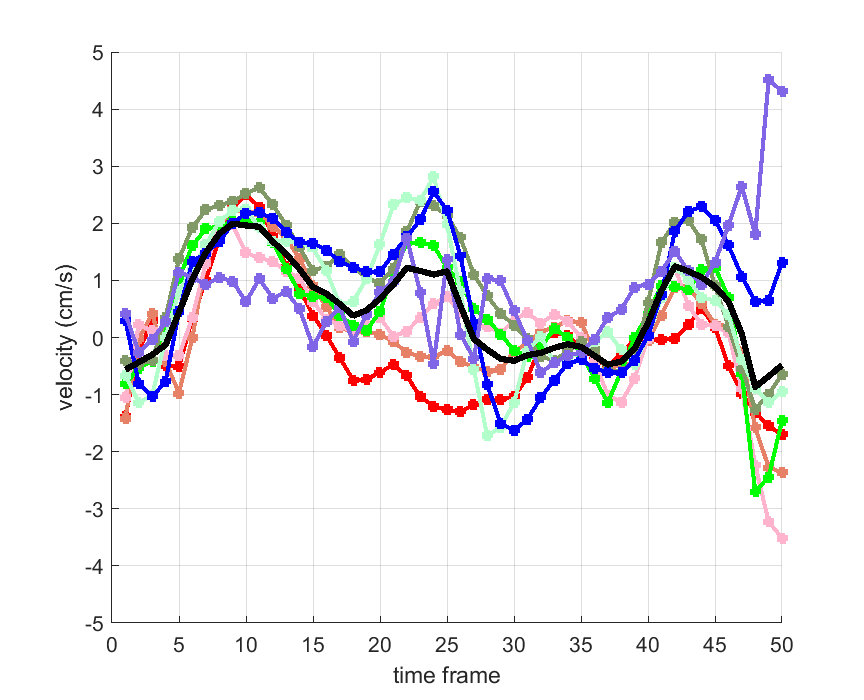}} 
\subfloat{\includegraphics[width = 1.65in, height = 0.9in]{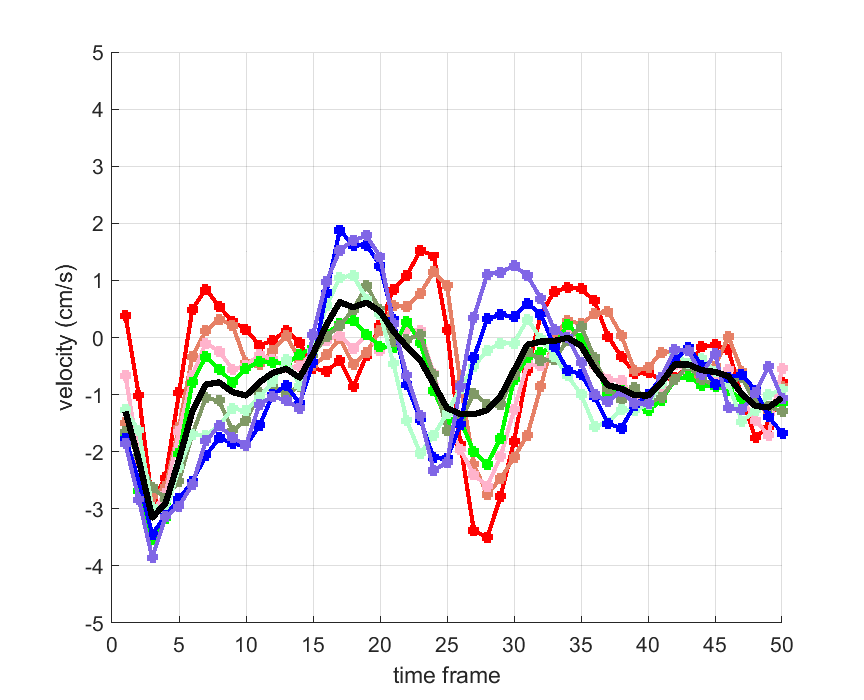}} 
\vspace{-10px}
\subfloat{\includegraphics[width = 1.65in, height = 0.9in]{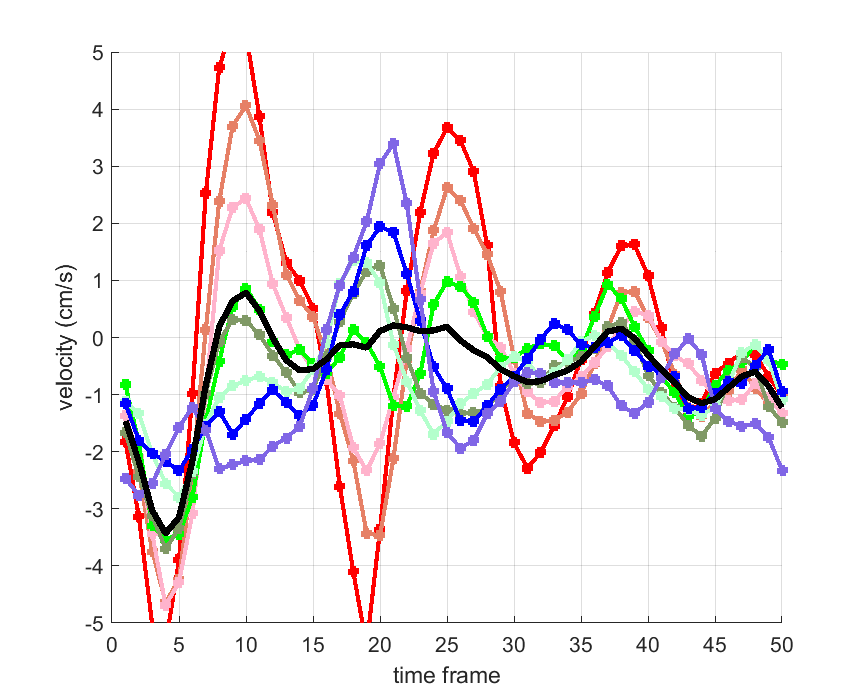}} 
\subfloat{\includegraphics[width = 1.65in, height = 0.9in]{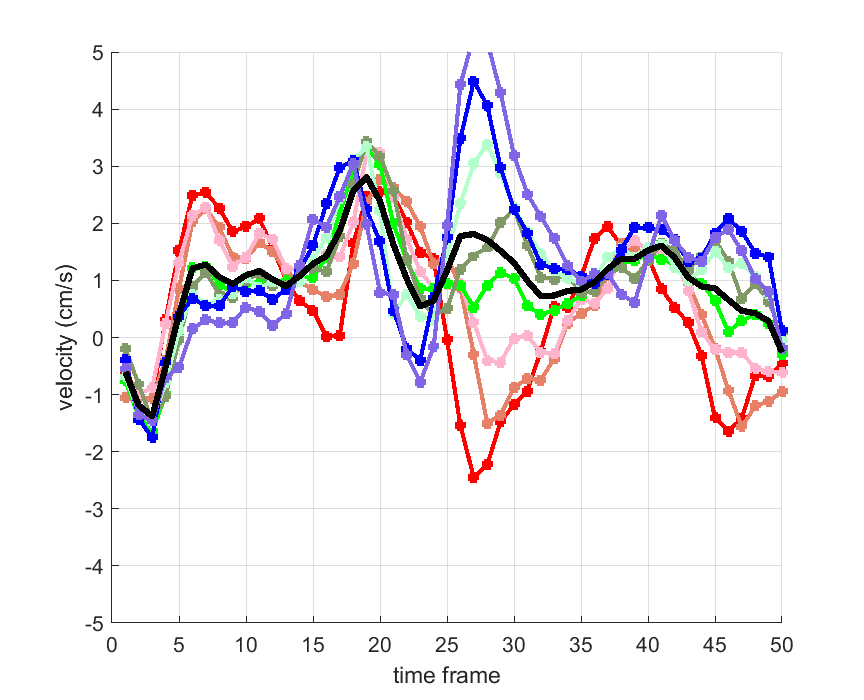}} 
\subfloat{\includegraphics[width = 1.65in, height = 0.9in]{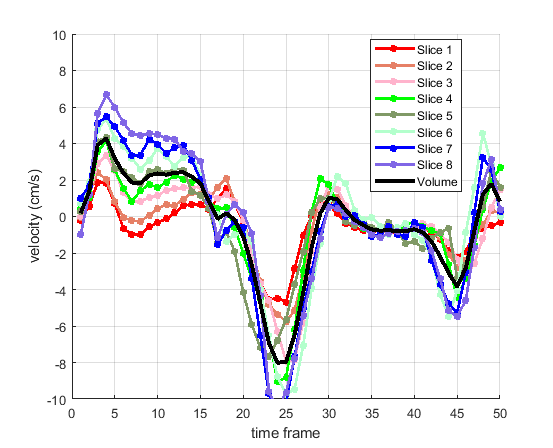}} 

\caption{Time courses of removed translational velocity component per slice with method 1 (coloured lines) and per volume with method 2 (black line) in y direction.}
\label{fig:time_coursesy}
\end{figure}

\begin{figure}
\subfloat{\includegraphics[width = 1.65in, height = 0.9in]{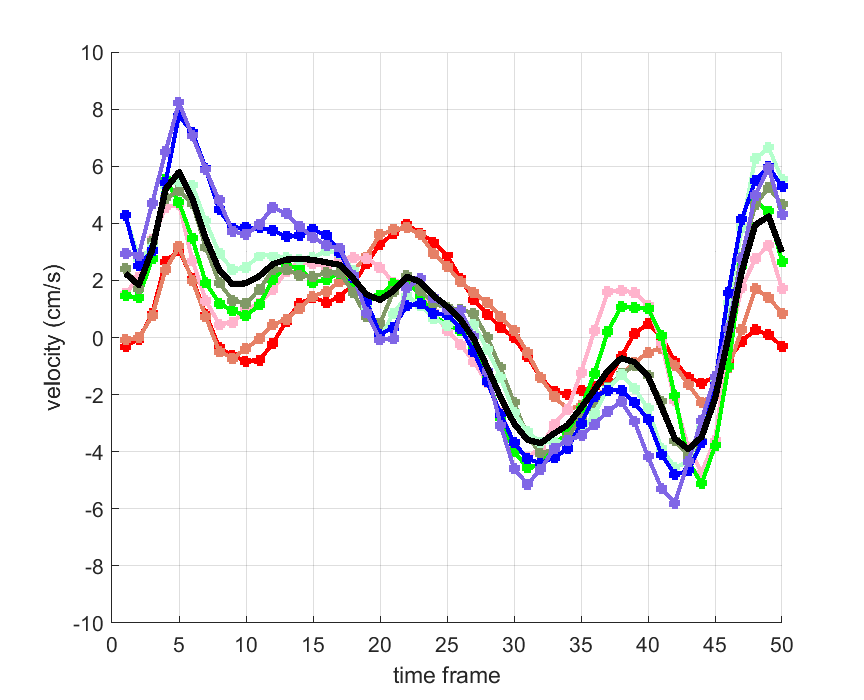}}
\subfloat{\includegraphics[width = 1.65in, height = 0.9in]{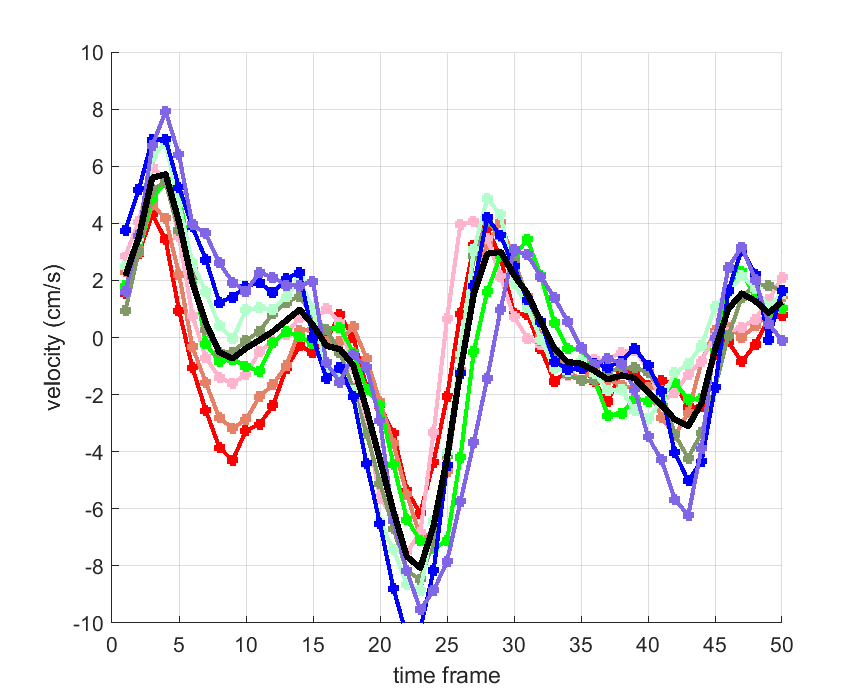}} 
\subfloat{\includegraphics[width = 1.65in, height = 0.9in]{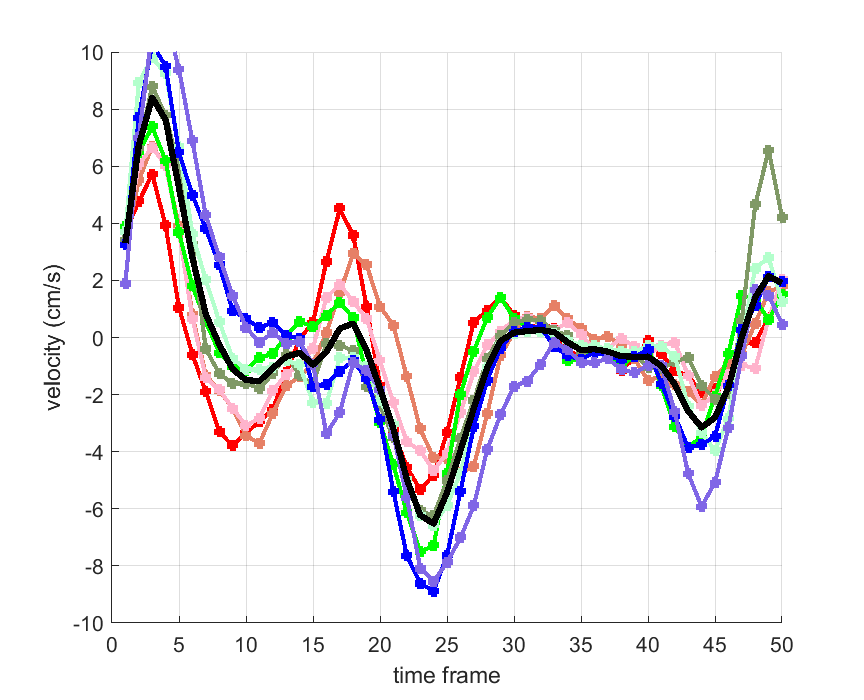}} 
\vspace{-10px}
\subfloat{\includegraphics[width = 1.65in, height = 0.9in]{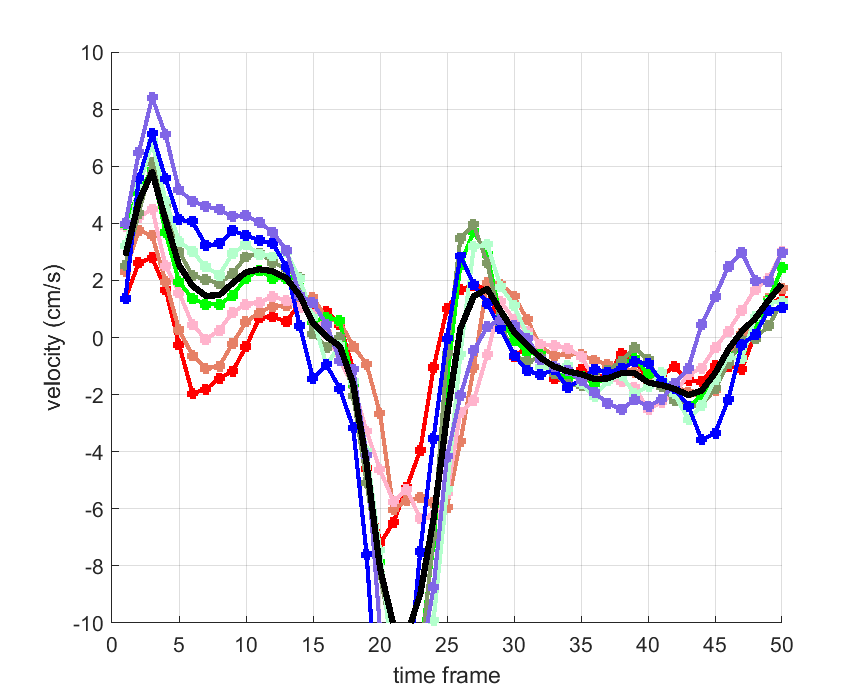}} 
\subfloat{\includegraphics[width = 1.65in, height = 0.9in]{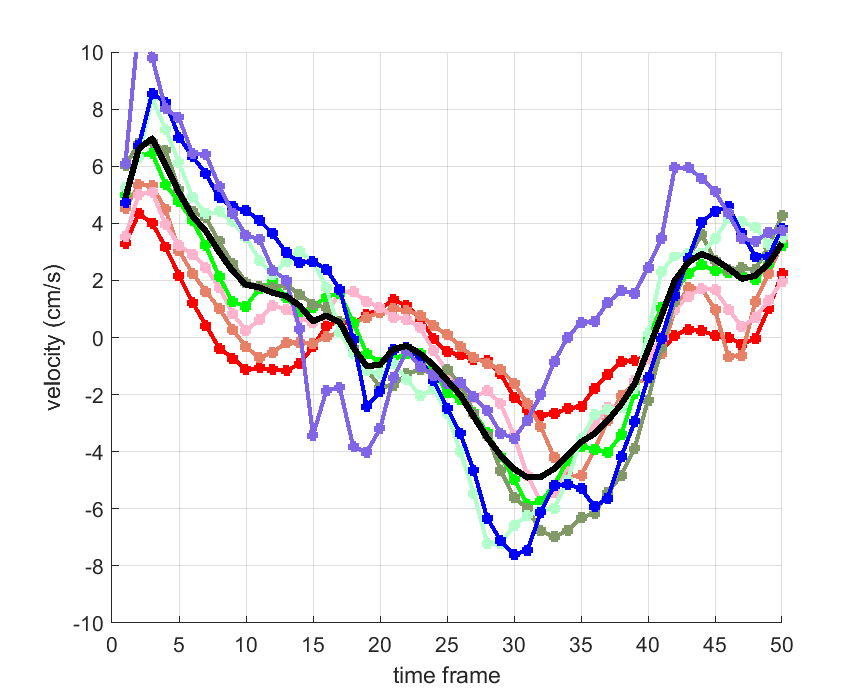}}
\subfloat{\includegraphics[width = 1.65in, height = 0.9in]{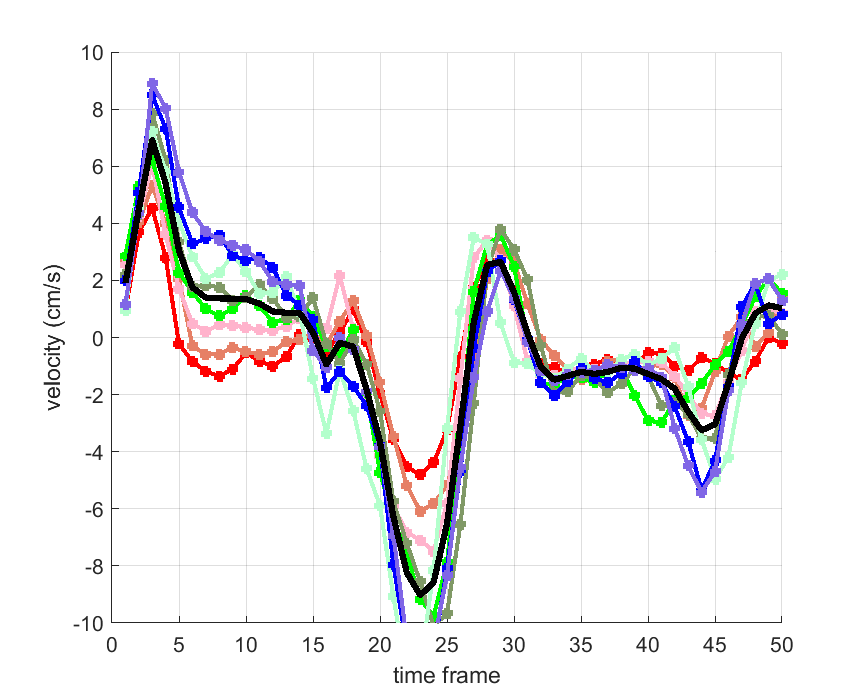}} 
\vspace{-10px}
\subfloat{\includegraphics[width = 1.65in, height = 0.9in]{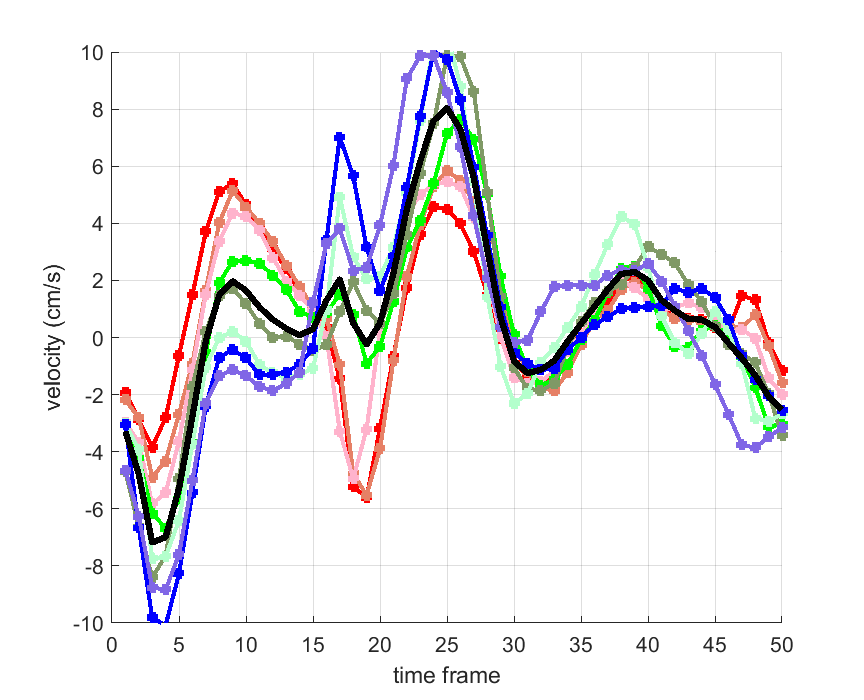}} 
\subfloat{\includegraphics[width = 1.65in, height = 0.9in]{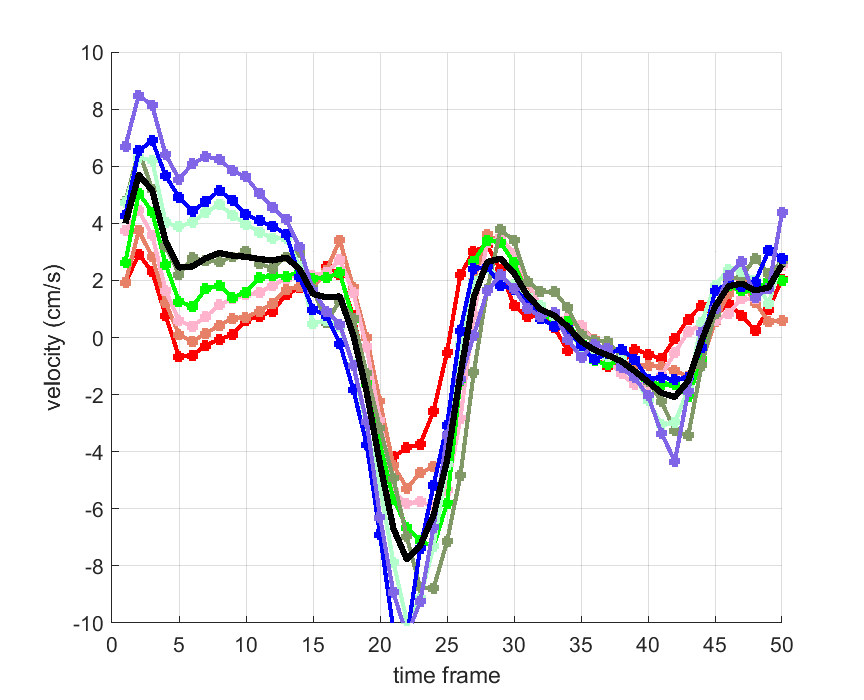}} 
\subfloat{\includegraphics[width = 1.65in, height = 0.9in]{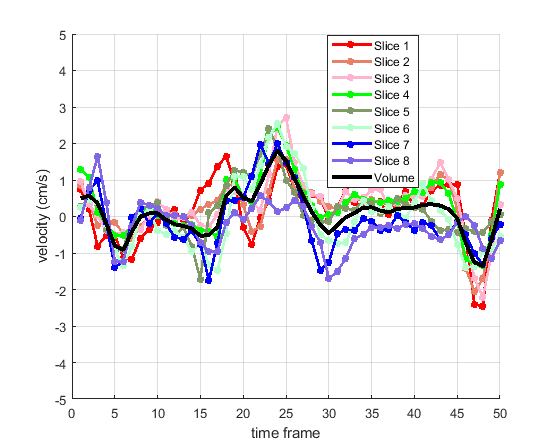}} 

\caption{Time courses of removed translational velocity component per slice with method 1 (coloured lines) and per volume with method 2 (black line) in z direction.}
\label{fig:time_coursesz}
\end{figure}

Time courses of the estimated translational component per subject, frame, slice and volume along the 3 velocity directions x, y and z are presented in Fig.\ref{fig:time_coursesx} - Fig.\ref{fig:time_coursesz}.

\section{Discussion and conclusions}
In this study we have proposed a novel, effective and robust method for motion correction of myocardial tissue velocities. Translational velocity of the left ventricle is estimated from a 2D or 3D region in the left ventricular myocardial wall and deducted from the total measured myocardial velocities. The selection of the region is critical in the efficiency of the method and it is important that it follows the deformation of the LV during the cardiac cycle. Furthermore, the selected region is independent from circumferential variations in the wall thickness that might occur in several myocardial conditions such as myocardial infarction or hibernation. The fundamental assumption in the proposed method states that the left ventricular deformation is governed by a degree of symmetry, which we have based on the observation of results from previously published studies on myocardial deformation. Accordingly, we have suggested two models for the distribution of the in-planar left ventricular myocardial velocities assuming planar symmetry for the one model and cylindrical symmetry for the other model. We used these models to generate synthetic velocity datasets and in-vivo TPM images to generate semi-synthetic datasets for the evaluation of the proposed motion correction method. The method showed excellent performance both on the synthetic and semi-synthetic datasets. The method was, subsequently, applied on TPM images from 10 healthy subjects and visual observation of the corrected velocity field 2D and 3D visualisations  indicated similarly remarkable performance.
 
 Motion correction was applied on a per slice and a per volume basis, which is equivalent to velocity transformation from the scanner's reference coordinate system to a local coordinate system that follows the slice in the first case, and the volume of the LV in the second case. The per slice and per volume estimated translational velocities can combine for the estimation of further motion parameters, for instance the rotation of the LV; rotation and per volume translation together describe the rigid body motion of the LV, if image artefacts are dismissed. After rigid body motion compensation, the residual velocities are attributed to myocardial tissue deformation. Although strain and strain rate are most established measures for the description of regional myocardial deformation, their derivation from the commonly used imaging modalities is followed by notable enhancement of the image noise and artefacts, which severely interferes with the generated strain and strain rate maps. For meaningful deformation maps, the strain and strain rate values are usually averaged over larger groups of pixels or segments and this has the consequence to overlie finer details of myocardial deformation and potential abnormalities. We propose the corrected velocity maps as an alternative intuitive representation of regional deformation at a per pixel level, which is significantly less affected by image noise and artefacts. Along with the translational motion, the proposed method corrects also for spatially slowly varying TPM image artefacts like the velocity offset errors and respiratory motion artefacts that affect primarily the free-breathing acquisitions. \newline

The proposed translational motion compensation method is non-invasive, computationally inexpensive and without prior information requirements. For this study, the method has been evaluated on cardiac MRI from the LV primarily developed for cardiac disease diagnosis. The applicability of the method is appraised quite broadly though, as it solves the generic engineering problem of motion correction. It is suited to any deforming soft tissue that has annular cross sections and fulfils the symmetry requirements. It also applies to velocity fields captured by any imaging modality or other technique beyond the bounds of TPM. The method can potentially benefit intraoperative image guidance and tracking, for instance in cardiac interventions, as it permits real-time processing. However, there is a certain amount of research to be done towards this end, in order to allow for real-time acquisitions of high quality tissue velocity images. 

\FloatBarrier
\bibliographystyle{IEEEtran} 

\begin{thebibliography}{10}
\providecommand{\url}[1]{#1}
\csname url@samestyle\endcsname
\providecommand{\newblock}{\relax}
\providecommand{\bibinfo}[2]{#2}
\providecommand{\BIBentrySTDinterwordspacing}{\spaceskip=0pt\relax}
\providecommand{\BIBentryALTinterwordstretchfactor}{4}
\providecommand{\BIBentryALTinterwordspacing}{\spaceskip=\fontdimen2\font plus
\BIBentryALTinterwordstretchfactor\fontdimen3\font minus
  \fontdimen4\font\relax}
\providecommand{\BIBforeignlanguage}[2]{{%
\expandafter\ifx\csname l@#1\endcsname\relax
\typeout{** WARNING: IEEEtran.bst: No hyphenation pattern has been}%
\typeout{** loaded for the language `#1'. Using the pattern for}%
\typeout{** the default language instead.}%
\else
\language=\csname l@#1\endcsname
\fi
#2}}
\providecommand{\BIBdecl}{\relax}
\BIBdecl

\bibitem{Mountney2010}
P.~Mountney, D.~Stoyanov, and G.~Z. Yang, ``{Three-dimensional tissue
  deformation recovery and tracking},'' \emph{IEEE Signal Process. Mag.},
  vol.~27, no.~4, pp. 14--24, 2010.

\bibitem{Meinzer2008}
H.~P. Meinzer, L.~Maier-Hein, I.~Wegner, M.~Baumhauer, and I.~Wolf,
  ``{Computer-assisted soft tissue interventions},'' \emph{2008 5th IEEE Int.
  Symp. Biomed. Imaging From Nano to Macro, Proceedings, ISBI}, no.~i, pp.
  1391--1394, 2008.

\bibitem{Yip2012}
M.~C. Yip, D.~G. Lowe, S.~E. Salcudean, R.~N. Rohling, and C.~Y. Nguan,
  ``{Tissue tracking and registration for image-guided surgery},'' \emph{IEEE
  Trans. Med. Imaging}, vol.~31, no.~11, pp. 2169--2182, 2012.

\bibitem{Nayak2015}
\BIBentryALTinterwordspacing
K.~S. Nayak, J.-F. Nielsen, M.~A. Bernstein, M.~Markl, P.~{D. Gatehouse},
  R.~{M. Botnar}, D.~Saloner, C.~Lorenz, H.~Wen, B.~{S. Hu}, F.~H. Epstein,
  J.~{N. Oshinski}, and S.~V. Raman, ``{Cardiovascular magnetic resonance phase
  contrast imaging},'' \emph{J. Cardiovasc. Magn. Reson.}, vol.~17, no.~1,
  p.~71, 2015. [Online]. 
\BIBentrySTDinterwordspacing

\bibitem{Ferreira2013}
\BIBentryALTinterwordspacing
P.~F. Ferreira, P.~D. Gatehouse, R.~H. Mohiaddin, and D.~N. Firmin,
  ``{Cardiovascular magnetic resonance artefacts},'' \emph{J. Cardiovasc. Magn.
  Reson.}, vol.~15, no.~1, p.~41, 2013. [Online]. 
\BIBentrySTDinterwordspacing

\bibitem{Koutsoumpa2015}
C.~Koutsoumpa, R.~Simpson, J.~Keegan, D.~Firmin, and G.-Z. Yang, ``{Restoration
  of Phase-Contrast Cardiovascular MRI for the Construction of Cardiac
  Contractility Atlases},'' in \emph{5th Int. Work. STACOM 2014, Held
  Conjunction with MICCAI 2014}, O.~C. Oscar.camara@upf.edu, T.~M.
  Tommaso.mansi@siemens.com, M.~P. Mihaela.pop@utoronto.ca, K.~R.
  Kawal.rhode@kcl.ac.uk, M.~S. Maxime.sermesant@inria.fr, and {Alistair Young
  a.young@auckland.ac.nz}, Eds.\hskip 1em plus 0.5em minus 0.4em\relax Boston,
  MA: Springer International Publishing, 2015, pp. 275--283.

\bibitem{Codreanu2014}
I.~Codreanu, T.~J. Pegg, J.~B. Selvanayagam, M.~D. Robson, O.~J. Rider, C.~A.
  Dasanu, B.~A. Jung, D.~P. Taggart, S.~J. Golding, K.~Clarke, and C.~J.
  Holloway, ``{Normal values of regional and global myocardial wall motion in
  young and elderly individuals using navigator gated tissue phase mapping},''
  \emph{Age (Omaha).}, vol.~36, no.~1, pp. 231--241, 2014.

\bibitem{Augustine2013}
\BIBentryALTinterwordspacing
D.~Augustine, A.~J. Lewandowski, M.~Lazdam, A.~Rai, J.~Francis, S.~Myerson,
  A.~Noble, H.~Becher, S.~Neubauer, S.~E. Petersen, and P.~Leeson, ``{Global
  and regional left ventricular myocardial deformation measures by magnetic
  resonance feature tracking in healthy volunteers: comparison with tagging and
  relevance of gender},'' \emph{J. Cardiovasc. Magn. Reson.}, vol.~15, no.~1,
  p.~8, 2013. [Online].
\BIBentrySTDinterwordspacing

\bibitem{Jung2012}
\BIBentryALTinterwordspacing
B.~Jung, K.~E. Odening, E.~Dall'Armellina, D.~F{\"{o}}ll, M.~Menza, M.~Markl,
  and J.~E. Schneider, ``{A quantitative comparison of regional myocardial
  motion in mice, rabbits and humans using in-vivo phase contrast CMR.}''
  \emph{J. Cardiovasc. Magn. Reson.}, vol.~14, p.~87, 2012. [Online].
 \BIBentrySTDinterwordspacing

\bibitem{Foll2009}
\BIBentryALTinterwordspacing
D.~F{\"{o}}ll, B.~Jung, F.~Staehle, E.~Schilli, C.~Bode, J.~Hennig, and
  M.~Markl, ``{Visualization of multidirectional regional left ventricular
  dynamics by high-temporal-resolution tissue phase mapping.}'' \emph{J. Magn.
  Reson. Imaging}, vol.~29, no.~5, pp. 1043--52, may 2009. [Online]. 
\BIBentrySTDinterwordspacing

\bibitem{Jung2006c}
\BIBentryALTinterwordspacing
B.~Jung, M.~Markl, D.~F{\"{o}}ll, and J.~Hennig, ``{Investigating myocardial
  motion by MRI using tissue phase mapping.}'' \emph{Eur. J. Cardiothorac.
  Surg.}, vol. 29 Suppl 1, pp. S150--7, apr 2006. [Online]. 
\BIBentrySTDinterwordspacing

\bibitem{Simpson2013a}
\BIBentryALTinterwordspacing
R.~Simpson, J.~Keegan, P.~Gatehouse, M.~Hansen, and D.~Firmin, ``{Spiral tissue
  phase velocity mapping in a breath-hold with non-cartesian SENSE},'' oct
  2013. [Online]. 
\BIBentrySTDinterwordspacing

\end{thebibliography}

\end{document}